\documentclass[fleqn,final,3p,11pt]{elsarticle}
\usepackage{soul} 
\usepackage{graphicx,color} 
\usepackage{subfigure}
\usepackage{hyperref}   
\hypersetup{
	colorlinks=true,
	linkcolor=blue,
	citecolor=blue,
	urlcolor=blue,
}
\usepackage{booktabs}
\usepackage{amsmath, amssymb, amsfonts}
\usepackage{graphicx, color, cleveref}
\usepackage{svg}
\usepackage{pgfplots}
\usepackage{comment}
\usepackage{multirow}
\usepackage{subcaption}
\pgfplotsset{compat=1.17}
\usepackage{tikz}
\usepackage{tikz-3dplot} 
\usepackage{pgfmath}
\usetikzlibrary{3d, calc}

\usepackage{amsmath, amssymb, array, booktabs, colortbl, xcolor, geometry}
\newcommand{\R}{\mathbb{R}}

\newcommand{\xx}{\mathbf{x}}
\newcommand{\yy}{\mathbf{y}}
\newcommand{\pp}{\mathbf{p}}

\newcommand{\uu}{\mathbf{u}}
\newcommand{\ff}{\mathbf{f}}
\newcommand{\FF}{\mathbf{F}}
\newcommand{\Bb}{\mathbf{b}}

\newcommand{\xiB}{\pmb{\xi}}

\newcommand{\etaB}{\pmb{\eta}}

\newcommand{\abs}[1]{\left\lvert#1\right\rvert}

\definecolor{antiquebrass}{rgb}{0.8, 0.58, 0.46}
 	\definecolor{mediumtaupe}{rgb}{0.4, 0.3, 0.28}
\journal{Computer Methods in Applied Mechanics and Engineering}
\begin{document}
\begin{frontmatter}
\title{Crushing, Comminution and Fracture: Extreme Particle Deformation in Three-Dimensional Granular Aggregates }

\author[grinnell]{Debdeep Bhattacharya}\fnref{fn1}
\author[lsu]{Davood Damircheli}\fnref{fn1}
\author[lsu]{Robert P. Lipton}\fnref{fn1}
\fntext[fn1]{Authors are listed in alphabetical order.}
\address[grinnell]{Department of Mathematics, Grinnell College, USA}
\address[lsu]{Department of Mathematics, Louisiana State University, USA}

\begin{abstract}
 We present a high-fidelity three dimensional computational framework for simulating the bulk mechanical behavior of granular aggregates composed of deformable  brittle grains. Departing from classical discrete element methods (DEM), our approach captures both inter-particle and intra-particle deformation using a nonlocal continuum formulation based on peridynamics. Each grain is individually meshed from level-set representations, enabling accurate modeling of elastic response and autonomous fracture evolution without requiring explicit crack tracking or fragment reconstruction. We validate the method through benchmark simulations, including the Kalthoff-Winkler fracture test, crushing of hollow spheres, and compound impact-crushing scenarios. The framework is further applied to large aggregates of up to 1000 sand grains of irregular shapes reconstructed from three dimensional X-ray computed tomography. Simulations reveal convergence of bulk stress response under compression, suggesting the feasibility of constructing representative volume elements (RVEs) for multiscale modeling. Finally, we investigate the role of grain geometry and topology on the macroscopic strength of the aggregate, providing insight into microstructure-driven failure mechanisms.  The framework exhibits excellent strong and weak scaling behavior, with simulations executed on up to 1600 cores, demonstrating its suitability for high-performance computing environments and large-scale modeling.
\end{abstract}

\begin{keyword}
 {Nonlocal model, Multiscale modeling, Damage, Granular media,  X-ray computed tomography, High performance computing} 
\end{keyword}
\end{frontmatter}

\section{Introduction}%
\label{sec:introduction}

While traditional DEM-based approaches capture the rigid motion of the boundary of each gravel particle, they do not account for the deformation of individual gravel particles. In addition, traditional DEM does not fully capture complex damage propagation based on each particle’s geometry, nor extreme deformation inside notched or pre-cracked particles. Hence the capability of capturing elastic and inelastic deformation of any individual particle will lead to improved models for granular media.  More recent work accounts for particle shapes on inter-particle interaction and rules for particle fracture using new enhanced DEM methods, see \cite{kawamoto2018all,HARMON2020112961}. 

In this treatment, we depart from DEM and model elastic as well as extreme inter and intra-particle  deformation using a nonlocal continuum formulation. 
For intra-particle modeling we follow a peridynamic modeling approach developed in \cite{SillingReformulation00}. In this approach both elastic deformation and fracture inside a particle are coupled implicitly and evolve autonomously.  Using this model the intra-particle deformation 
can exhibit extreme deformation resulting in fracture. Of course once particles fracture into two or more particles these new particles interact with the surrounding particles and our methodology handles this autonomously, eliminating the need for explicit reconstruction of fragment boundary. 

In related work peridynamic models have been used for mesoscale modeling of granular media, especially for capturing elastic and inelastic deformation and intra-granular force within individual grains. Shock wave  decay in particle beds of circular particles is analized in \cite{BehzadinasabEtAlPeridynamics18}.  Zhu and Zhao \cite{ZhuZhaoModeling19,ZhuZhaoPeridynamic19} investigate the crushing of sand piles using a combination of Weibull statistics and peridynamic forces.  The current work is motivated by the recent work of the authors and colleagues that combines the direct element method (DEM) with peridynamics (PeriDEM) to the modeling of granular flows, Jha et. al. \cite{JhaEtAlPeridynamicsbased21} and  Bhattacharya and Lipton \cite{bhattacharyalipton2023},   {and most recently extended to investegate vehicle mobility over granular road beds by Bhattacharya and Lipton \cite{bhattacharyalipton2025}. Recent work introduces an inelastic peridynamic formulation of yield and creep for molecular crystal particles in three dimensions, Silling et. al. \cite{Silling-Ban}.} 

  {In this work the aggregate is described by a Lagrangian system of $N$ particles, where at each instant $t$, each particle $P_i(t)$, $i=1,\dots,N$, is represented by a set of points $\mathbf{x} + \mathbf{u}(\mathbf{x}, t)= \mathbf{p}(\mathbf{x}, t) \in P_i(t)$ with initial particle configuration $\xx+\uu_0(\xx)\in P_i(0)$. Here $\uu(\xx,t)$ is the deformation and $\uu_0(\xx)=\uu(\xx,0)$. 
We show how to compute extreme particle deformation and intra-particle fracture and communition for convex and nonconvex particles in three dimensions.  The goal is to provide high fidelity modeling of aggregates of thousands of particles each of different shape. And with an eye towards designing aggregates we examine the aggregate effect of fixed particle geometries such as jacks, hollow spheres, and cubes.}

  {The first goal is to model impacts between three dimensional particles resulting in intra-particle fracture. To validate our modeling we conduct a simulation in Section \ref{Kalhtoff} in which one particle is a three dimensional impactor colliding with a second particle given by a thin 3 dimensional plate. The simulation is seen to recover the prototypical  Kalthoff Winkler fracture profile inside the thin plate, see Figure \ref{fig:kaltf-1}. The next goal is to squash a hollow spherical particle between two container walls see Section \ref{Particle Breakige}  Figures \ref{fig:communi-3r/4} and \ref{fig:communi-r/2} . The third simulation combines impacts between particles and the squashing of particles between container walls, this is illustrated in Figure \ref{fig:double-hallow-sphere} in Section \ref{double-hollow-sphere} for a container containing two hollow spheres. }

  {Next we dramatically increase the level of complexity in Section \ref{sub:shape-setup} and crush aggregates of differently shaped nonconvex particles that are confined to containers. Here aggregates ranging from two hundred to a thousand particles are considered. Each particle geometry is different and the particle geometries are those of  actual particles obtained from three dimensional X-ray computed tomography \cite{vlahinic2014towards}. The particle shapes are stored as point cloud data and then converted to an FEM mesh for use in computation.  Simulations given in Section \ref{sub:shape-setup} show that the force per unit area response of the aggregate to compression and crushing converges as the number of particles comprising the aggregate is increased. It points toward an opportunity of using an RVE to model local elastic and inelastic response. The local information can then be employed in a multiscale numerical analysis of soils and sands at the field scale.  }

  {We explore the role of grain shape and topology on the overall strength of the aggregate. This is the objective of Section   \ref{grain shape topology} where prototypical grain shapes both convex and complex are considered and their effect on the bulk strength of the aggregate is examined for different grain shapes. In Section \ref{sec:hollow} we consider larger scale simulations of hollow granular materials. We investigate the role of sample size for aggregates of hollow spheres on the bulk strength of the aggregate.}

 {To demonstrate the computational scalability and parallel efficiency of the proposed PeriDEM framework, we conducted a comprehensive scaling analysis on the SuperMIC cluster at Louisiana State University. This included both strong and weak scaling studies on up to 80 cores, as well as production-scale simulations utilizing up to 1600 cores. The results confirm that the framework maintains high CPU and memory efficiency across a wide range of particle counts and system sizes. These performance benchmarks not only validate the feasibility of large-scale three-dimensional simulations but also establish PeriDEM as a robust tool for modeling the collective behavior of granular media under high-resolution conditions. A detailed account of these scaling results is provided in the computational performance in section \ref{sec:scale}.}

  {The paper is organized as follows: We introduce our overall formulation in Section \ref{Overview}. In Sections \ref{sec:peridynamic_model} and \ref{sec:damage_model} we describe the elastic and damage forces inside each particle. The forces acting between particles are described as follows: the repulsive force between particles is described in Section \ref{Repulsive}, the  stick-slip friction force between particles is described in Section \ref{stickslip} and the damping force between particles is described in Section \ref{damping}. The force interaction between grains and the container walls is described in Sections \ref{rigid wall} and \ref{sub:Contact with multiple wall boundaries} . The method used to coordinate particle geometry and internal forces inside nonconvex particles is given in Section \ref{section:self-force}. The results of this section are also used to identify particles that split off a parent particle during the course of the evolution. In Section \ref{section: bond removal} specific bond removal strategies are employed to ensure that bonds outside the domain of interest are removed when fracture is occurring in non convex regions. The numerical experiments are outlined and given in Sections \ref{Kalhtoff}  through \ref{grain shape topology} .}

\section{Overview of the PeriDEM Model}
\label{Overview}

The goal is to compute the dynamics of particle aggregates, incorporating both elastic and inelastic effects within each particle.
 The model extends previous work to three dimensions and models the  effects particle crushing, aggregates confined to time-dependent domains, and is numerically stabile for sliding friction in three dimensions. 
\subsection{Domain and Particle Representation}
   {
Let $D \subset \mathbb{R}^d$ ($d = 2$ or 3) represent the domain containing the particle assemblage, consisting of $N$ particles $P_i \subset D$, where $i = 1, 2, \dots, N$. 
For a material point $\mathbf{x} \in P_i$, we define the deformation of $\xx$ at time $t$ by $\uu(\xx,t)$ and velocity as $\mathbf{v}(\xx,t)$. At time $t$, the location of the material point is given by $\mathbf{p}(\mathbf{x}, t) = \mathbf{x} + \mathbf{u}(\mathbf{x}, t)$, with velocity $\mathbf{v}(\mathbf{x}, t) = \dot{\mathbf{u}}(\mathbf{x}, t)$.
Suppose $D \in \R^d$ is a container holding the particle aggregates at all times. Then we have $\pp(\xx,t) \in D$ for all $\xx \in P_i$ for all $i = 1, \dots, N$ and for all $t \in [0, T]$.
At time $t$, the location of the $i$-th particle is denoted by $P_i(t)$, and is defined as
$
    P_i(t) = \{ \pp(\xx, t) : \xx \in P_i \}.
$ 
}

\bigskip

In this model we consider three types of interactions:
\begin{itemize}
\item Intraparticle interaction: Each particle reacts to forces on its boundary, driving the evolution of internal forces.
\item Interparticle interaction: Particles exchange forces at their interfaces via contact.
\item External interaction: Particles interact with domain walls and externally imposed forces such as gravity.
\end{itemize}
The intraparticle interaction is described using peridynamics, while the interparticle interaction is modeled using a combined peridynamics and DEM approach. This allows for the simulation of arbitrarily shaped particles and different particle topologies, e.g., particles with holes.
   {The dynamics of every point of each particle can be described using a force density equation due to Newton's law.
The intraparticle force density acting at each point $\mathbf{x}$ due to the deformation $\mathbf{u}$ in the particle aggregate at time $t$ is $ \mathbf{F}^{\text{intra}} (\mathbf{x},\mathbf{u};t)$. 
The interparticle force density acting at each point $\mathbf{x}$ due to the deformation $\mathbf{u}$ in the particle aggregate at time $t$ is $ \mathbf{F}^{\text{inter}} (\mathbf{x}, \mathbf{u};t)$. The sum of these two force densities is the force density internal to the aggregate and is written $\mathbf{F}^{\text{int}} (\mathbf{x}, \mathbf{u};t)=\mathbf{F}^{\text{intra}} (\mathbf{x},\mathbf{u};t)+\mathbf{F}^{\text{inter}} (\mathbf{x},\mathbf{u};t)$. External force densities gravity and confinement by container walls is written $\mathbf{F}^{\text{ext}} (\mathbf{x},\mathbf{u};t)$.
\\
The equation of motion for the aggregate of particles is given by
\begin{align}\label{equ: motion}
\rho(\mathbf{x})\ddot{\mathbf{u}}(\mathbf{x},t)=\mathbf{F}^{\text{int}}(\mathbf{x},\mathbf{u};t) + \mathbf{F}^{\text{ext}} (\mathbf{x},\mathbf{u};t), \hbox{ $\mathbf{x}\in\,  \cup_{i=1}^N P_i$}
\end{align}
where $\rho(\mathbf{x})$ is the material density at position $\mathbf{x}$ and supplemented by initial conditions
\begin{equation}
\mathbf{u}(\mathbf{x}, 0) = \mathbf{u}_0(\mathbf{x}), \quad \dot{\mathbf{u}}(\mathbf{x}, 0) = \mathbf{v}_0(\mathbf{x}).
\end{equation}
Here, $\rho$ is the common mass density of the particles,
$\mathbf{F}^{\text{int}}$ represents force density internal to the aggregate and 
$\mathbf{F}^{\text{ext}}$ represents external force density acting on the particle aggregate, including container wall interactions and external forces such as gravity.
The internal and external force densities have units of force per unit volume.
}

A summary of the internal and external forces is provided in Table~\ref{Table:forces}. Below, we discuss the explicit forms of $\mathbf{F}^{\text{int}}$ and $\mathbf{F}^{\text{ext}}$. In \Cref{section:internal-forces,section:self-force}, we focus on the forces acting within a particle or between different parts of a particle. Section \ref{section:external-forces} describes the various forces between particles.

\section{Internal Forces}\label{section:internal-forces}
In this model, the internal forces govern the interactions and dynamics within and between particles. These forces include intraparticle effects modeled by peridynamics, repulsive forces to prevent particle overlap, friction forces capturing sliding resistance, and damping forces to account for energy dissipation during interactions. Together, these forces define how particles interact internally and influence each other’s behavior.
\subsection{Peridynamic Intraparticle Force Model}\label{sec:peridynamic_model}
Our approach to modeling uses a field theory that considers both elastic and inelastic deformation within particles, extending across space and time in a nonlocal manner. For any point $\xx$ within the $i^{th}$ particle's domain $P_i$ in $\R^d$, we calculate the internal force using this integro-differential equation:
\begin{align}
\label{eq:peri}
\boldsymbol{F}^{\text{intra}}(\xx,\uu;t) = \int\limits_{H_\epsilon(\xx) \cap P_i}^{} \ff^{\text{peri}}(\uu(\xx', t), \uu(\xx, t), \xx', \xx, t) d\xx',
\end{align}
Here, $\ff^{\text{peri}}$ represents the pairwise force function,    {which is measured in force per unit volume squared}.
We define the  \textit{horizon} as $\epsilon$, which is the maximum distance for nonlocal interactions. For any point $\xx$, its neighborhood $H_\epsilon(\xx)$ includes all points within $\epsilon$ distance. 

   {
Now, let us explain how force  relates to strain through a constitutive law. We are using a bond-based peridynamics approach, which considers two-point interactions \cite{SillingReformulation00}. A \textit{bond} between two points $\xx$ and $\xx'$ in the domain $P_i$ is defined as the vector $\xiB = \xx' - \xx$, where $\xx'$ is within the horizon of $\xx$.  We also define $\etaB$ as the difference between the displacements of these points, i.e. $\etaB = \uu(\xx', t) - \uu(\xx, t)$. The \textit{stretch} $s$ of a bond is calculated as:
$
s=s(\uu(\xx', t), \uu(\xx,t), \xx', \xx):= \frac{\abs{\xiB + \etaB } - \abs{\xiB} }{\abs{\xiB} }.
$
}

We're using a microelastic model \cite{BobaruEtAl09} where the pairwise force density function $\ff$ (a.k.a. bond force) is defined as:
\begin{align}
\label{pmb}
\ff^{\text{peri}} =
\begin{cases}
c_w\ w(\abs{\xiB} )\ s\  \frac{\xiB + \etaB}{\abs{\xiB + \etaB} }  & \text{ if } \abs{\xiB} < \epsilon \\
0 & \text{ otherwise},
\end{cases}
\end{align}

In this equation, $w(r)$ is the  \textit{micromodulus function} - a non-negative, non-increasing function of $r$. The constant $c_w$ is chosen so that the integral operator matches the Cauchy-Navier operator up to the second order. The specific values for the peridynamic spring constant and two micromodulus functions are provided in a separate \Cref{tab:peri-combined}.

   {
Note that dependence of the bond force density on the bond stretch is linear here and abruptly breaks once the stretch reaches the critical value. One can consider a more complex bond deformation profile to accommodate different material behavior. See, for example \cite{lipton2025energy},\cite{lipton2024energy}, \cite{coskdarilipt2025}.
}

\subsubsection{Deformations Inside a Grain both Elastic and Extreme and the Appearance of Damage and Memory}%
\label{sec:damage_model}
The prototype microelastic brittle (PMB) material model introduced by Silling and Askari \cite{SillingAskariMeshfree05} is used here to capture reversible and irreversible stretches between points within a grain that is due to both elastic deformations as well as extreme deformations that trigger material failure. In this model damage is initiated when a bond $\xiB$ between points $\xx$ and $\xx'$ breaks. This happens at time $t$ if the absolute value of the stretch $s(\xx', \xx, t)$ exceeds a critical value $s_0$, i.e. the strength of the material. Mathematically, this condition is expressed as $\abs{s} > \abs{s_0}$.
The critical stretch $s_0$ is determined by equating the critical energy release rate $G_c$ with the total energy needed to sever all bonds across a unit area of crack surface.

\begin{table}[ht]
\centering
\caption{Peridynamic spring constant for various choices of the micromodulus function in 2D and 3D.}
\label{tab:peri-combined}
\begin{tabular}{ c  c  c  c  c }
\hline
\multirow{2}{*}{Type} & \multicolumn{2}{c}{$c_w w(\abs{\xiB})$} & \multicolumn{2}{c}{$s_0$} \\
\cline{2-5}
& 2D & 3D & 2D & 3D \\
\hline
\hline
Constant & $\frac{6 E}{\pi \epsilon^3(1 - \nu)}$ & $\frac{12 E}{\pi \epsilon^4(1 - \nu)}$ & $\sqrt{\frac{4 \pi G_c}{9 E \epsilon} }$ & $\sqrt{\frac{5 \pi G_c}{12 E \epsilon} }$ \\
\hline
Conic & $  \frac{24 E}{\pi \epsilon^4(1 - \nu)}\left(\epsilon - \abs{\xiB} \right)$ & $\frac{60 E}{\pi \epsilon^5(1 - \nu)}\left(\epsilon - \abs{\xiB} \right)$ & $\sqrt{\frac{5 \pi G_c}{9 E \epsilon} }$ & $\sqrt{\frac{7 \pi G_c}{12 E \epsilon} }$ \\
\hline
\end{tabular}
\end{table}


The table above lists the values of the critical stretch for both constant and conic micromodulus functions \cite{HaBobaruStudies10}. Once a bond breaks at time $t = t_0$, it remains broken for all subsequent times $t > t_0$. We define the \textit{damage} of a material point $\xx$ as the ratio of intact bonds connected to $\xx$ at time $t$ to the total number of bonds connected to $\xx$ in the reference configuration (at $t=0$).
   { We explicitly express the history dependence of  damage inside the particle by
defining the \textit{bond memory} $\gamma$ as the indicator function of no history of over-stretching by
\begin{equation}
    \gamma(\xx', \xx, t) = 
    \begin{cases}
        0 & \text{ if } \exists \tau \in [0, t] \text{ such that } s(\xx', \xx, \tau) > s_0
        \\
        1 & \text{ otherwise }
    \end{cases}
\end{equation}
The intra-particle force within $P_i$ with history of damage is therefore given by
\begin{align}
\label{eq:peri-damage}
\boldsymbol{F}^{\text{intra}}(\xx,\uu;t) = \int\limits_{H_\epsilon(\xx) \cap P_i}^{} \gamma(\xx', \xx, t) \tilde{\ff}^{\text{peri}}(\uu(\xx', t), \uu(\xx, t), \xx', \xx, t) d\xx',
\end{align}
where 
\begin{align}
\label{eq:peri2}
\tilde{\ff}^{\text{peri}}(\uu(\xx', t), \uu(\xx, t), \xx', \xx, t)= c_w\, w(\abs{\xiB} )\, s\,  \frac{\xiB + \etaB}{\abs{\xiB + \etaB} }.
\end{align}}
It's worth noting that bond-based peridynamics, being a two-point interaction model, imposes a fixed Poisson ratio of $\frac{1}{3}$ in 2D (plane strain) and $\frac{1}{4}$ in 3D  \cite{TrageserSelesonBondBased20}. This limitation can be overcome by using a state-based model, as proposed in \cite{SillingEtAlPeridynamic07}. However here we explore what can be done with the simplest model exploiting its computational efficiency over state-based methods.


\subsection{Repulsive Contact Force Between Particles}
\label{Repulsive}
In the study of granular materials and particulate systems, accurately modeling the interactions between particles is crucial for understanding the system's behavior. This section presents a short-range contact force model that captures intergrain interactions in peridynamic bodies. The model is based on the work of \cite{SillingAskariMeshfree05} and \cite{BehzadinasabEtAlPeridynamics18}, and provides a robust framework for simulating particle dynamics.

Consider two distinct particles, denoted as $P_i$ and $P_j$ $(i \neq j)$, each represented by a peridynamic body. Within these particles, we focus on material points $\mathbf{p}(\mathbf{x},t) \in P_i(t)$ and $\mathbf{p}(\mathbf{y},t) \in P_j(t)$. The interaction between these points is governed by a \textit{contact radius} $R_c$, which defines the maximum distance at which two points are considered to be in contact.

To describe the interaction between particles, we first define the normal interaction direction from point $\mathbf{x}$ to point $\mathbf{y}$ as a unit vector
\begin{equation}
    \mathbf{e}(\mathbf{y},\mathbf{x},t) = \frac{\mathbf{p}(\mathbf{y},t) - \mathbf{p}(\mathbf{x},t)}{|\mathbf{p}(\mathbf{y},t) - \mathbf{p}(\mathbf{x},t)|}.
\end{equation}

This vector plays a crucial role in determining the direction of the contact force between particles.
We consider a short-range repulsive bond force $\mathbf{f}_r(\mathbf{y},\mathbf{x},t)$ exerted on $\mathbf{p}(\mathbf{x},t)$ by $\mathbf{p}(\mathbf{y},t)$. For future reference all bond forces between points inside different particles have units of force per unit volume squared. The repulsive bond force  is defined as
   {
\begin{equation}\label{eq:repulsive-force}
    \mathbf{f}_r(\mathbf{y},\mathbf{x},t) = 
    \begin{cases} 
    -K_n\left( R_c - |\mathbf{p}(\mathbf{y},t) - \mathbf{p}(\mathbf{x},t)| \right) \mathbf{e}(\mathbf{y},\mathbf{x},t) & \text{if } |\mathbf{p}(\mathbf{y},t) - \mathbf{p}(\mathbf{x},t)| < R_c, \\ 
    0 & \text{otherwise}
    \end{cases}
\end{equation}
Here, $K_n$ represents the normal contact stiffness, given by $K_n = \frac{18k}{\pi \varepsilon^5}$, where $k$ is the bulk modulus and $\varepsilon$ is a small parameter related to the peridynamic horizon. 
It is worth noting that when the interacting peridynamic bodies have different bulk moduli $k_1$ and $k_2$, we use an effective bulk modulus given by the harmonic mean
$
    k = \frac{2k_1 k_2}{k_1 + k_2}.
$
This approach allows for a more accurate representation of interactions between particles with different material properties. 
\\
In a multi-particle system, a single point may interact with multiple neighboring particles. Therefore, we define the total repulsive force density on a point $\mathbf{p}(\mathbf{x},t) \in P_i(t)$ as the sum of all pairwise interactions given by
\begin{equation}
     \mathbf{F}_r(\mathbf{x},\mathbf{u};t) = \sum_{j \neq i} \int_{\{ \mathbf{y} \in P_j : |\mathbf{p}(\mathbf{y},t) - \mathbf{p}(\mathbf{x},t)| < R_c \}} \mathbf{f}_r(\mathbf{y},\mathbf{x},t) d\yy, \hbox{ $\mathbf{x}\in P_i$}.
\end{equation}
}

The presented model assumes a linear relationship between the repulsive force and the distance between points in contact. While nonlinear relations are possible, the linear approximation provides comparable results over short distances, as noted by \cite{DesaiEtAlRheometry19}.

\subsection{Nonlocal Stick Slip Friction}
\label{stickslip}
A tangential damping force is applied to nodes in contact, incorporating energy dissipation due to friction following Coulomb's law. Such forces are introduced into DEM models through virtual spring displacement (see, for example, \cite{luding2008introduction}). To model the stick/slip transition in our dynamic friction framework, we adopt the regularized Coulomb's model \cite{martinsOden1983numerical, campos1982numerical}, adapted to a nonlocal context. Let the relative velocity of $\mathbf{p}(\mathbf{x},t) \in P_i(t)$ with respect to $\mathbf{p}(\mathbf{y},t) \in P_j(t)$ be given as $\mathbf{v}(\mathbf{x},\mathbf{y},t) = \dot{\mathbf{u}}(\mathbf{x},t) - \dot{\mathbf{u}}(\mathbf{y},t)$. The tangential component of the relative velocity of $\mathbf{p}(\mathbf{x},t)$ with respect to $\mathbf{p}(\mathbf{y},t)$ is given by
\begin{equation*}
\mathbf{v}_\perp(\mathbf{x},\mathbf{y},t) = \mathbf{v}(\mathbf{x},\mathbf{y},t) - (\mathbf{v}(\mathbf{x},\mathbf{y},t) \cdot \mathbf{e}(\mathbf{y},\mathbf{x},t)) \mathbf{e}(\mathbf{y},\mathbf{x},t).
\end{equation*}
   {
Next, we define the tangential contact direction as $\mathbf{e}_\perp(\mathbf{x},\mathbf{y},t) = \frac{\mathbf{v}_\perp(\mathbf{x},\mathbf{y},t)}{|\mathbf{v}_\perp(\mathbf{x},\mathbf{y},t)|}$, along which the friction acts. 
Note that, in this formulation, the relative velocity of points in contact determine the direction of friction. Therefore, one does not need to know or compute that local particle boundary information, and is suitable for particle simulations with breakage.
\\
The magnitude of the friction force $\mathbf{f}_f(\mathbf{y},\mathbf{x},t)$ on $\mathbf{p}(\mathbf{x},t)$ due to $\mathbf{p}(\mathbf{y},t)$ depends on the magnitude of the tangential force $\tau_\perp(\mathbf{x},\mathbf{y},t)$ acting along the direction $\mathbf{e}_\perp(\mathbf{x},\mathbf{y},t)$. In the sticking regime, when $|\tau_\perp(\mathbf{x},\mathbf{y},t)| \leq \mu |\mathbf{f}_r(\mathbf{y},\mathbf{x},t)|$, the friction force acts as a restoring force, preventing relative motion between $\mathbf{p}(\mathbf{x},t)$ and $\mathbf{p}(\mathbf{y},t)$. In the slipping regime, when $|\tau_\perp(\mathbf{x},\mathbf{y},t)| > \mu |\mathbf{f}_r(\mathbf{y},\mathbf{x},t)|$, the friction force reaches a constant magnitude of $\mu |\mathbf{f}_r(\mathbf{y},\mathbf{x},t)|$, opposing the direction of $\mathbf{v}_\perp(\mathbf{x},\mathbf{y},t)$. The transition between the sticking and slipping regimes is determined by the prescribed tangential force. As discussed in \cite{martinsOden1983numerical}, the stick/slip transition in our model is characterized by the relative speed $|\mathbf{v}_\perp(\mathbf{x},\mathbf{y},t)|$ exceeding a speed threshold $v_{\text{thr}}$. This threshold is related to the numerical time step $\Delta t$ and is given by
\begin{equation*}
v_{\text{thr}}(\mathbf{x},\mathbf{y},t,\Delta t) = \frac{\mu}{\rho } |\mathbf{f}_r(\mathbf{y},\mathbf{x},t)| \Delta t,
\end{equation*}
where $v_{\text{thr}}(\mathbf{x},\mathbf{y},t,\Delta t)$ is derived from approximating the maximum possible magnitude of $|\mathbf{v}_\perp(\mathbf{x},\mathbf{y},t)|$ resulting from the prescribed tangential force $\tau_\perp(\mathbf{x},\mathbf{y},t) = \mu |\mathbf{F}_r(\mathbf{y},\mathbf{x},t)| \mathbf{e}_\perp(\mathbf{x},\mathbf{y},t)$ in the absence of friction. Note that $v_{\text{thr}}(\mathbf{x},\mathbf{y},t,\Delta t) \to 0$ as $\Delta t \to 0$.
The frictional force  $\mathbf{f}_f$ on $\mathbf{p}(\mathbf{x},t) \in P_i(t)$ exerted by $\mathbf{p}(\mathbf{y},t) \in P_j(t)$ is given by
\begin{equation}
    \label{eq:regularized-friction}
\mathbf{f}_f(\mathbf{y},\mathbf{x},t) =
\begin{cases}
    -\mu |\mathbf{f}_r(\mathbf{y},\mathbf{x},t)| \mathbf{e}_\perp(\mathbf{y},\mathbf{x},t) & \text{if } |\mathbf{v}_\perp(\mathbf{x},\mathbf{y},t)| > v_{\text{thr}}(\mathbf{x},\mathbf{y},t,\Delta t), \\
    -\frac{1}{\varepsilon_f} |\mathbf{v}_\perp(\mathbf{x},\mathbf{y},t)| \mathbf{e}_\perp(\mathbf{y},\mathbf{x},t) & \text{if } 0 \leq |\mathbf{v}_\perp(\mathbf{x},\mathbf{y},t)| \leq v_{\text{thr}}(\mathbf{x},\mathbf{y},t,\Delta t),
\end{cases}
\end{equation}
where $\varepsilon_f$ is a regularization parameter dependent on $\Delta t$, given by
$
\varepsilon_f = \frac{\Delta t}{\rho}.
$
This value of $\varepsilon_f$ ensures that once $|\tau_\perp(\mathbf{x},\mathbf{y},t)| \leq \mu |\mathbf{F}_r(\mathbf{y},\mathbf{x},t)|$, the relative velocity of $\mathbf{p}(\mathbf{x},t)$ becomes zero in the next time step. Similar regularized nonlocal tangential friction models for particles are introduced in \cite{kamensky2019peridynamic}.
\\
Consequently, the friction force acting on the point $\mathbf{p}(\mathbf{x},t) \in P_i(t)$ from all neighboring particles is computed as
\begin{equation*}
\FF_f(\xx, \uu;t) = \sum_{j \neq i} \int_{\{ \mathbf{y} \in P_j : |\mathbf{p}(\mathbf{y},t) - \mathbf{p}(\mathbf{x},t)| < R_c \}} \mathbf{f}_f(\mathbf{y},\mathbf{x},t) d\yy, \hbox{ $\mathbf{x}\in P_i$}.
\end{equation*}
}

\subsection{Normal Damping Force}
\label{damping}
In this formulation, damping is introduced to facilitate energy dissipation during normal contact between particles. This damping mechanism shortens relaxation times and consequently reduces computational cost in simulations targeting mechanical equilibrium. The damping force  acting on $\mathbf{p}(\mathbf{x},t) \in P_i(t)$ due to $\mathbf{p}(\mathbf{y},t) \in P_j(t)$ is defined as
   
{
\begin{equation}
    \label{eq:damping-force}
 \mathbf{f}_d(\mathbf{y},\mathbf{x},t) =
\begin{cases}
    -\beta_d \left( \frac{\mathbf{v}(\mathbf{y},\mathbf{x},t)}{|\mathbf{v}(\mathbf{y},\mathbf{x},t)|} \cdot \mathbf{e}(\mathbf{y},\mathbf{x},t) \right) \mathbf{e}(\mathbf{y},\mathbf{x},t) & \text{if } |\mathbf{p}(\mathbf{y},t) - \mathbf{p}(\mathbf{x},t)| < R_c, \\
    0 & \text{otherwise},
\end{cases}   
\end{equation}
where $\beta_d$ is the damping coefficient.
The total damping force density on the point $\mathbf{p}(\mathbf{x},t) \in P_i(t)$ due to all other neighboring particles is then expressed as
\begin{equation*}
\FF_d(\xx,\uu; t) = 
\sum_{j \neq i} \int_{\{ \mathbf{y} \in P_j : |\mathbf{p}(\mathbf{y},t) - \mathbf{p}(\mathbf{x},t)| < R_c \}} \mathbf{f}_d(\mathbf{y},\mathbf{x},t) d\yy, \hbox{ $\mathbf{x}\in P_i$}.
\end{equation*}
The damping force acts opposite to $\mathbf{e}(\mathbf{y},\mathbf{x},t)$, and its magnitude depends on the normal projection of the relative velocity. While nonlinear damping models such as \cite{JankowskiAnalytical06} could be used, we focus on the linear case here. 
The combined bond forces in (\ref{eq:repulsive-force}) and (\ref{eq:damping-force}) is analogous to a viscoelastic spring that connects $\mathbf{p}(\mathbf{x},t)$ and $\mathbf{p}(\mathbf{y},t)$, with a reference length of $R_c$.
}

\section{External Forces}\label{section:external-forces}
In this section, we describe forces that represent interactions between particles and their environment, such as gravity and reaction from the container walls.

\subsection{Contact Between Peridynamic Body and Rigid Wall}
\label{rigid wall}
   {
Here, we assume that the inner walls of the container holding the particle aggregate are rigid (i.e. non-deformable).
This treatment allows us to save computation time by avoiding meshing the container walls, as in \cite{BehzadinasabEtAlPeridynamics18, JhaEtAlPeridynamicsbased21}. Additionally, by consider simple container wall geometry (such as planar walls), we are able to compute analytical expressions of the total reaction force from the wall boundaries acting on each point of each particle that is in contact with the wall. 
The method described here can be generalized to any particle container shape by approximating the inner boundary of the wall into $M$ piecewise planer segments. Let the inner wall of the boundary is given by the surface $S = \bigcup_{i=1}^M  l_i$, where $l_i$ is a planar wall segment.
\\
For simplicity, the inner boundary of the container is assumed to be rectangular with six sides, made up of planar segments, i.e., $M = 6$. 
\\
The point $\mathbf{p}(\mathbf{x},t)$ inside the particle $P_k(t)$ experiences a contact force from the wall if there exists a planar wall segment $l \in L$ such that the perpendicular distance between $\mathbf{p}(\mathbf{x},t)$ and $l$ is smaller than the contact radius $R_c$. The set of all points in the container wall exerting contact forces on $\mathbf{p}(\mathbf{x},t)$ is denoted by $S_l(\mathbf{x},t)$, which forms a spherical segment of the ball $B_{R_c}(\mathbf{p}(\mathbf{x},t))$. The repulsive force density exerted by the wall on $\mathbf{p}(\mathbf{x},t)$ is then given by
\begin{equation*}
    \mathbf{F}_r^{l}(\mathbf{x},\uu;t) =
\begin{cases}
    -K_n (R_c - |c_l(\mathbf{x},t) - \mathbf{p}(\mathbf{x},t)|) |S_l(\mathbf{x},t)| \mathbf{e}_l(\mathbf{x},t) & \text{if } d_l(\mathbf{x},t) < R_c, \\
    0 & \text{otherwise},
\end{cases}
\end{equation*}
where $|S_l(\mathbf{x},t)|$ is the area of the contact region $S_l(\mathbf{x},t)$, $c_l(\mathbf{x},t)$ is the centroid of $S_l(\xx, t)$, and $d_l(\mathbf{x},t)$ represents the perpendicular distance from $\mathbf{p}(\mathbf{x},t)$ to the planar segment $l$. The vector $\mathbf{e}_l(\mathbf{x},t)$ is a unit vector along the direction from $c_l(\mathbf{x},t)$ to $\mathbf{p}(\mathbf{x},t)$, defined as:
\begin{equation*}
\mathbf{e}_l(\mathbf{x},t) = \frac{c_l(\mathbf{x},t) - \mathbf{p}(\mathbf{x},t)}{|c_l(\mathbf{x},t) - \mathbf{p}(\mathbf{x},t)|},
\end{equation*}
which specifies the direction of the repulsive force exerted by the wall on $\mathbf{p}(\mathbf{x},t)$. 
Since the wall boundaries are non-deformable, the contact areas $|S_l|$ and $|S_{l_i, l_j}|$ can be computed analytically, reducing the computational cost as wall discretization is unnecessary during simulations. For clarity, we provide the analytical expressions below.
For the contact with a single planar wall boundary $l$, the area of the contact region is given by
\begin{equation*}
    |S_l(\mathbf{x},t)| = \frac{\pi}{3} (R_c - d_l)^2 (2R_c - d_l)
\end{equation*}
and the distance from $\mathbf{p}(\mathbf{x},t)$ to the centroid of $S_l(\mathbf{x},t)$ is given by
\begin{equation*}
|c_l(\mathbf{x},t) - \mathbf{p}(\mathbf{x},t)| = \frac{\pi}{4} \frac{1}{|S_l(\mathbf{x},t)|} \left( R_c^2 - d_l(\mathbf{x},t)^2 \right)^{2}.
\end{equation*}
Note that when $\mathbf{p}(\mathbf{x},t)$ touches the wall boundary at segment $l$ (i.e., $d_l(\mathbf{x},t) = 0$), the effective contact area becomes $|S_l(\mathbf{x},t)| = \frac{2\pi R_c^3}{3}$, which corresponds to the area of a half-sphere.
\\
The friction and damping forces $\mathbf{F}_f^{l}(\mathbf{x},t)$ and $\mathbf{F}_d^{l}(\mathbf{x},t)$ due to the wall boundary can be computed similarly, by replacing $\mathbf{e}(\mathbf{y},\mathbf{x},t)$ with $\mathbf{e}_l(\mathbf{x},t)$, $V_y$ with $|S_l(\mathbf{x},t)|$, and $\mathbf{p}(\mathbf{y},t)$ with $c_l(\mathbf{x},t)$ in \Cref{eq:damping-force,eq:regularized-friction}.
}

\subsubsection{Contact with Multiple Wall Boundaries}%
\label{sub:Contact with multiple wall boundaries}
   {
In the case where a point $\mathbf{p}(\mathbf{x},t) \in P_i(t)$ is in contact with multiple planar wall boundary segments, inclusion-exclusion principle can be used to compute the total reaction (and friction and damping) force on $\pp(\xx,t)$ due to the entire wall boundary surface.
This is fairly uncommon, but can happen when $\pp(\xx,t)$ is near a corner of the rectangular container, where it may experience contact forces from at most three adjacent planar wall segments, say $l_1, l_2, l_3$. In this case, the repulsive force on $\mathbf{p}(\mathbf{x},t)$ due to the container wall is
\begin{equation*}
    \mathbf{F}_r^{\text{wall}}(\mathbf{x},\uu;t) = \sum_{m = 1}^3 \mathbf{F}_r^{l_m}(\mathbf{x},t)  - \sum_{i \ne j} \mathbf{F}_r^{l_{i}, l_{j}}+ \mathbf{F}_r^{l_1, l_2, l_3}(\mathbf{x},t),
\end{equation*}
where for any two $i, j \in \{1, 2, 3\}$,
\begin{equation*}
    \mathbf{F}_r^{l_{i}, l_{j}}(\mathbf{x},t) 
    = -K_n (R_c - |c_{l_{i}, l_{j}}(\mathbf{x},t) - \mathbf{p}(\mathbf{x},t)|) 
    \abs{(S_{l_{i}} \cap S_{l_{j}})(\mathbf{x},t)}
    \mathbf{e}_{l_{i}, l_{j}}(\mathbf{x},t),
\end{equation*}
where $c_{l_{i}, l_{j}}(\mathbf{x},t)$ is the centroid of $S_{l_{i}} \cap S_{l_{j}}$ and $\mathbf{e}_{l_{i}, l_{j}}(\mathbf{x},t)$ is the unit vector defined as
\begin{equation*}
\mathbf{e}_{l_{i}, l_{j}}(\mathbf{x},t) = \frac{c_{l_{i}, l_{j}}(\mathbf{x},t) - \mathbf{p}(\mathbf{x},t)}{|c_{l_{i}, l_{j}}(\mathbf{x},t) - \mathbf{p}(\mathbf{x},t)|}.
\end{equation*}
Similarly, 
\begin{equation*}
    \mathbf{F}_r^{l_1, l_2, l_3}(\mathbf{x},t) 
    = -K_n (R_c - |c_{l_1, l_2, l_3}(\mathbf{x},t) - \mathbf{p}(\mathbf{x},t)|) 
    V_x \abs{S_{l_1, l_2, l_3}(\mathbf{x},t)}
    \mathbf{e}_{l_1, l_2, l_3}(\mathbf{x},t),
\end{equation*}
where $S_{l_1, l_2, l_3} = S_{l_1} \cap S_{l_2} \cap S_{l_3}$ representing the overlapping contact area, $c_{l_1, l_2, l_3}(\mathbf{x},t)$ is its centroid, and $\mathbf{e}_{l_1, l_2, l_3}(\mathbf{x},t)$ is the unit vector defined as
\begin{equation*}
\mathbf{e}_{l_1, l_2, l_3}(\mathbf{x},t) = \frac{c_{l_1, l_2, l_3}(\mathbf{x},t) - \mathbf{p}(\mathbf{x},t)}{|c_{l_1, l_2, l_3}(\mathbf{x},t) - \mathbf{p}(\mathbf{x},t)|}.
\end{equation*}
\\
The friction and damping force densities $\FF^{\text{wall}}_f(\xx, t)$ and $\FF^{\text{wall}}_d(\xx, t)$ on $\pp(\xx,t)$ by the wall boundary can be computed similarly. Combined all interaction from the container wall, we define the total force density on $\pp(\xx, t) \in P_i(t)$ as
\begin{align}
    \FF^{\text{wall}} (\xx,\uu; t)=
    \FF^{\text{wall}}_r (\xx,\uu; t)
    +
    \FF^{\text{wall}}_f (\xx,\uu; t)
    +
    \FF^{\text{wall}}_d (\xx,\uu; t).
\end{align}
}

\section{Self-contact Forces}\label{section:self-force}
The existence of a peridynamic bond between two material points within the same parent particle provides the necessary repulsive force to prevent material overlap. However, when this peridynamic bond is broken (due to breakage), these repulsive forces are no longer present. Therefore, it is essential to model a self-contact law between points within the same parent particle that are not connected by a peridynamic bond, but are brought close together due to significant deformations. This is particularly important to avoid numerical interpenetration of various sections of nonconvex particle shapes and to model interactions between different disconnected segments of a parent particle when peridynamic forces are absent.

   {
In this discussion, we assume that $\epsilon \le R_c$. Assume that two points $\pp(\xx, t), \pp(\yy, t) \in P_i(t)$ are in contact, i.e., $\abs{\pp(\xx, t) - \pp(\yy, t)}  < R_c$, with no peridynamic bond between them at time $t$. If the reference bond length $|\mathbf{y}-\mathbf{x}| < R_c$, it is implied that they were connected by a now-broken peridynamic bond at $t=0$. In this case, the normal repulsion force on $\pp(\xx,t)$ due to $\pp(\yy, t)$ can be modeled by an typical repulsive-only  peridynamic bond force. On the other hand, if  the reference length satisfies $R_c < |\mathbf{y}-\mathbf{x}|$ (i.e., $R_c < |\mathbf{y}-\mathbf{x}| < \epsilon$ with a previously intact peridynamic bond, or $|\mathbf{y}-\mathbf{x}| > \epsilon$ with no previous peridynamic bond), the repulsion force could be modeled simply using a spring of reference length $R_c$.
The self-contact law is based on the distance between nodes in the reference (undeformed) configuration. If no peridynamic bond exists between two nodes $\mathbf{x}$ and $\mathbf{y}$ from the same parent particle $P_i$ at time $t$, and the current distance between them falls within $R_c$ (i.e., if $\abs{\pp(\xx, t) - \pp(\yy, t)} < R_c$), the \textit{normal repulsive force} exerted on $\mathbf{x}$ by $\mathbf{y}$ is given by
\begin{equation*}
\mathbf{f}_r^{\text{self}} (\mathbf{x},\mathbf{y},t) =
\begin{cases}
    c_w \frac{\abs{\pp(\xx, t) - \pp(\yy, t)} - |\mathbf{y}-\mathbf{x}|}{|\mathbf{y}-\mathbf{x}|} \mathbf{e}(\mathbf{y},\mathbf{x},t)\chi_{ \{\abs{\pp(\xx, t) - \pp(\yy, t)} < |\mathbf{y}-\mathbf{x}|\} } & \text{if } |\mathbf{y}-\mathbf{x}| < R_c, \\
    \\
    c_w \frac{\abs{\pp(\xx, t) - \pp(\yy, t)} - R_c}{R_c} \mathbf{e}(\mathbf{y},\mathbf{x},t) \chi_{ \{\abs{\pp(\xx, t) - \pp(\yy, t)} < R_c \} } & \text{if } R_c < |\mathbf{y}-\mathbf{x}|,
\end{cases}
\end{equation*}
where $\chi_S$ is the characteristic function of the set $S$. In cases where the reference distance is small ($|\mathbf{y}-\mathbf{x}|< R_c < \epsilon$), the repulsive contact force is modeled using a purely repulsive peridynamic bond force. This ensures that two points with reference distance $|\mathbf{y}-\mathbf{x}| < R_c$ do not experience any repulsive force unless they come closer than their reference distance $|\mathbf{y}-\mathbf{x}|$. For $|\mathbf{y}-\mathbf{x}| > R_c$, the contact force between nodes from the same parent particle is treated the same as the contact force between nodes from different parent particles, with $c_w / R_c = K_n$.
\\
\begin{align}
    \ff^{\text{self}} = 
    \ff^{\text{self}}_r
    +
    \ff^{\text{self}}_f
    +
    \ff^{\text{self}}_d
\end{align}
The modified intraparticle force for $\mathbf{x}\in P_i$ is therefore given by
\begin{align}
\label{eq:peri-damage-self}
\boldsymbol{F}^{\text{intra}}(\xx,\uu; t) 
 = & \int\limits_{H_\epsilon(\xx) \cap P_i}^{} \gamma(\xx', \xx, t) \ff^{\text{peri}}(\uu(\xx', t), \uu(\xx, t), \xx', \xx, t) d\xx'
\\
 & + \int\limits_{P_i}^{} (1 - \gamma(\xx', \xx, t)) \ff^{\text{self}}(\xx', \xx, t) d\xx'.
\end{align}
Note that, in this formulation, the contact force between particle fragments is agnostic to which parent particle it came from.
}


\begin{table}[htbp]
\centering
\resizebox{\textwidth}{!}{
\begin{tabular}{>{\raggedright\arraybackslash}p{3.2cm} >{\raggedright\arraybackslash}p{5.2cm} >{\raggedright\arraybackslash}p{7.5cm}}
\toprule
\rowcolor{gray!20}
\textbf{Force Type} & \textbf{Description} & \textbf{Expression} \\
\midrule
\multicolumn{3}{>{\columncolor{gray!10}}c}{\textbf{Internal Forces}} \\
Intraparticle Force Density & Elastic/inelastic effects within particles (peridynamics) &
$\boldsymbol{F}^{\text{intra}}(\mathbf{x}, \mathbf{u}; t) = \int\limits \gamma(\mathbf{x}', \mathbf{x}, t)\, \tilde{\mathbf{f}}^{\text{peri}}(\mathbf{u}(\mathbf{x}', t), \mathbf{u}(\mathbf{x}, t), \mathbf{x}', \mathbf{x}, t)\, d\mathbf{x}'$ \\
Repulsive Contact Force & Short-range repulsive bond force preventing particle overlap &
\begin{minipage}[t]{\linewidth}\small
$\displaystyle
\mathbf{f}_r(\mathbf{y}, \mathbf{x}, t) =
\begin{cases}
-K_n(R_c - |\boldsymbol{\delta}_{yx}|)\, \mathbf{e}, & \text{if } |\boldsymbol{\delta}_{yx}| < R_c \\
0, & \text{otherwise}
\end{cases}$
\end{minipage} \\
Nonlocal Stick Slip Friction & Tangential friction force modeling sliding resistance &
$\mathbf{f}_f(\mathbf{y}, \mathbf{x}, t) =
\begin{cases}
-\mu |\mathbf{f}_r|\, \mathbf{e}_\perp, & \text{if } |\mathbf{v}_\perp| > v_{\text{thr}} \\
\displaystyle -\frac{1}{\epsilon_f} |\mathbf{v}_\perp|\, \mathbf{e}_\perp, & \text{otherwise}
\end{cases}$ \\
Normal Damping Force & Energy dissipation during interaction &
$\mathbf{f}_d(\mathbf{y}, \mathbf{x}, t) =
\begin{cases}
-\beta_d \left( \dfrac{\mathbf{v}}{|\mathbf{v}|} \cdot \mathbf{e} \right) \mathbf{e}, & \text{if } |\boldsymbol{\delta}_{yx}| < R_c \\
0, & \text{otherwise}
\end{cases}$ \\
\midrule
\multicolumn{3}{>{\columncolor{gray!10}}c}{\textbf{External Forces}} \\
Body Force Density (e.g., Gravity) & External force acting on all particles &
$\mathbf{b}(\mathbf{x}, t)$ \\
Wall Repulsive Force Density & Prevents penetration into container walls &
$\mathbf{F}_r^{\text{wall}}(\mathbf{x}, t) =
\begin{cases}
-K_n(R_c - |\mathbf{c}_l - \mathbf{p}|)\, |S_l|\, \mathbf{e}_l, & \text{if } d_l(\mathbf{x}, t) < R_c \\
0, & \text{otherwise}
\end{cases}$ \\
\midrule
\multicolumn{3}{>{\columncolor{gray!10}}c}{\textbf{Self-Contact Forces}} \\
Self-Repulsive Bond Force & Avoids overlap within the same particle &
$\displaystyle
\scriptsize\mathbf{f}_r^{\text{self}}(\mathbf{x}, \mathbf{y}, t) =
 \begin{cases}
\displaystyle c_w \frac{|\boldsymbol{\delta}_{xy}| - |\mathbf{y} - \mathbf{x}|}{|\mathbf{y} - \mathbf{x}|} \mathbf{e}(\mathbf{y},\mathbf{x},t)
\chi_{\{|\boldsymbol{\delta}_{xy}| < |\mathbf{y} - \mathbf{x}|\}}, & |\mathbf{y} - \mathbf{x}| < R_c \\
\displaystyle c_w \frac{|\boldsymbol{\delta}_{xy}| - R_c}{R_c} \mathbf{e}(\mathbf{y},\mathbf{x},t)
\chi_{\{|\boldsymbol{\delta}_{xy}| < R_c\}}, & \text{otherwise}
\end{cases}$ \\
\bottomrule
\end{tabular}}
\caption{Categorization of force terms in the PeriDEM particle dynamics model. All bond forces are expressed per unit volume squared. Here, $\boldsymbol{\delta}_{yx} = \mathbf{p}(\mathbf{y}, t) - \mathbf{p}(\mathbf{x}, t)$ and $\boldsymbol{\delta}_{xy} = \mathbf{p}(\mathbf{x}, t) - \mathbf{p}(\mathbf{y}, t)$ denote the relative positions used in bond evaluations.}
\label{Table:forces}
\end{table}

\subsection{Combined Model: Equation of Motion for Particle Aggregates}\label{sub:combineP_model}
We summarize reiterating that the aggregate is described by a Lagrangian system of $N$ particles, where each particle $P_i(t)$, $i=1,\dots,N$, is represented by a set of points $\mathbf{x} + \mathbf{u}(\mathbf{x}, t)= \mathbf{p}(\mathbf{x}, t) \in P_i(t)$ with initial particle configuration $\xx+\uu_0(\xx)\in P_i(0)$ with $\xx\in P_i$, $i=1,\ldots,N$. The equation of motion of each of these points is given by:
   {
\begin{equation}
    \rho(\mathbf{x}) \ddot{\mathbf{u}}(\mathbf{x}, t) = \mathbf{F}^{\text{int}}(\mathbf{x},t) + \FF^{\text{ext}} (\xx,t)
    \label{eq:motion}
\end{equation}
\noindent where $\rho(\mathbf{x})$ is the material density at position $\mathbf{x}$, $\ddot{\mathbf{u}}(\mathbf{x}, t)$ is the acceleration, $\mathbf{F}^{\text{int}}(\mathbf{x},t)$ and $\mathbf{F}^{\text{ext}}(\xx, t)$ represent the sum of all force densities interior and exterior to the particle aggregate, respectively.
The total force density on the particle interior can be decomposed into three main components:
\begin{equation}
    \mathbf{F}^{\text{int}}(\mathbf{x},t) = \mathbf{F}^\text{intra}(\mathbf{x},t) + \mathbf{F}^\text{inter}(\mathbf{x},t).
    \label{eq:force_components}
\end{equation}
The total external force density is given by
\begin{equation}
    \mathbf{F}^\text{ext}(\mathbf{x},t) = \FF^\text{wall}(\xx, t) + \Bb(\xx, t)
    \label{eq:intra_particle}
\end{equation}
where $\Bb(\xx,t)$ denote all other body forces such as gravity.
}

%
%
%
%
%
%
%

This comprehensive equation of motion (\Cref{eq:motion,eq:force_components,eq:intra_particle}) complememnted with initial conditions on the displacement, provides a robust framework for modeling particle dynamics.  The flexibility of this formulation makes it a versatile tool for a wide range of particulate system simulations involving extreme intraparticle deformations.
 {\section{Numerical Implementation}
After establishing our combined model for particle aggregate dynamics, we now discuss the numerical implementation necessary to solve these equations (\ref{eq:motion}). Our approach involves two key aspects: spatial discretization and time integration.
\subsection{Spatial Discretization}
To solve the equation of motion, we discretize both space and time. The total internal and external force densities consist of different physical components and are discretized accordingly. The intraxparticle forces, which model deformation and fracture within each grain, are computed using a nonlocal peridynamic formulation. The associated volume integrals are approximated by discrete summations over neighboring material points within a finite horizon. This enables the simulation of large deformations and fracture evolution without the need for remeshing or crack tracking.
The interparticle contact forces and external forces such as wall contact and gravity are handled using a discrete element method (DEM)-style framework. Here, forces are computed between distinct particles or between particles and boundary surfaces based on their relative positions.
\subsection{Time Integration}
To evolve the system in time, we implement the Velocity-Verlet scheme \cite{HairerEtAlGeometric03}, a symplectic integrator commonly used in molecular dynamics and particle simulations. This second-order method provides excellent energy conservation properties for long-duration simulations, which is crucial for accurately capturing the dynamics of particle systems.
The Velocity-Verlet algorithm advances both positions and velocities of all nodes in a two-step process. First, it updates positions using current velocities and accelerations. Then, it calculates new accelerations based on these updated positions before completing the velocity update. This leapfrog-like structure makes the algorithm particularly stable for the oscillatory behavior common in particle systems.
We selected this time integration scheme due to its balance of computational efficiency and numerical accuracy. For our simulations, we typically use a time step of 2 microseconds, which we determined through stability analysis to be sufficiently small to capture the relevant dynamics while allowing for efficient computation.\\
\\
The combination of our spatial discretization approach and the Velocity-Verlet time integration creates a robust numerical framework capable of handling complex particle geometries, contact interactions, and fragmentation events without requiring explicit tracking of crack surfaces or fragment boundaries.}

\subsection{Bond Removal Strategies for Non-Convex Geometries}
\label{section: bond removal}
In peridDEM simulations, the initial configuration of particles and their interactions must accurately reflect the physical geometry of the problem. For non-convex geometries, such as pre-notched plates or hollow spheres, specific bond removal strategies are employed to ensure that bonds outside the domain of interest are removed. This subsection outlines the bond removal strategies for two specific cases: the Kalthoff-Winkler experiment and hollow spheres.

\subsubsection{Kalthoff-Winkler Experiment}\label{subsec:bnd-kalthof}
In the Kalthoff-Winkler experiment, a plate is pre-notched in a V or U shape, and a plane is defined through the middle of the notch. Bonds that cross this plane from both sides of the notch are removed in the initial configuration to accurately represent the physical conditions of the experiment. Let the plane defining the notch be represented by the equation:
\begin{align*}
ax + by + cz + d = 0,
\end{align*}
where \(a\), \(b\), \(c\), and \(d\) are constants defining the orientation and position of the plane. Consider two particle point \(P_1(x_1, y_1, z_1)\) and \(P_2(x_2, y_2, z_2)\) connected by a bond. The parametric equation of the line connecting \(P_1\) and \(P_2\) is:
\begin{align*}
\mathbf{r}(t) = \mathbf{P}_1 + t (\mathbf{P}_2 - \mathbf{P}_1),
\end{align*}
where \(t \in [0, 1]\). The bond crosses the plane if there exists a \(t\) such that:
\begin{align*}
a(x_1 + t(x_2 - x_1)) + b(y_1 + t(y_2 - y_1)) + c(z_1 + t(z_2 - z_1)) + d = 0.
\end{align*}
Solving for \(t\):
\begin{align*}
t = -\frac{a x_1 + b y_1 + c z_1 + d}{a(x_2 - x_1) + b(y_2 - y_1) + c(z_2 - z_1)}.
\end{align*}
If \(0 < t < 1\), the bond crosses the plane and is removed from the initial configuration.

\subsubsection{Hollow Spheres}
For hollow spheres, bonds that lie entirely within the inner sphere or connect a point inside the inner sphere to a point outside must be removed. This ensures that no interactions occur between particles in the void region and those in the material region. 
Let the inner sphere be defined by its center \(\mathbf{c} = (c_x, c_y, c_z)\) and radius \(R_{\text{inner}}\). A bond between two particles \(P_1(x_1, y_1, z_1)\) and \(P_2(x_2, y_2, z_2)\) is removed if:
\begin{enumerate}
    \item Both particles lie inside the inner sphere:
    \begin{align*}
    (x_1 - c_x)^2 + (y_1 - c_y)^2 + (z_1 - c_z)^2 < R_{\text{inner}}^2
    \end{align*}
    and
    \begin{align*}
    (x_2 - c_x)^2 + (y_2 - c_y)^2 + (z_2 - c_z)^2 < R_{\text{inner}}^2.
    \end{align*}
    \item One particle lies inside the inner sphere, and the other lies outside:
    \begin{align*}
    (x_1 - c_x)^2 + (y_1 - c_y)^2 + (z_1 - c_z)^2 < R_{\text{inner}}^2
    \end{align*}
    and
    \begin{align*}
    (x_2 - c_x)^2 + (y_2 - c_y)^2 + (z_2 - c_z)^2 \geq R_{\text{inner}}^2,
    \end{align*}
    or vice versa.
\end{enumerate}

These bond removal strategies ensure that the initial configuration of the simulation accurately reflects the physical geometry of the problem, enabling accurate modeling of deformation and fracture propagation in non-convex geometries.

\section{Simulation Setup and Initial Condition}\label{sec:sim-setup}

\begin{table}[h!]
\centering
\caption{Common material properties used across simulations.}
\begin{tabular}{cccc}
\hline
\textbf{Material} & \textbf{Young's modulus ($E$)} & \textbf{Bulk modulus ($k$)} & \textbf{Density ($\rho$)} \\ \hline
$M_1$ & $191 \times 10^9$ Pa & $159.2 \times 10^9$ Pa & 8000 kg/m$^3$ \\ \hline
$M_2$ & $1.23 \times 10^9$ Pa & $2 \times 10^9$ Pa & 1200 kg/m$^3$ \\ \hline
\end{tabular}
\end{table}


\subsection{Kalthoff-winkler 3D}
\label{Kalhtoff}
\begin{figure}[!htb]
\minipage{0.45\textwidth}
  \includegraphics[width=\linewidth]{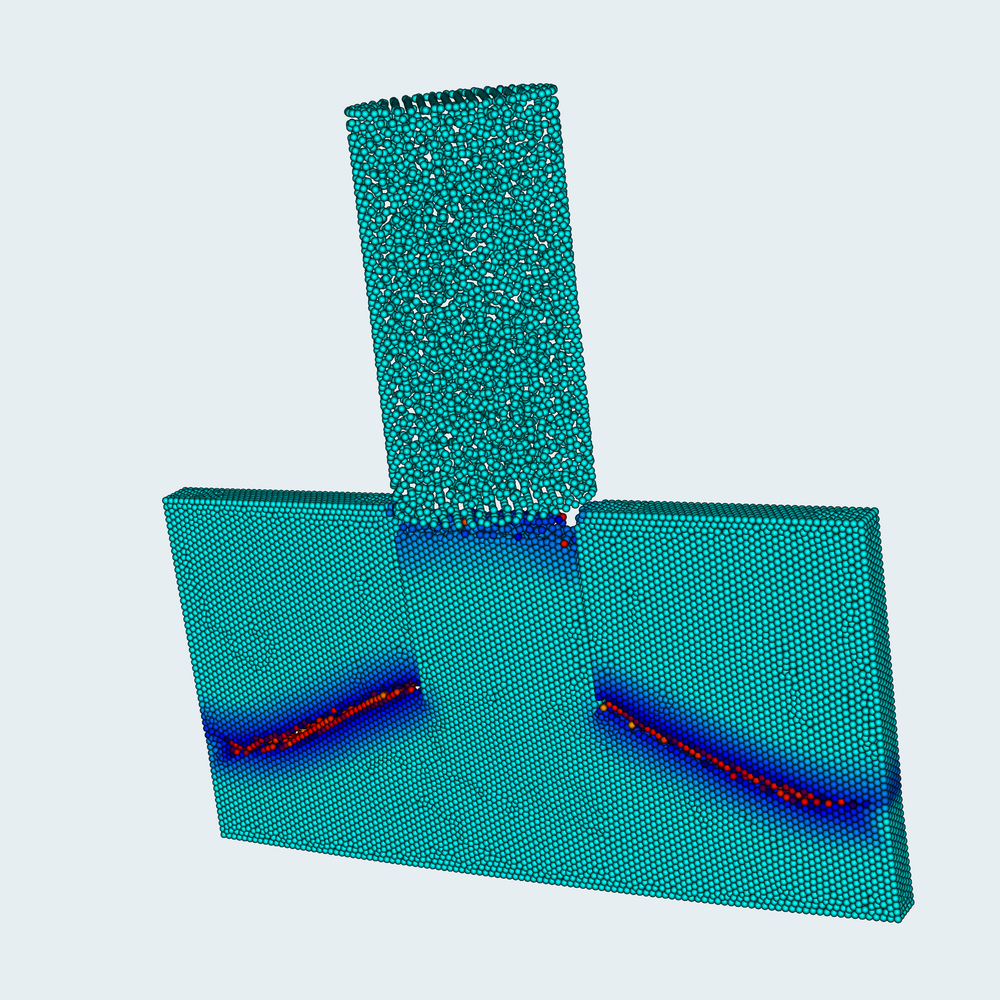}
   \endminipage\hfill
\minipage{0.45\textwidth}
  \includegraphics[width=\linewidth]{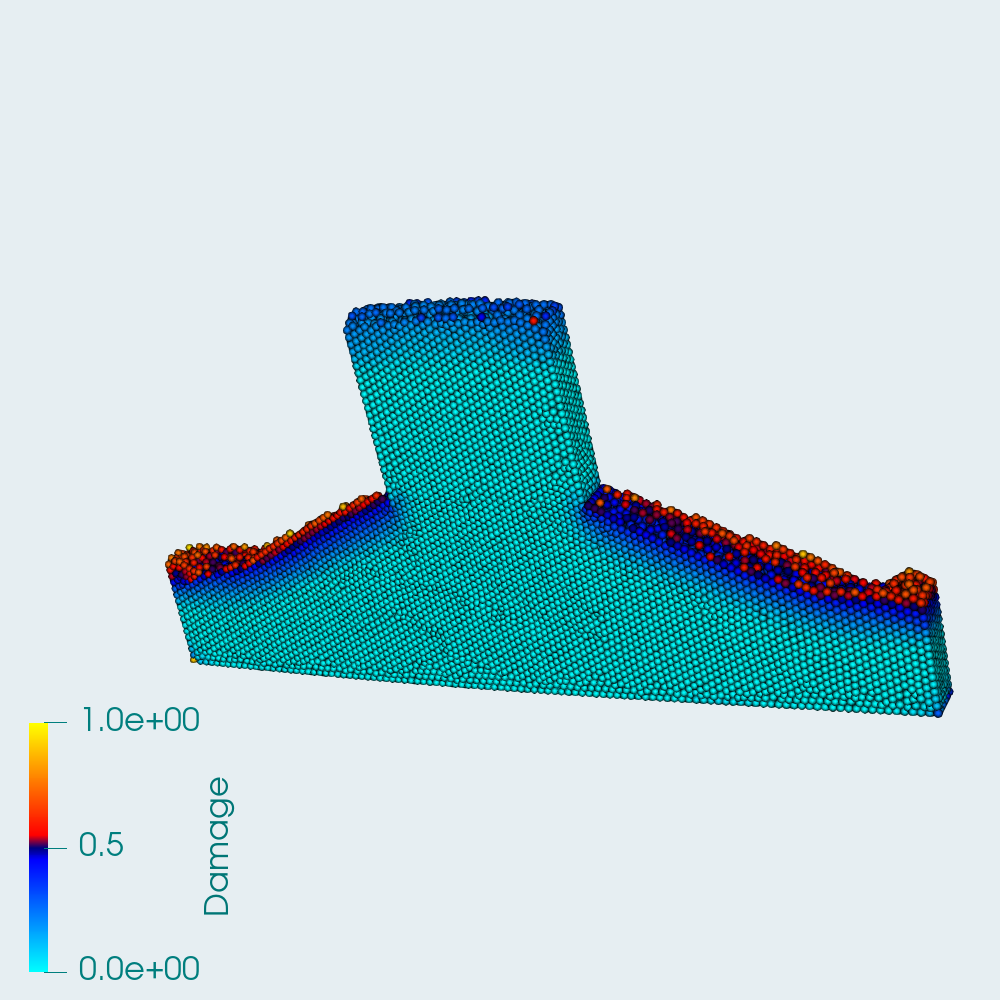}
   \endminipage\hfill
\caption{Three-dimensional cracking pattern for the Kalthoff--Winkler simulation for a thin 3D plate.  }\label{fig:kaltf-1}
\end{figure}

The Kalthoff-Winkler experiment, first conducted in 1988, serves as a well-established benchmark for dynamic fracture problems. In this experiment, a cylindrical impactor strikes a pre-notched steel plate, resulting in a reproducible crack pattern. The setup typically involves a rectangular plate (100 mm width, 200 mm height, $9-19$ mm thickness) with two parallel notches, impacted by a cylindrical projectile (100 mm length, 50 mm diameter) at a constant speed of 30 m/s. The plate and projectile are made of high-strength maraging steel X2 NiCoMo 1895, with material properties including Young's modulus E = 190 GPa, Poisson's ratio $\nu = 0.3$, density $\rho = 8000 kg/m^3$, and critical intensity factor $KI = 68 MPam^{\frac{1}{2}}$. The small crack tip plastic zone of this material makes it particularly suitable for linear elastic or small-scale yielding fracture mechanics analysis.
In this work, we validate our dynamic fracture and contact model for nonconvex domains using the PeriDEM method to simulate the Kalthoff-Winkler experiment in three dimensions. Unlike previous studies that used mode II dynamic displacement conditions to represent the impactor's effect, we model the experiment as a two-particle collision problem, solving the equation of motion in three dimensions \eqref{equ: motion} for $N = 2$.  {Our simulation models a cylindrical impactor initially traveling at 32m/s towards a stationary notched plate, with the left and right ligaments held fixed at the top. While our framework can accommodate different notch geometries, such as U-shaped or V-shaped notches, this study specifically uses V-shaped notches to match the configuration of the Kalthoff–Winkler experiment.} We remove the peridynamic bonds across the notches using the approach described in our methodology section  {\ref{subsec:bnd-kalthof}}.  {This 3D PeriDEM simulation (Fig.~\ref{fig:kaltf-1}) allows us to capture the complex dynamics of the impact and resulting crack propagation, providing a more comprehensive representation of the experimental conditions compared to previous 2D models. Notably, we observe a converged fragment shape that reproduces the experimentally observed $68^{\circ}$ crack angle at the pre-notch tip, consistent with reported results from the Kalthoff–Winkler experiment.}



\subsection{Particle Breakage and Comminution }
\label{Particle Breakige}
Particle  crushing (also known as particle breakage) and comminution are critical processes in numerous industrial applications involving granular materials, including soil mechanics, powder processing, mining, and pharmaceuticals. These phenomena, which involves the fracture and disintegration of individual particles under applied mechanical stresses, significantly impact macroscopic properties such as strength, dilatancy, and permeability. Quantifying the extent of comminution and breakage is crucial, leading to the proposal of various particle breakage factors. However, modeling these processes remains challenging due to complexities at both the particle and representative volume element (RVE) levels. Current approaches, often based on Discrete Element Method (DEM), face limitations such as computational inefficiencies, oversimplified assumptions about particle composition and crushing conditions, and neglect of particle shape effects.
The PeriDEM method in 3D offers a promising solution to these challenges, enabling a natural way to capture particle breakage evolution without additional assumptions. Unlike traditional methods that rely on predetermined child particle formations or cluster approaches, PeriDEM can simulate the five modes of grain crushing identified by Nakata et al.\cite{nakata2001microscopic}: no visible damage, single asperity crushing, multiple asperity fracture, major splitting, and further crushing of sub-particles. This approach allows for a more realistic and comprehensive simulation of particle fracturing, addressing the need for discrete simulation of particle breakage and continuous evolution of particle properties during loading. By overcoming the limitations of current models, PeriDEM provides a more rigorous and efficient tool for studying particle crushing in various industrial and natural systems.

\begin{figure}[htbp]
    \centering
    {
        \includegraphics[width=0.3\textwidth]{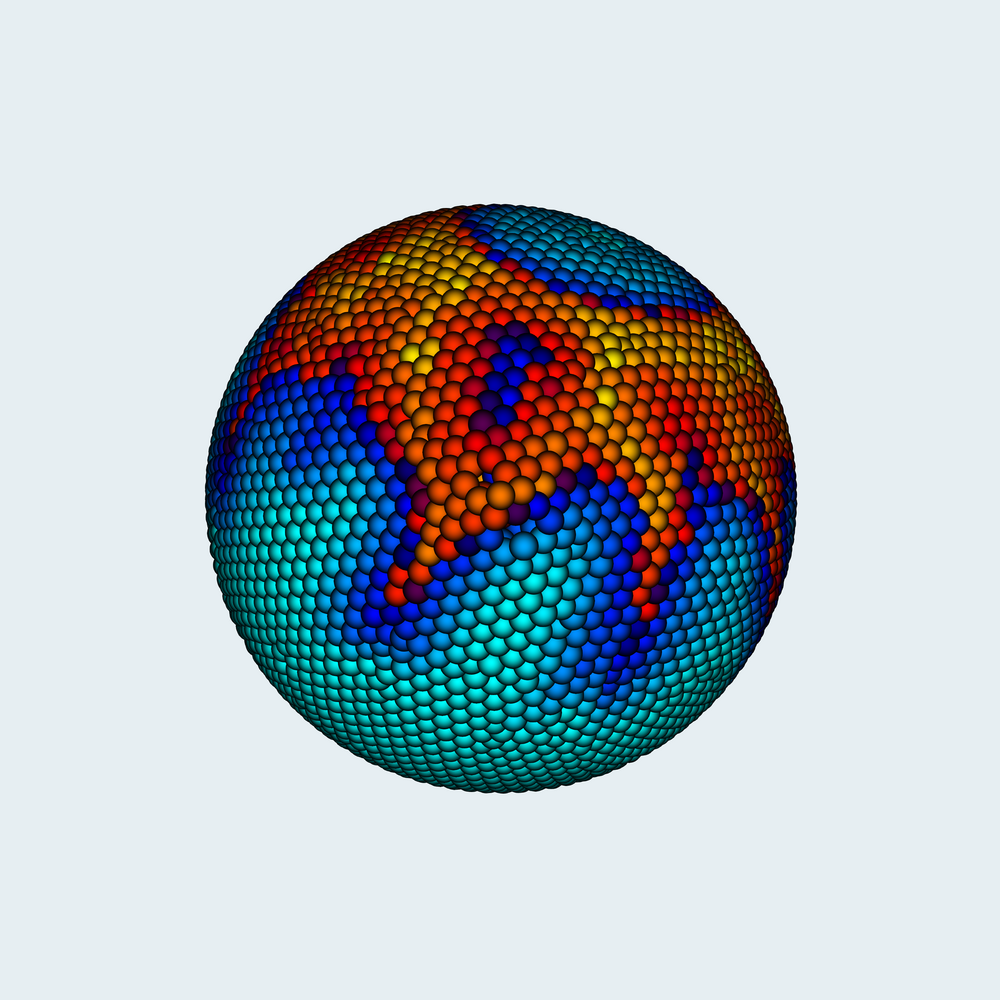}
    }
    {
        \includegraphics[width=0.3\textwidth]{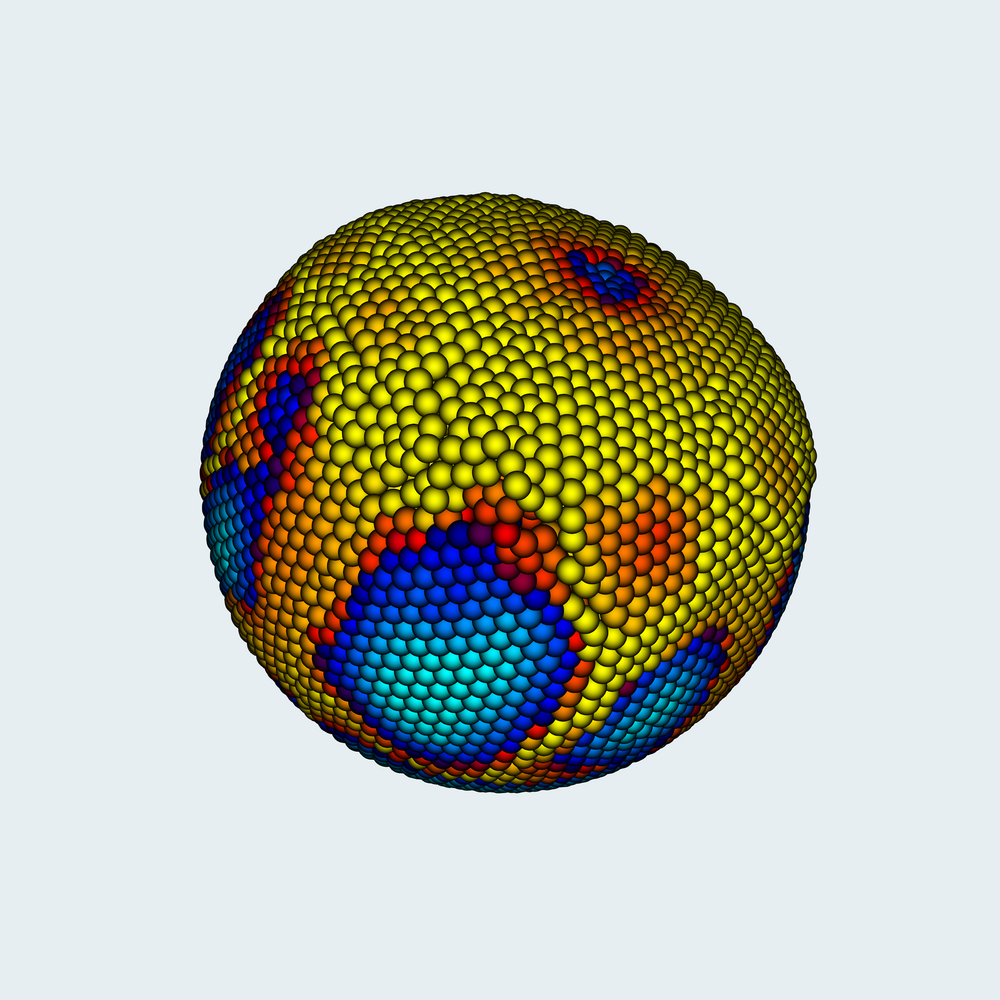}
    }
    {
        \includegraphics[width=0.3\textwidth]{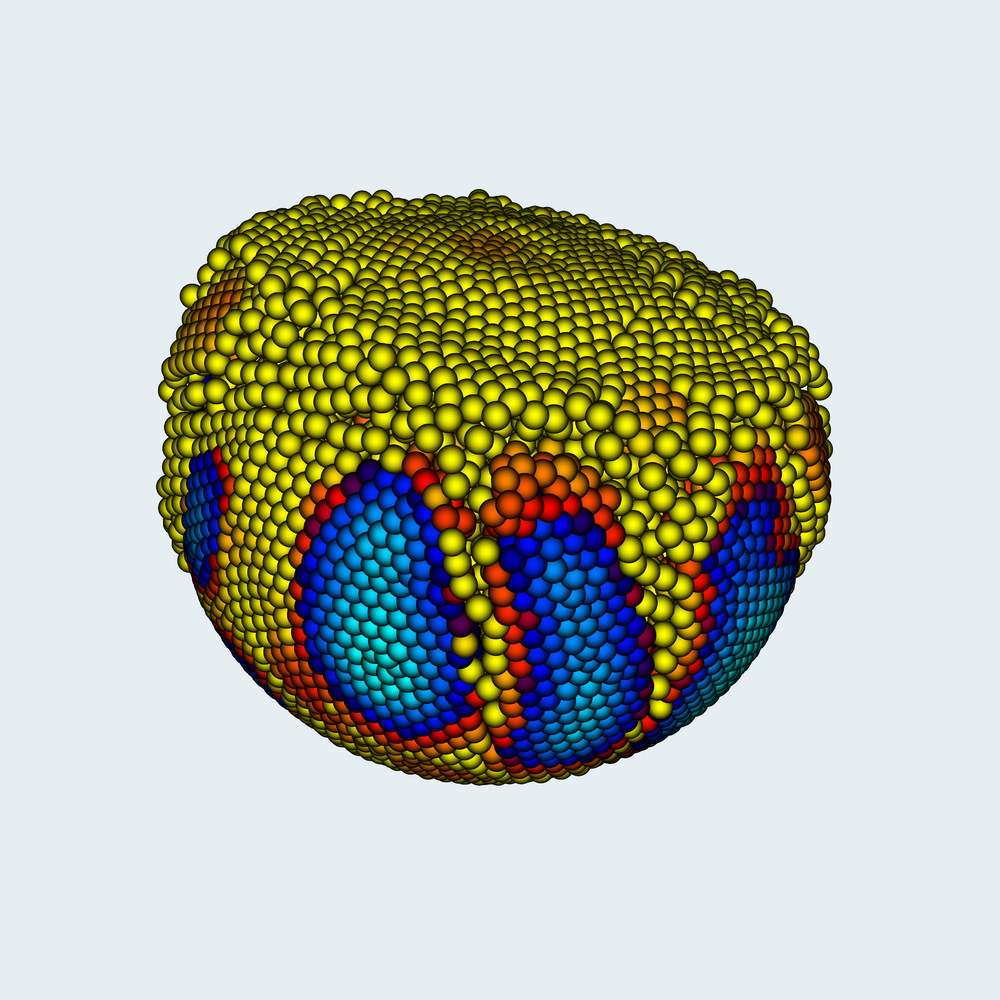}
    }
    \subfigure[$t=50$]{
        \includegraphics[width=0.3\textwidth]{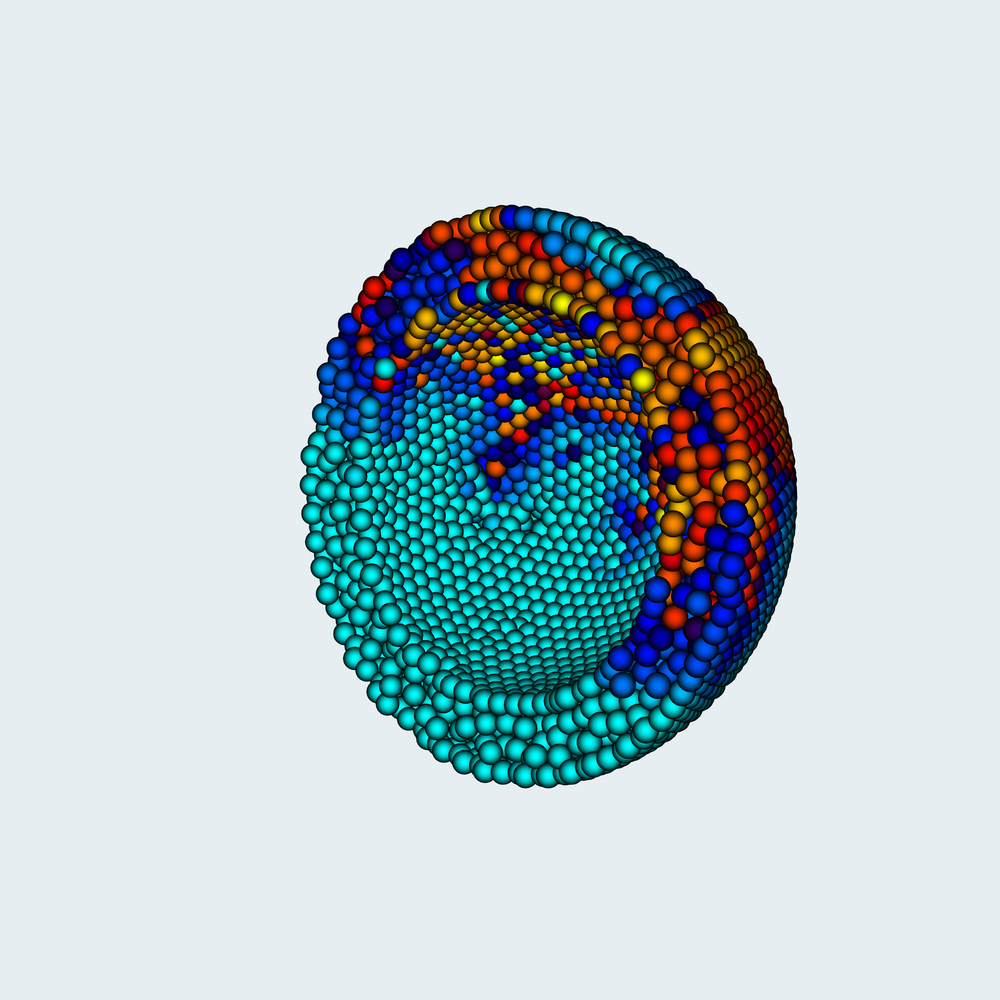}
    }
    \subfigure[$t=200$]{
        \includegraphics[width=0.3\textwidth]{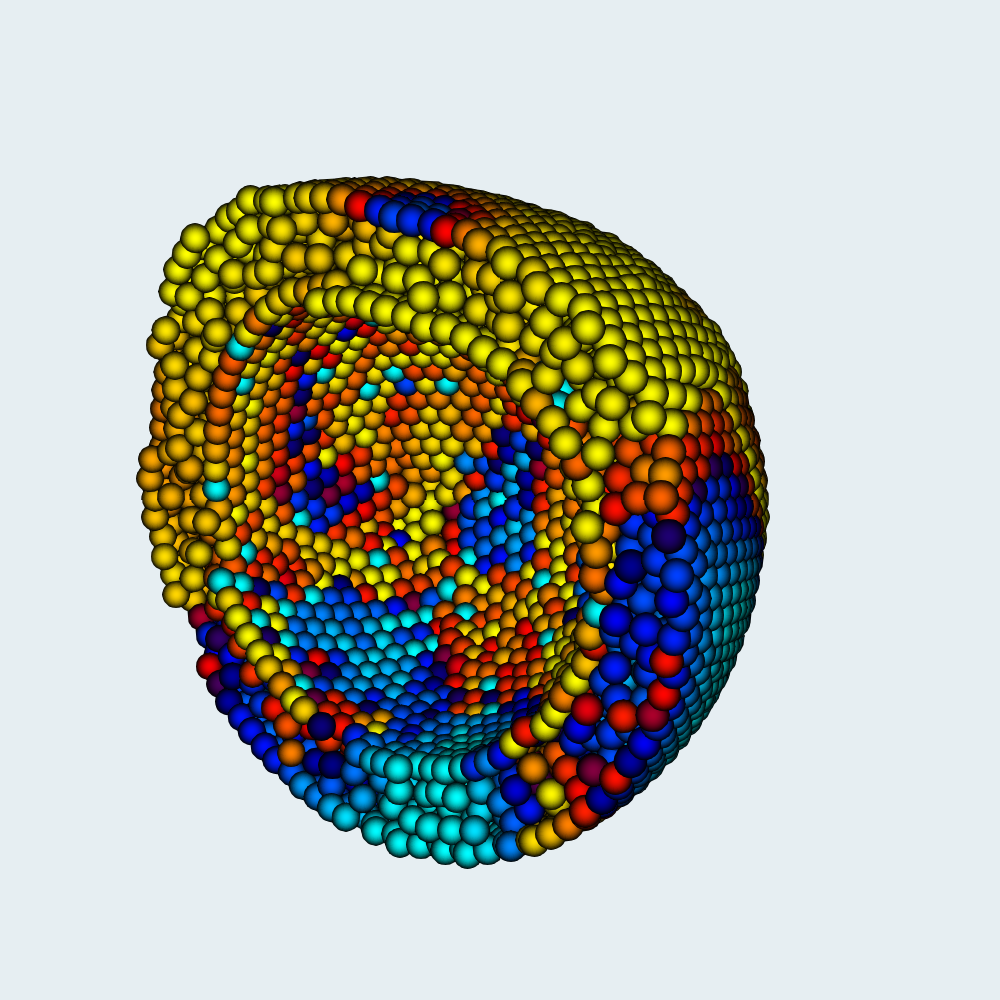}
    }
    \subfigure[$t=500$]{
        \includegraphics[width=0.3\textwidth]{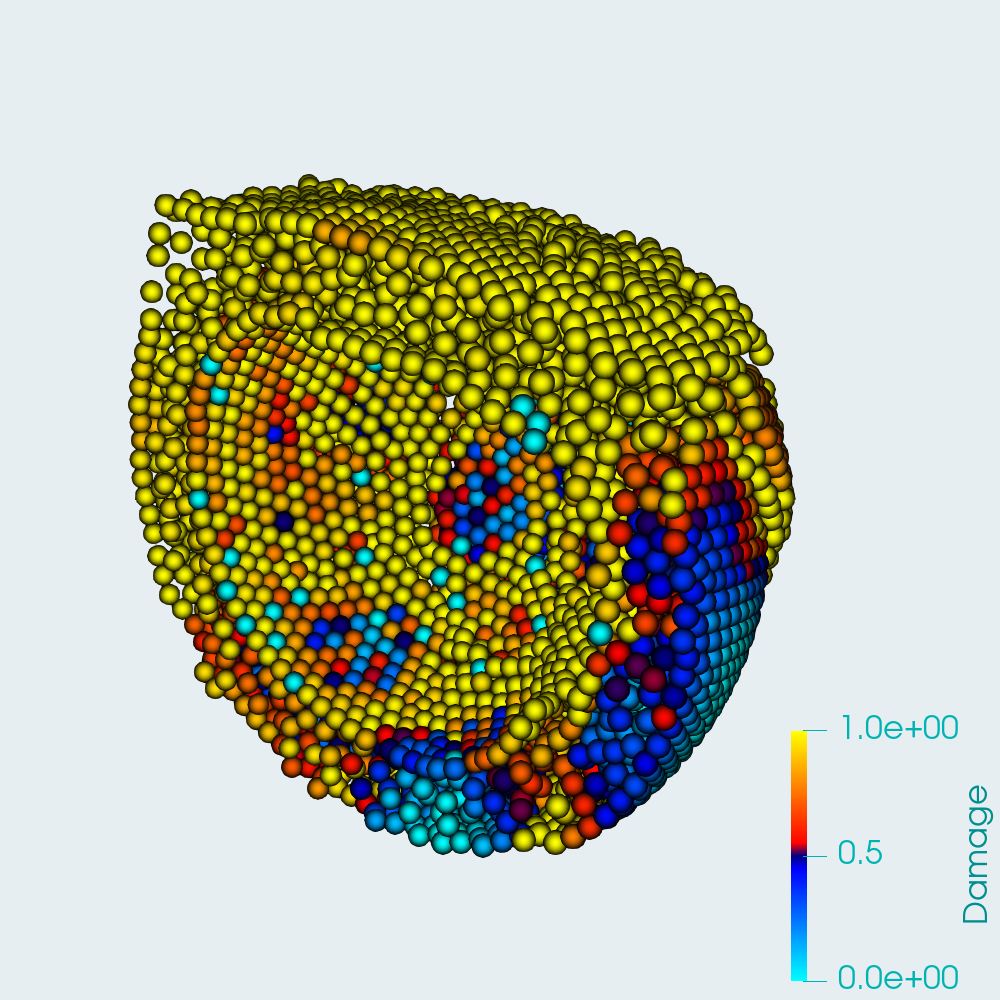}
    }
    \caption{ {PeriDEM simulation results for a hollow sphere with inner radius $r=R/2$. The top row shows the full 3D outer surface of the sphere, while the bottom row displays cross-sectional views of the same configurations. From left to right, the subplots correspond to time steps $t=50$, $t=200$, and $t=500$ in units of $5 \mu$s, illustrating the progressive deformation and damage under compression between two rigid plates.}}\label{fig:communi-3r/4}
\end{figure}

\begin{figure}[htbp]
    \centering
       {
        \includegraphics[width=0.3\textwidth]{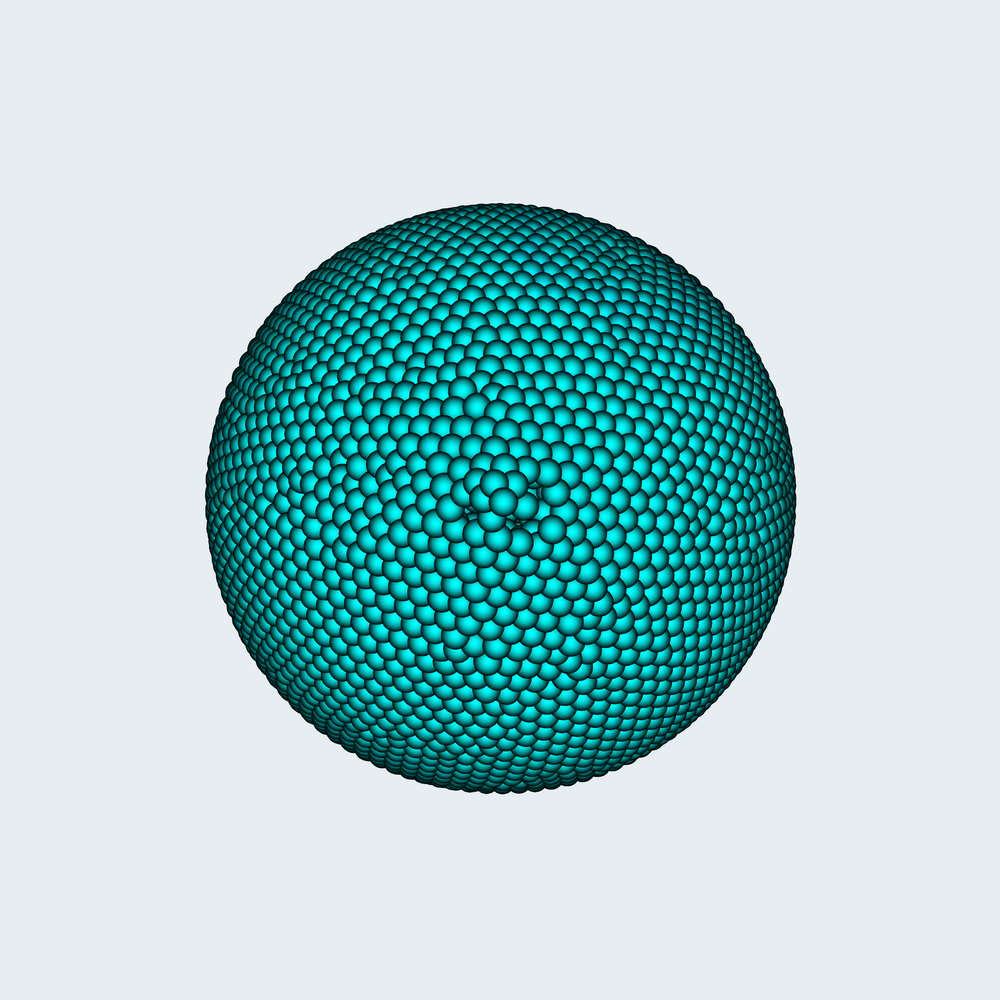}
    }
    {
        \includegraphics[width=0.3\textwidth]{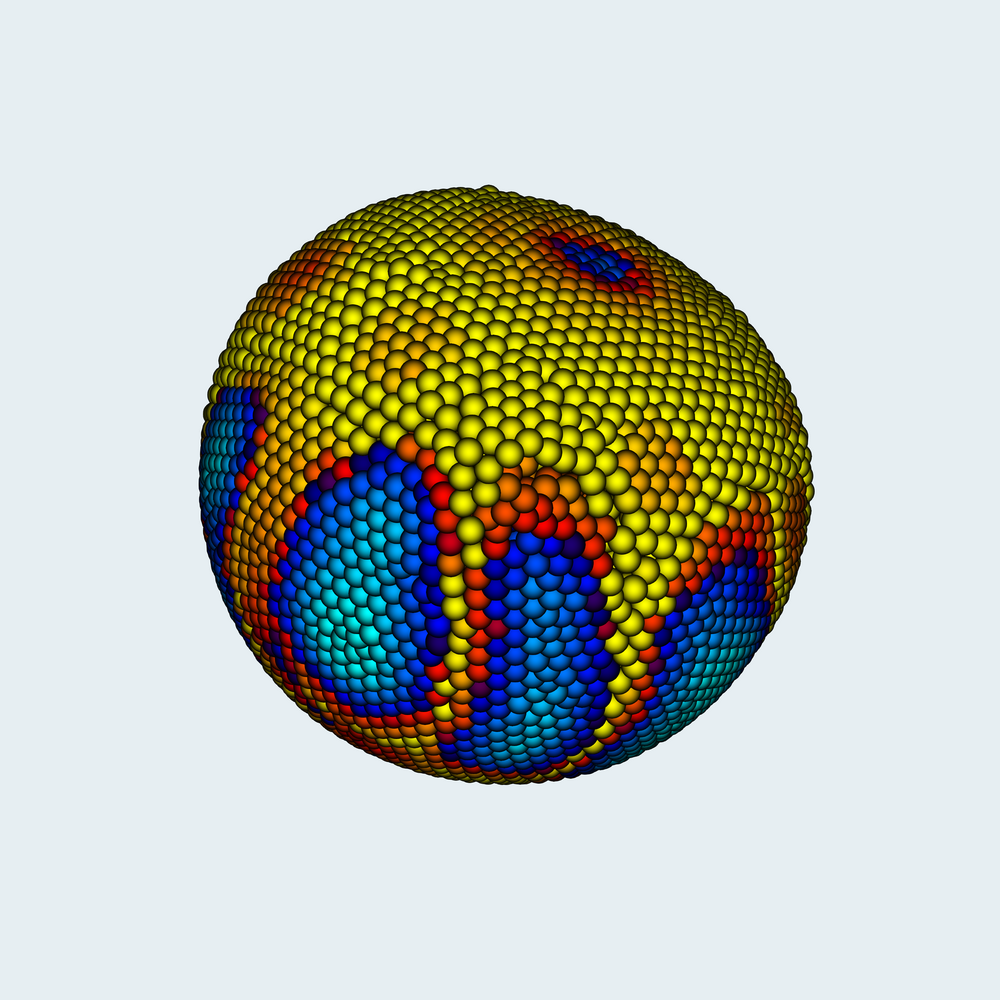}
    }
    {
        \includegraphics[width=0.3\textwidth]{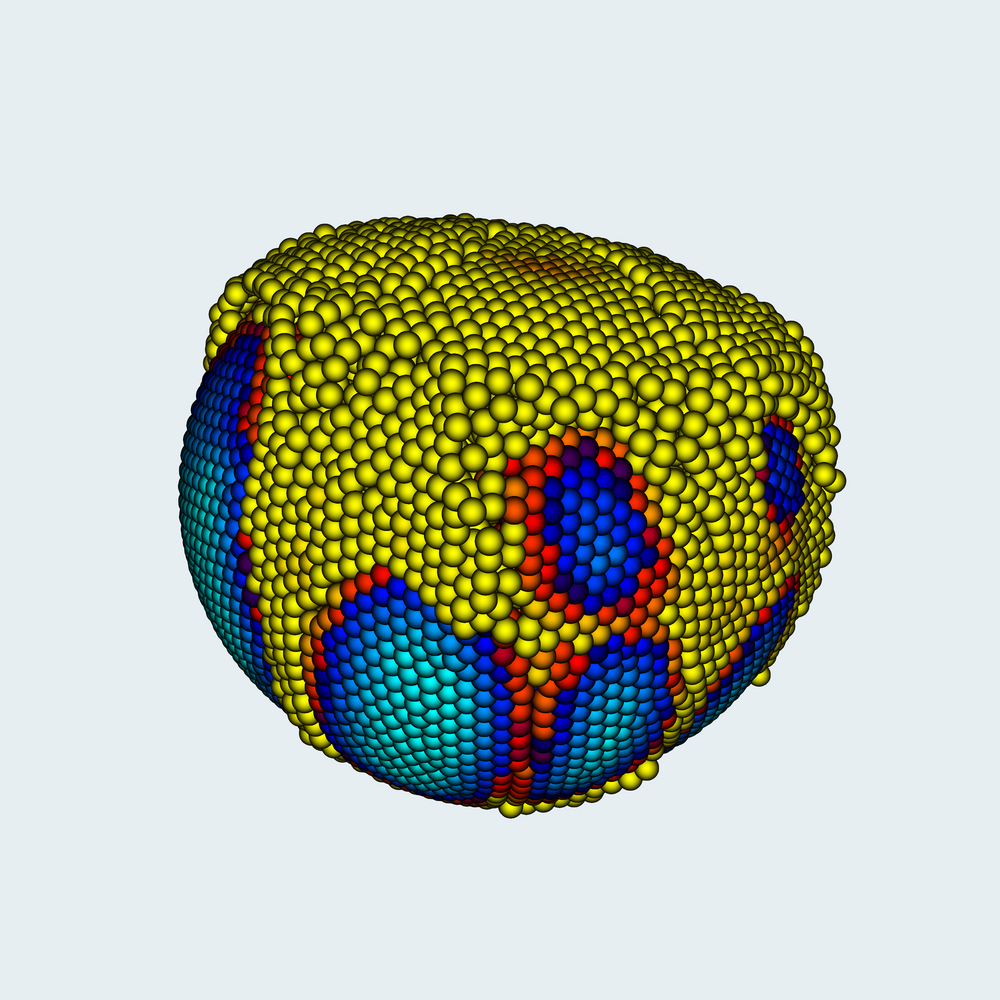}
    }
    \subfigure[$t=50$]{
        \includegraphics[width=0.3\textwidth]{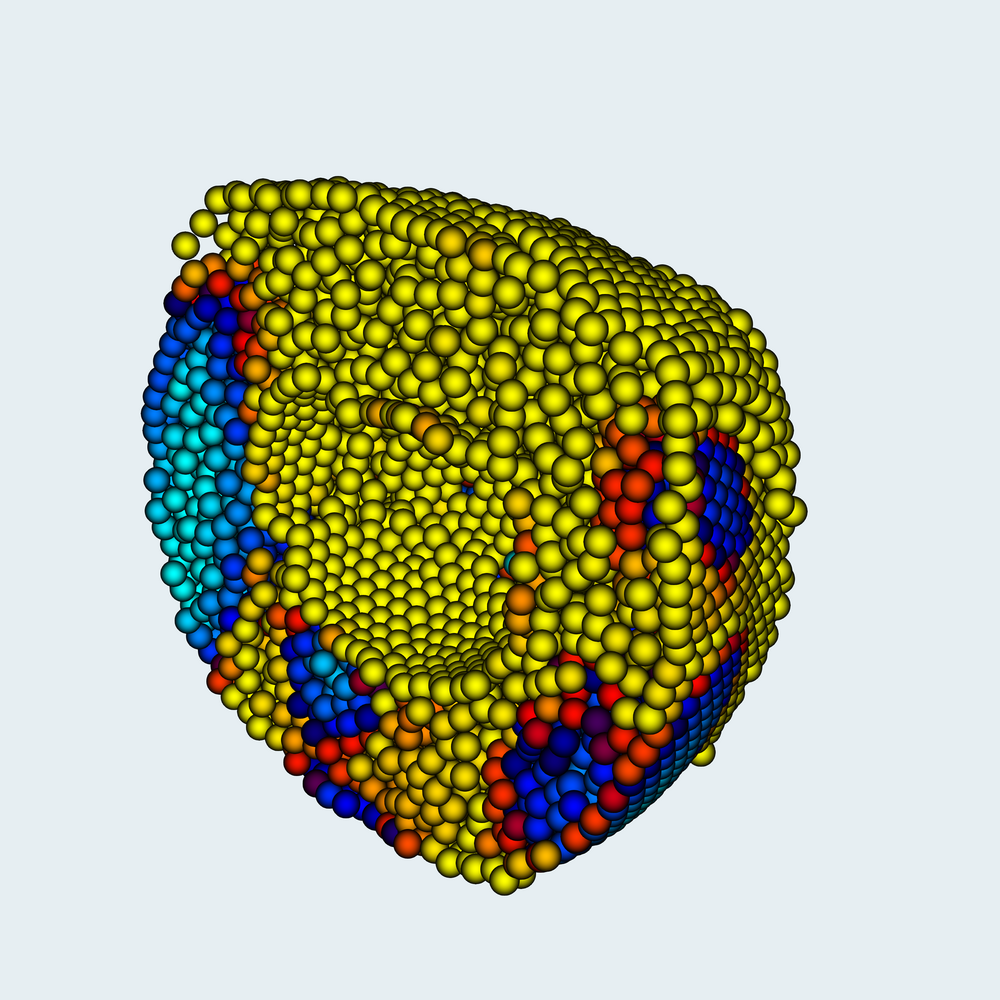}
    }
    \subfigure[$t=200$]{
        \includegraphics[width=0.3\textwidth]{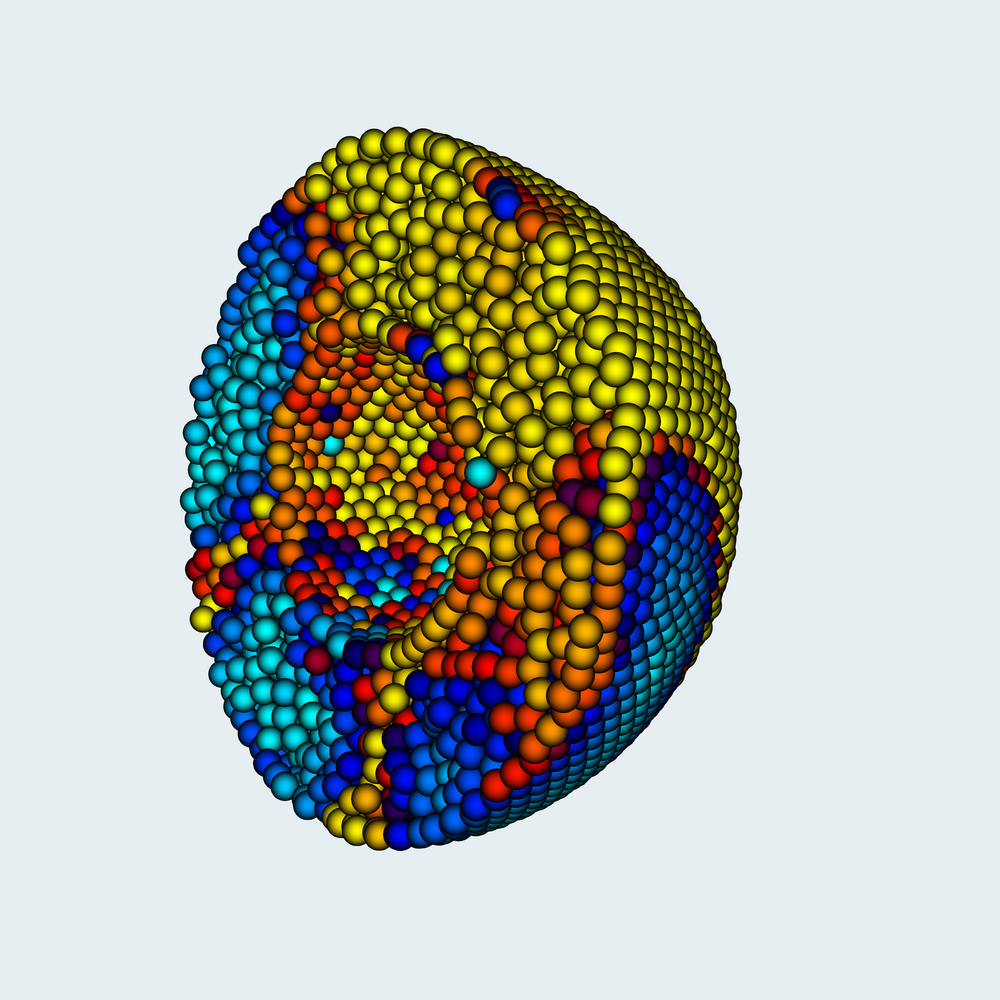}
    }
    \subfigure[$t=500$]{
        \includegraphics[width=0.3\textwidth]{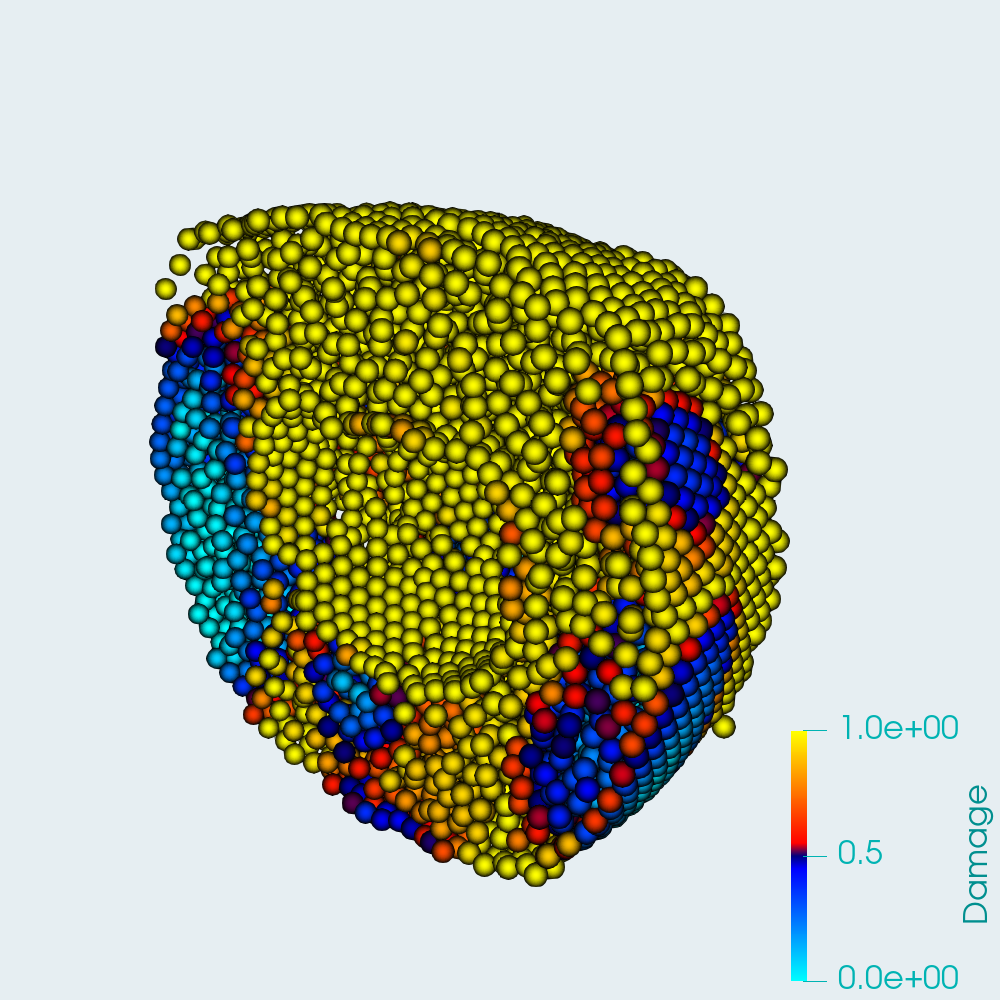}
    }
    \caption{ {PeriDEM simulation results for a hollow sphere with inner radius $r=3R/4$. The layout and time steps are identical to Fig.~\ref{fig:communi-r/2}, with the top row showing the full spheres and the bottom row showing cross-sections. The results highlight the influence of shell thickness on fracture and deformation behavior during compression.}}
    \label{fig:communi-r/2}
\end{figure}

Fig.\ref{fig:communi-r/2} and Fig.\ref{fig:communi-3r/4} present the results of PeriDEM compression simulations for a single hollow sphere.  {The simulations are conducted inside a cubic container where all walls are held rigid, except for the top wall, which moves downward at a constant velocity of 10 m/s to apply compressive loading. }We compare the results for two different hollow spheres with inner radii of $r=R/2$ and $r=3R/4$, respectively.  These figures demonstrate the outcome of tests performed by compressing a hollow sphere between two rigid plates. As can be observed, PeriDEM captures the realistic behavior of the experiment, accurately simulating the crushing process for two different hollow sphere geometries. Simulations show the thinner shell, the hollow sphere breaks easier than the hollow sphere with thicker shell under the same tension.  This is seen in Fig.~\ref{fig:communi-r/2} and Fig.~\ref{fig:communi-3r/4}  showing the results of PeriDEM simulations for two hollow spheres with different internal radii  $r=R/2$ and $r=3/4R$. Each figure contains six subplots illustrating both the outer surface and cross-sections of the hollow spheres, demonstrating how the material responds to compression between two rigid plates.

In Fig.~\ref{fig:communi-r/2}, the sphere, with a larger interior void, exhibits a more pronounced crushing behavior under compression. The subplots reveal significant internal and external breakage, as seen from both the surface patterns and the cross-sections. The higher degree of fragmentation is evident from the irregular distribution of stresses, leading to larger and more dispersed damaged regions. The particle structure with the thinner outer shell cannot withstand the applied tension as effectively, resulting in more extensive breakage throughout the shell. This is particularly noticeable in the cross-sectional views, where cracks propagate deeply into the shell, breaking it into smaller fragments.

In Fig.~\ref{fig:communi-3r/4} , the sphere with smaller interior void, which has a thicker spherical shell, demonstrates more resistance to the applied forces. While the spherical shell still deforms and experiences damage, the extent of breakage is significantly reduced compared to the hollow sphere with larger spherical void. The cross-sectional views show that the damage remains more localized, with fewer cracks penetrating deeply into the spherical shell. The thicker walls of the spherical shell allow for better distribution of the stress, preventing widespread failure. 

Overall, these figures confirm that the shell thickness of the hollow sphere plays a crucial role in determining its resistance to crushing. The thinner spherical shell experiences more severe damage and breaks more easily under the same loading conditions, while the thicker shell is more resilient, exhibiting less fragmentation. The PeriDEM method accurately captures these behaviors, closely mirroring the expected physical responses of hollow spheres under compressive loading.


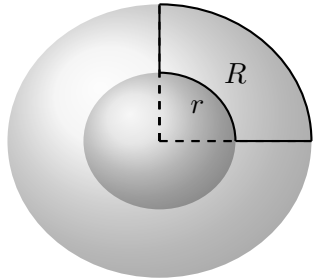
\begin{figure}[htbp]
\centering

\tdplotsetmaincoords{25}{0} 

\begin{tikzpicture}[tdplot_main_coords, scale=1.0]


\def\r{1} 
\def\R{2} 

\shade[ball color=gray!20, opacity=0.5] (0,0,0) circle (\R);

\shade[ball color=gray!40, opacity=0.5] (0,0,0) circle (\r);

\begin{scope}[canvas is xy plane at z=0]
    \draw[thick] (\R,0) arc (0:90:\R);
    \draw[thick] (\r,0) arc (0:90:\r);
    \draw[thick] (0,\r) -- (0,\R);
    \draw[thick] (\r,0) -- (\R,0);
\end{scope}

\draw[thick, dashed] (0,0,0) -- (\R,0,0); 
\draw[thick, dashed] (0,0,0) -- (0,\R,0); 
\draw[thick, dashed] (0,0,0) -- (0,0,\R); 

\draw[thick, dashed] (0,0,0) -- (\r,0,0); 
\draw[thick, dashed] (0,0,0) -- (0,\r,0); 
\draw[thick, dashed] (0,0,0) -- (0,0,\r); 

\node at (0.5*\r, 0.5*\r, 0) {$r$};
\node at (0.5*\R, 0.5*\R, 0) {$R$};

\end{tikzpicture}

\caption{Hollow spherical shell of inner radius $r$ and outer radius $R$.}
\label{fig:hollow}
\end{figure}

\subsection{Force Evolution}

\begin{figure}[htbp]
    \centering
    \subfigure[Hollow spherical shells $r=\frac{1}{2}R$]{
        \includegraphics[width=0.45\textwidth]{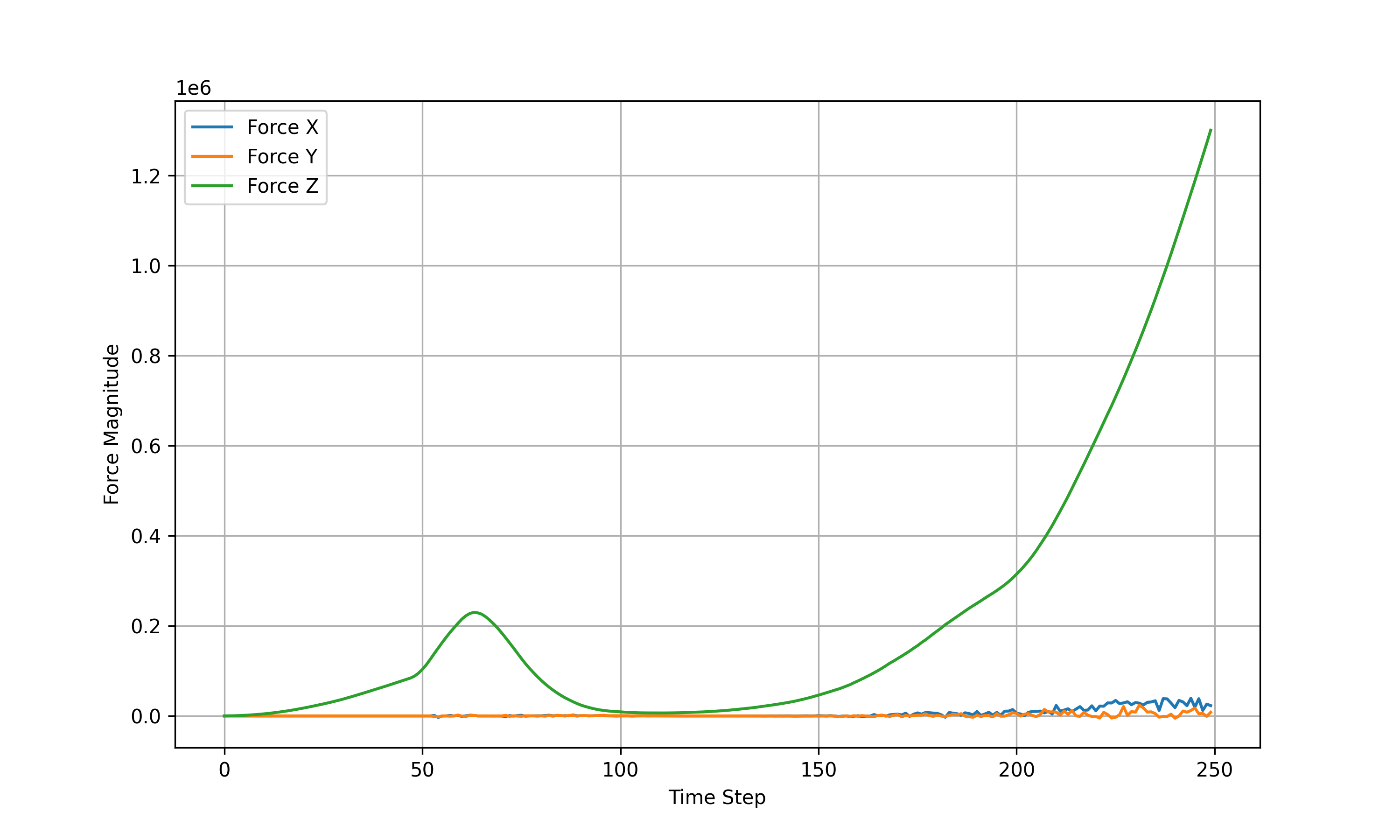}
        
    }
    \subfigure[hollow spherical shells $r=\frac{3}{4}R$]{
        \includegraphics[width=0.45\textwidth]{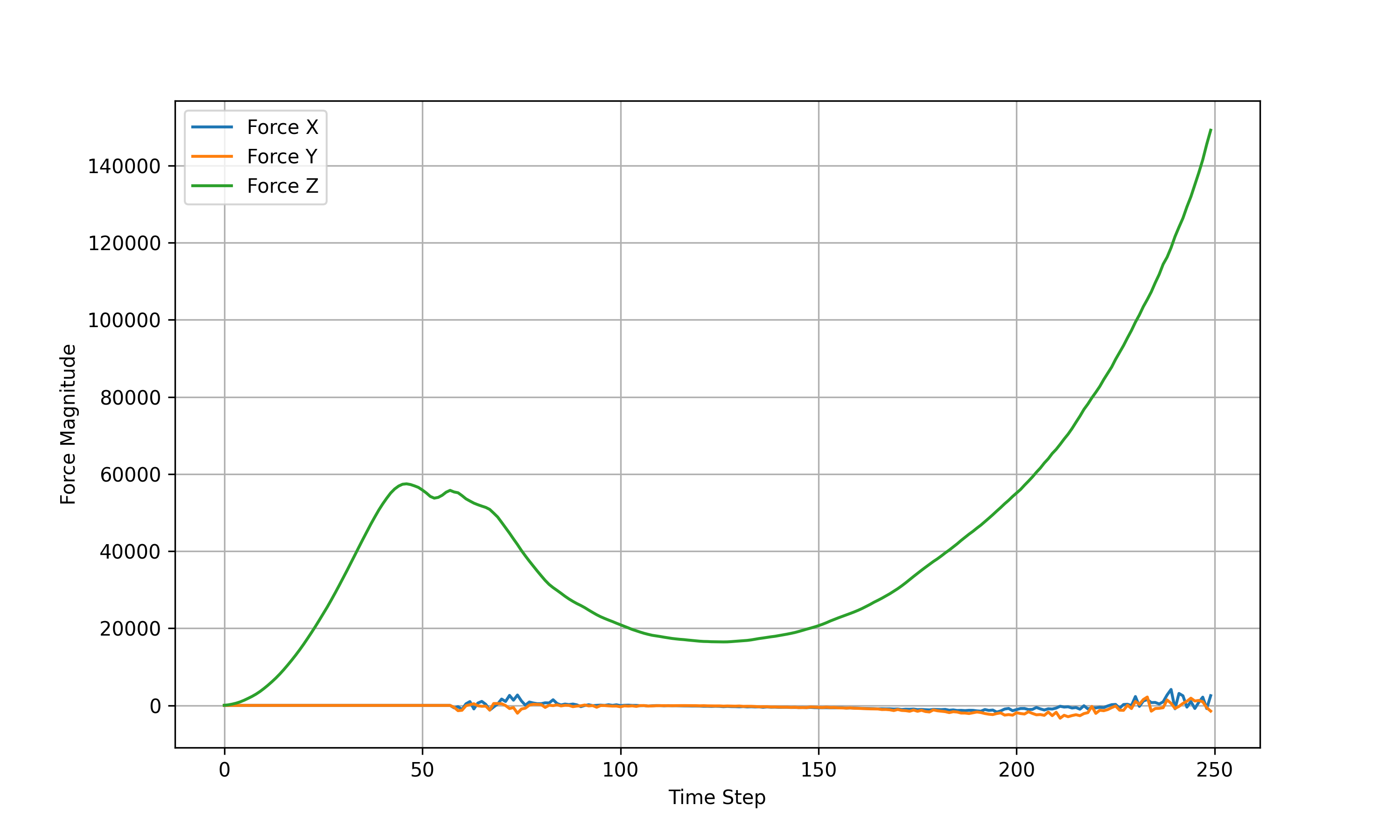}
    }
     \caption{ {Time evolution of the reaction force exerted by the top wall during the compression of hollow spherical shells. The left plot corresponds to a shell with inner radius $r=\frac{1}{2}R$, while the right plot shows the case with $r=\frac{3}{4}R$. The force is decomposed into its X, Y, and Z components, with the vertical Z-component dominating the response due to the applied uniaxial compression. }}\label{fig:force_evolution}
\end{figure}

The resistance force of the particle against  the wall is measured.  The top wall time evolution of the force associated with the compression of the hollow sphere is shown in Fig.~\ref{fig:force_evolution}. The force in the $Z$-direction is dominant in both cases, as expected, since the wall applies a downward compressive force. Initially, the force increases gradually as the spheres deform elastically. Around $t \approx 50$, a peak in the force curve is observed, followed by a drop, which indicates the onset of structural failure and breakage of the shell. After this initial breakage, the force increases again as the wall continues to compress the remaining fragments.

A comparison between the two cases reveals that the thicker spherical shell(\( r = \frac{1}{2} R \)) exhibits a significantly higher force peak, reaching approximately $1.2 \times 10^6$, whereas the thinner spherical shell (\( r = \frac{3}{4} R \)) reaches only around $1.4 \times 10^5$. This suggests that the thicker shell can sustain higher loads before failing. Additionally, after breakage, the force associated with the sphere with smaller void increases more steeply compared to the sphere with larger void. This difference indicates that the fragments of the thick spherical shell interlock and resist further compression, while the fragments of the thin spherical shell rearrange more freely, leading to a more gradual force increase. The force components in the $X$ and $Y$ directions remain relatively small throughout the simulation, with only minor fluctuations, which may be attributed to slight asymmetries in the breakage process or numerical effects in the simulation.

These results highlight the impact of shell thickness on the mechanical response of hollow spheres under compression. The particle with thicker spherical shell exhibits greater resistance to deformation, sustaining a higher peak force before failure and showing stronger post-failure resistance due to fragment interlocking. In contrast, the particle with thinner spherical shell fails more easily and undergoes a more gradual force increase after breakage, suggesting that its fragments are more free to rearrange under continued compression. This analysis demonstrates that shell thickness plays a crucial role in determining the structural integrity and failure characteristics of hollow spheres subjected to compressive loading.

\subsection{Double Hollow Sphere}
\label{double-hollow-sphere}

\begin{figure}[htbp]
    \centering
    {
        \includegraphics[width=0.3\textwidth]{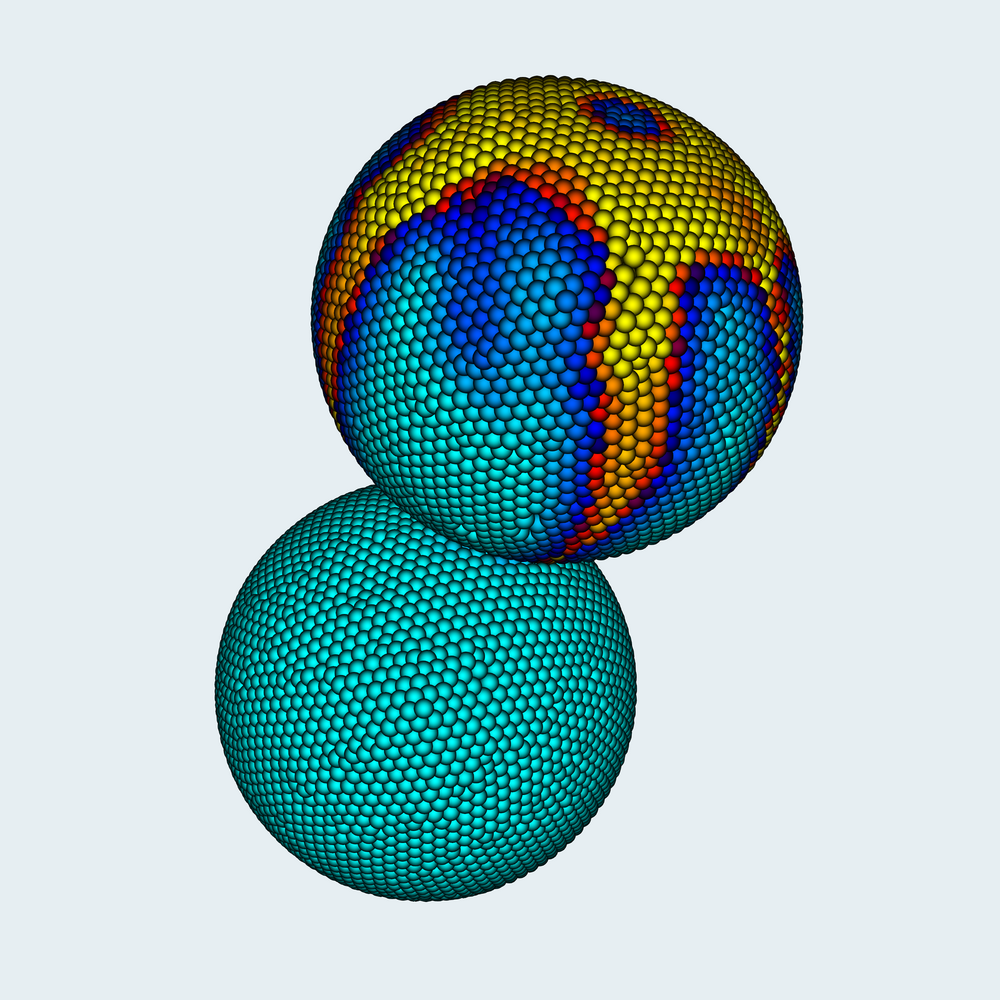}
    }
    {
        \includegraphics[width=0.3\textwidth]{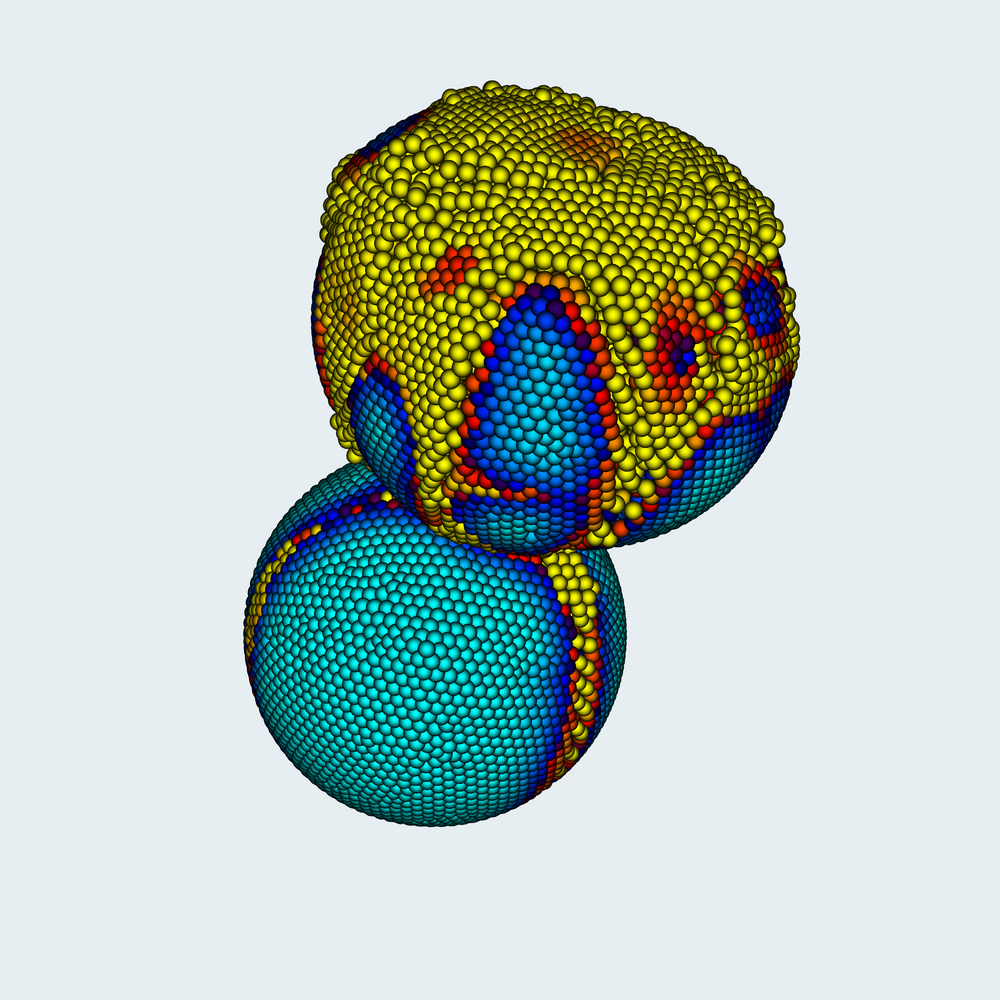}
    }
    {
        \includegraphics[width=0.3\textwidth]{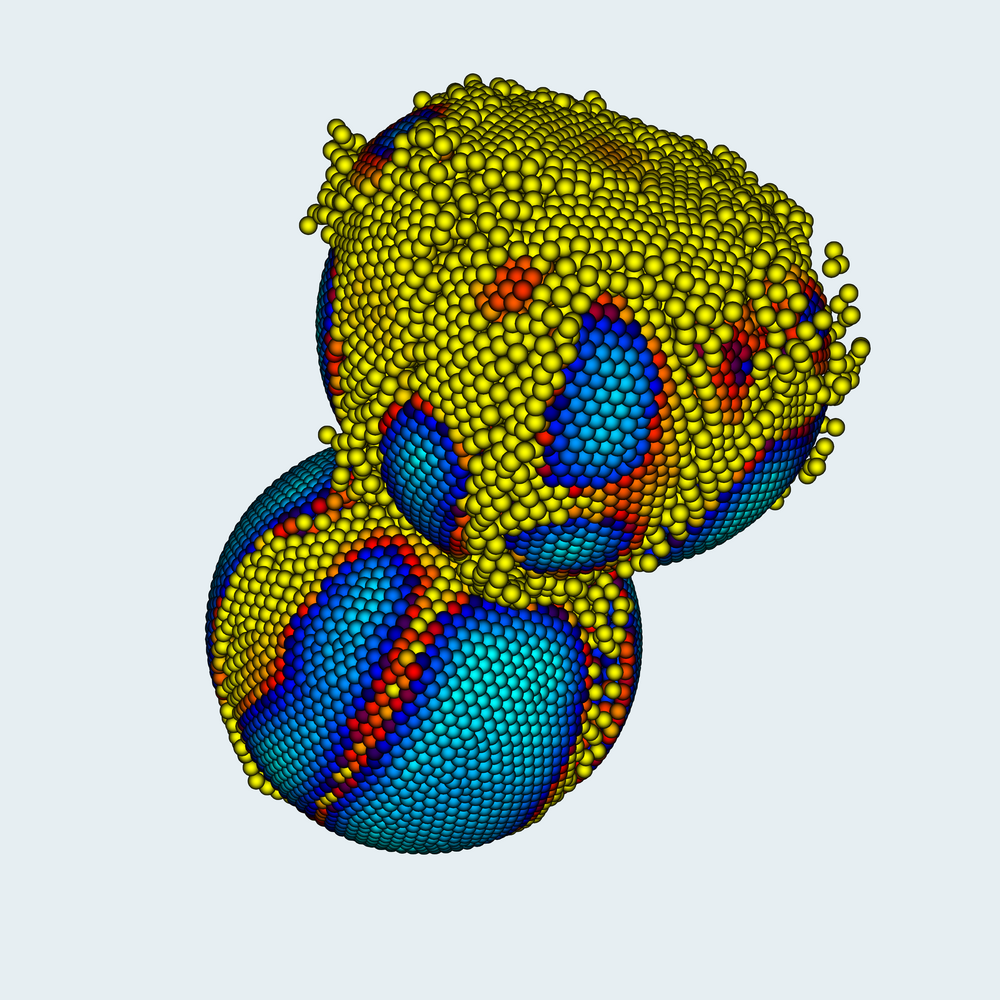}
    }
    \subfigure[][$t=50$]{
        \includegraphics[width=0.3\textwidth]{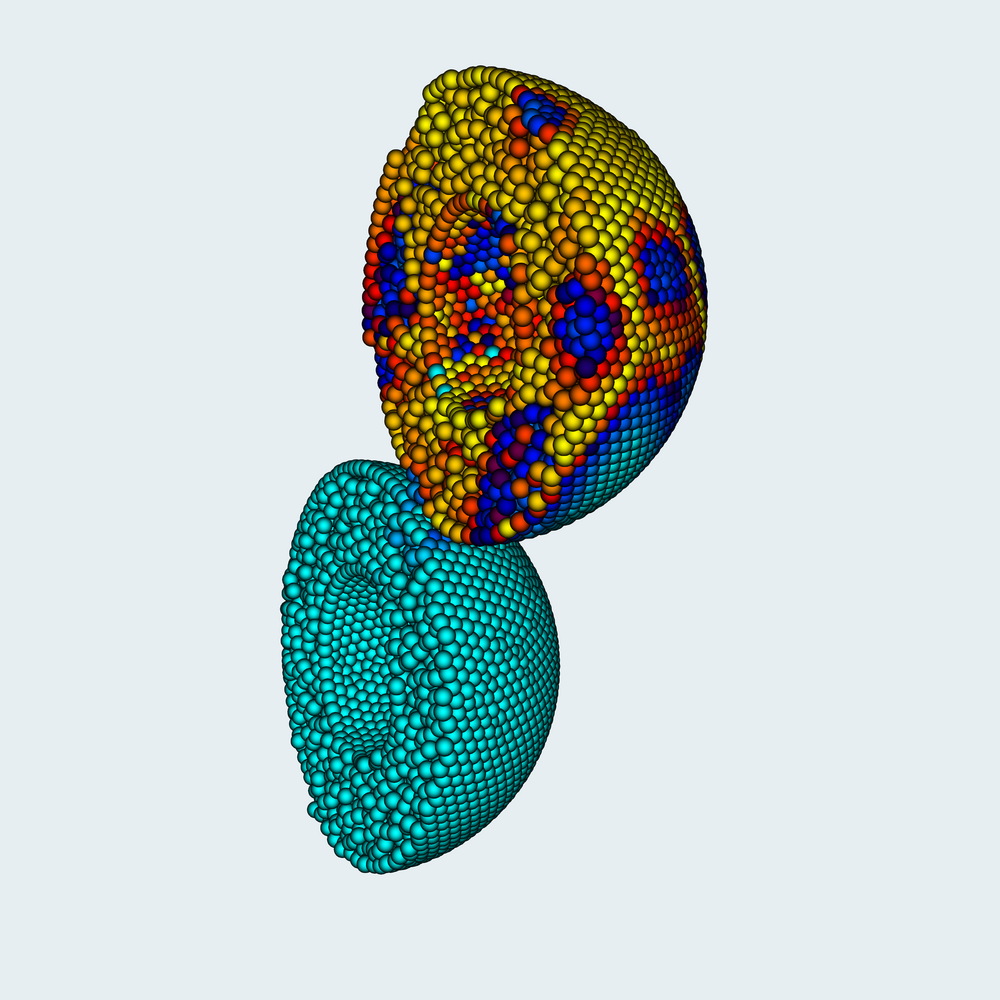}
   \label{fig:DHS-1} }
    \subfigure[][$t=200$]{
        \includegraphics[width=0.3\textwidth]{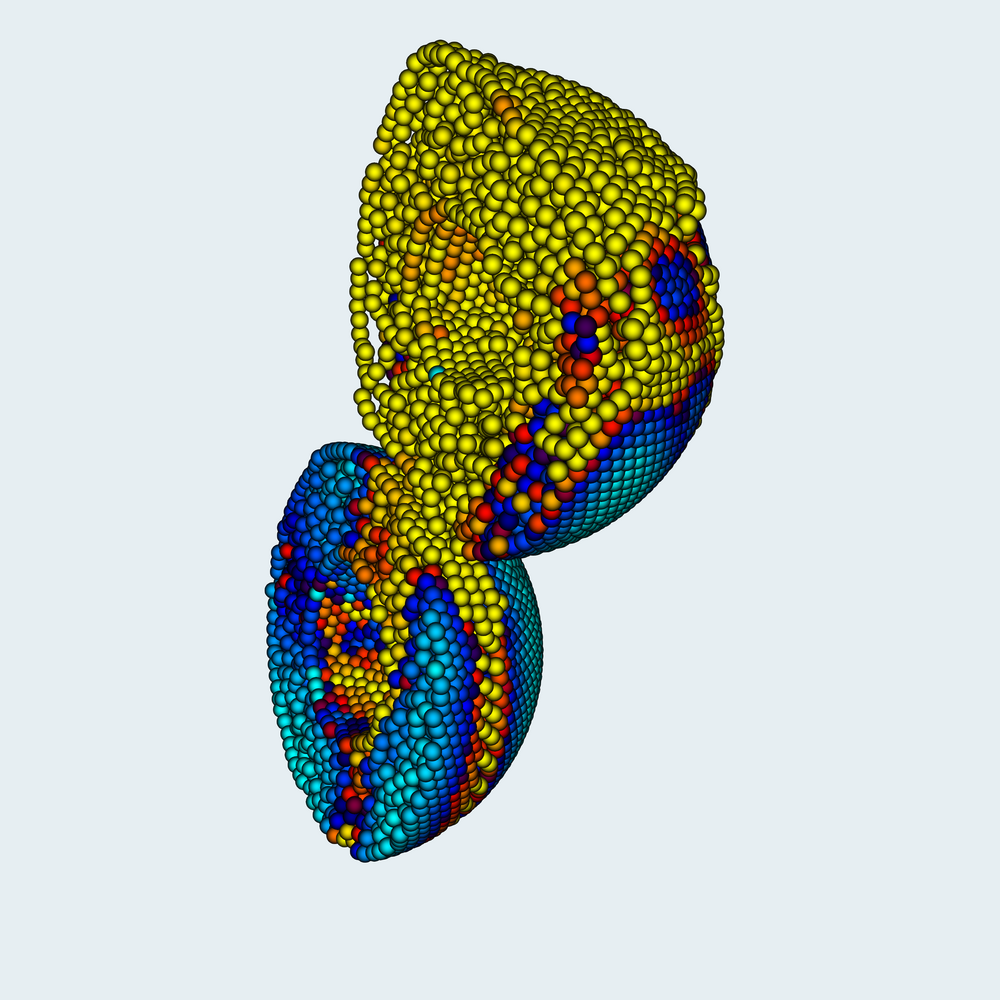}\label{fig:DHS-2} }
    \subfigure[][$t=500$]{
        \includegraphics[width=0.3\textwidth]{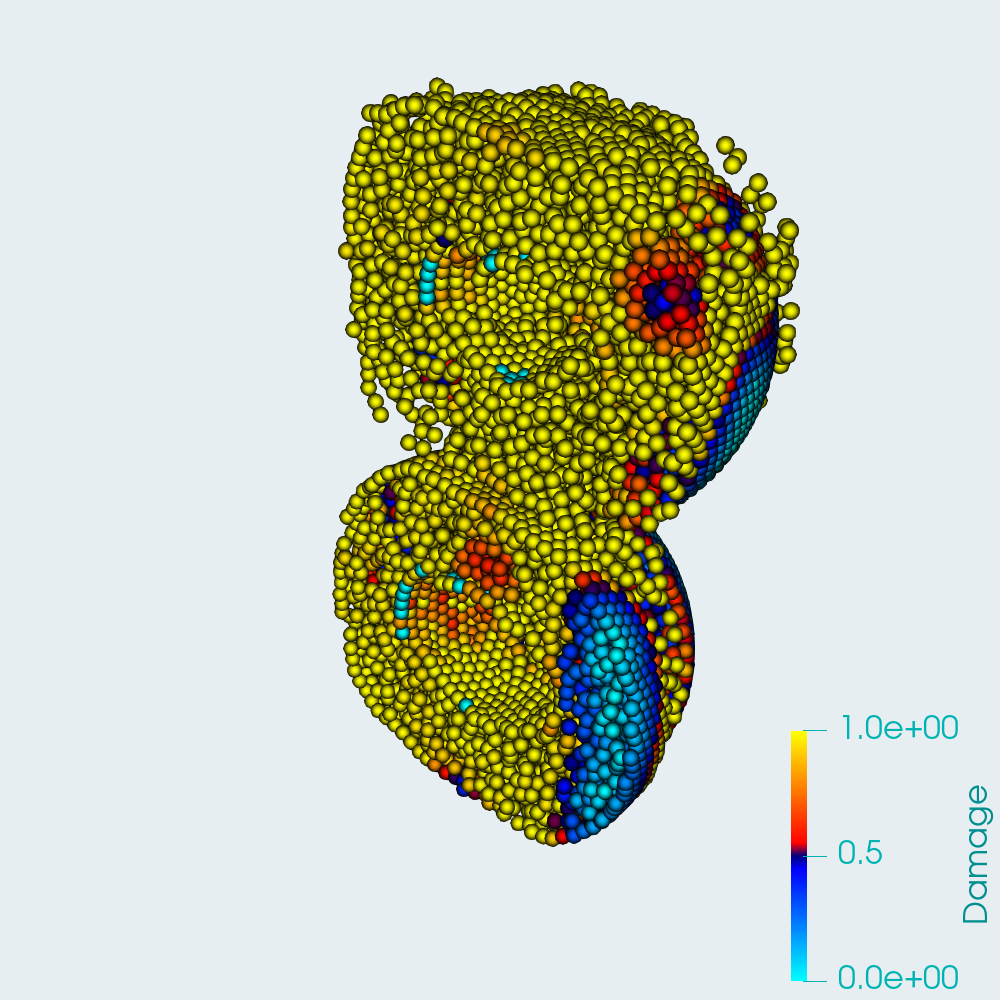}
    \label{fig:DHS-3} }
    \caption{ {PeriDEM simulation results showing the dynamic compression of two stacked hollow spheres at three different time steps: $t=50$, $t=200$, and $t=500$ (left to right). The top row displays the full outer surfaces of the spheres, while the bottom row shows corresponding cross-sectional views at each time step. The damage field is visualized to highlight the progression of fracture and material failure during compression.}}\label{fig:double-hallow-sphere}
\end{figure}

 {
Fig.~\ref{fig:double-hallow-sphere} illustrates the results of PeriDEM simulations for two hollow spheres positioned on top of each other and compressed from the top by a rigid plate. The setup involves a cubic container with fixed lateral and bottom walls, while the top boundary acts as a moving rigid wall descending at a constant speed of 10m/s to induce compression. The six subplots depict both the outer surfaces and cross-sections of the spheres, providing insight into the interaction, collision, and breakage of the particles under loading.
In the top row of Fig.~\ref{fig:double-hallow-sphere}, the outer surfaces of the spheres are shown at three different time steps: $t=50$, $t=200$, and $t=500$ (left to right). These plots illustrate the progression of compression as the upper sphere is driven downward by a rigid plate and begins to interact with the lower sphere. The damage field visualization clearly shows how stress accumulates near the contact interface, with increasing severity over time. The onset and growth of damage are most pronounced at the collision region, where the compressive forces are concentrated.
The bottom row presents corresponding cross-sectional views at the same time steps, capturing the internal evolution of damage and fracture. At early stages, localized cracking begins near the contact zone. As the simulation progresses, cracks propagate into the interior of both spheres, leading to significant fragmentation. By $t=500$, the internal structure shows extensive damage, with multiple fracture paths indicating complex failure behavior under compressive loading.
}

The PeriDEM method successfully captures the realistic behavior of the spheres under these conditions, particularly in terms of how the material responds to both direct compression and the interaction between the two spheres. The simulation accurately replicates the physical phenomena that occur when two particles collide under load, including stress localization at contact points, crack propagation, and eventual particle breakage. These behaviors are critical for understanding how hollow spheres interact in real-world scenarios, such as in packed bed reactors, granular materials, or other systems involving particle collisions under compression.

This simulation demonstrates the robustness of the PeriDEM approach in modeling the physics of particle interactions, with a high degree of realism. It effectively captures both the deformation of the spheres and the internal fragmentation resulting from the interaction, providing valuable insights into the crushing process in multi-particle systems.
\newpage

\subsection{Simulation of  Aggregates Made up of Particles Obtained X-ray Computed Tomography}
\label{sub:shape-setup}
\begin{figure}[htbp]
    \centering
          \includegraphics[width=0.24\textwidth]{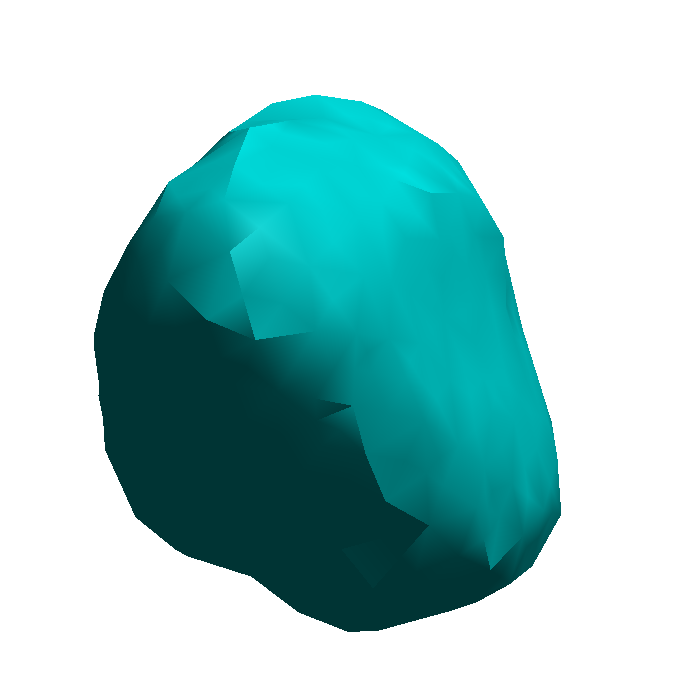}
          \includegraphics[width=0.24\textwidth]{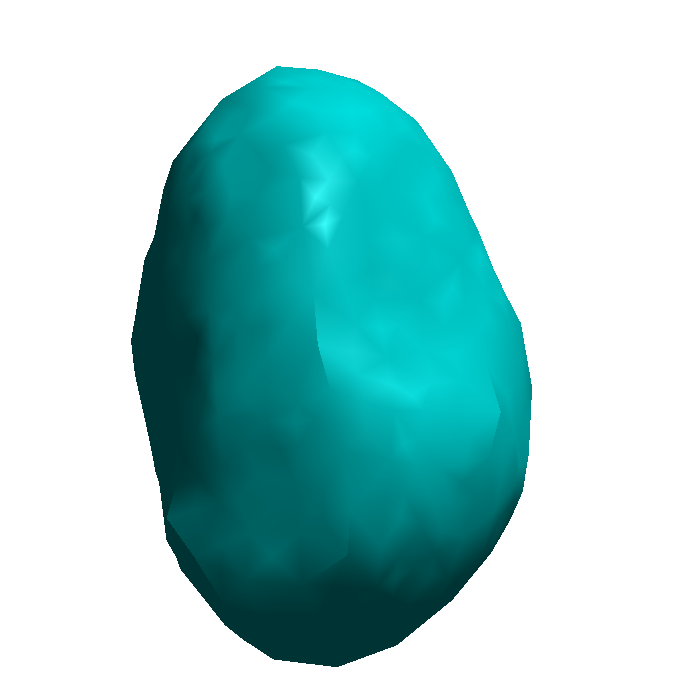}
          \includegraphics[width=0.24\textwidth]{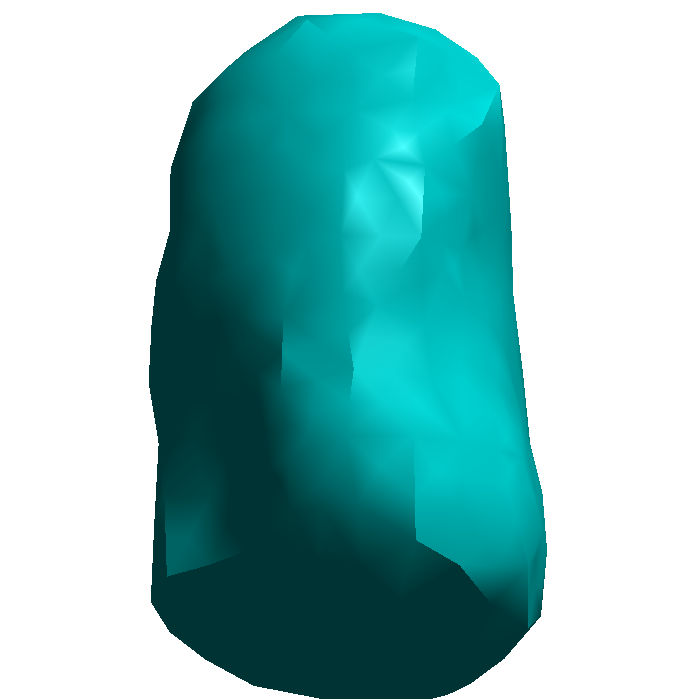}
          \includegraphics[width=0.24\textwidth]{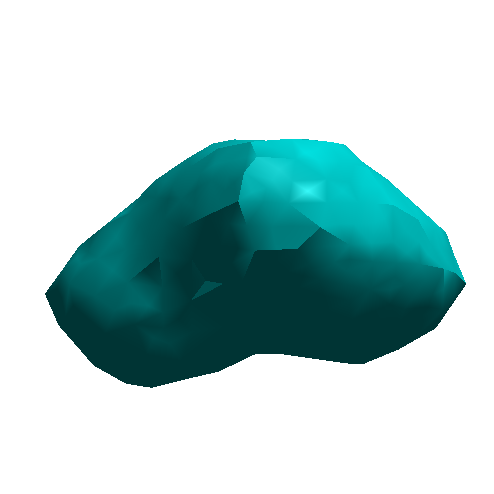}
    \caption{ {Representative examples of reconstructed sand grains obtained from 3D X-ray computed tomography data. Each grain’s surface is reconstructed using the ball-pivoting algorithm, revealing the natural irregularity and morphological diversity captured from the original point cloud data.}}\label{fig:grain-vox}
\end{figure}

   {Here, we consider 1000 arbitrarily shaped sand grains reconstructed from three dimensional X-ray computed tomography \cite{vlahinic2014towards}.}
The particle is described by point cloud data. A surface reconstruction algorithm is performed on the point cloud data to construct a volumetric mesh. 
To reduce the surface detail, a subsampling algorithm is applied to uniformly sample from the point cloud while maintaining enough surface detail.
The outward unit normal at each level set point is estimated using 10 nearest neighbors. Then, a minimal bounding surface is reconstructed using a ball-pivoting algorithm.
The resulting reconstructing surface for a grain is shown in \Cref{fig:grain-vox} after various levels of subsampling.


 {To generate the initial grain configuration inside a cubic container, a rigid motion—comprising rotation about the centroid and translation of the centroid—is applied to each grain’s volume mesh, consistent with the experimental setup. All walls of the container are rigidly fixed, except for the top wall, which is prescribed a constant downward velocity of 10m/s to initiate compression in the later stage of the simulation. To avoid initial overlaps between particles—caused by numerical inaccuracies in surface reconstruction—and to prevent the activation of contact forces at the start, the grains are slightly scaled down in the radial direction.} 

\begin{figure}[htbp]
    \centering
    \subfigure[400 particles]{
        \includegraphics[width=0.45\textwidth]{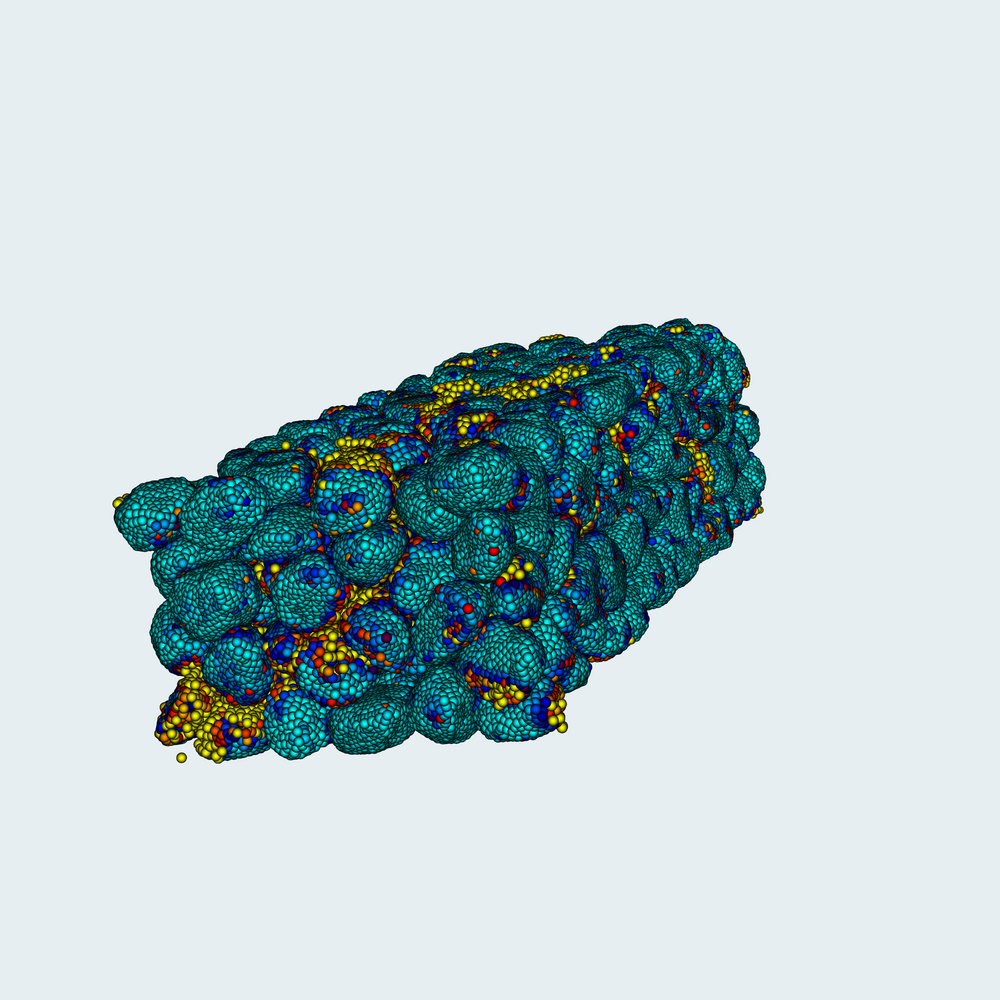}
    }
    \subfigure[600 particles]{
        \includegraphics[width=0.45\textwidth]{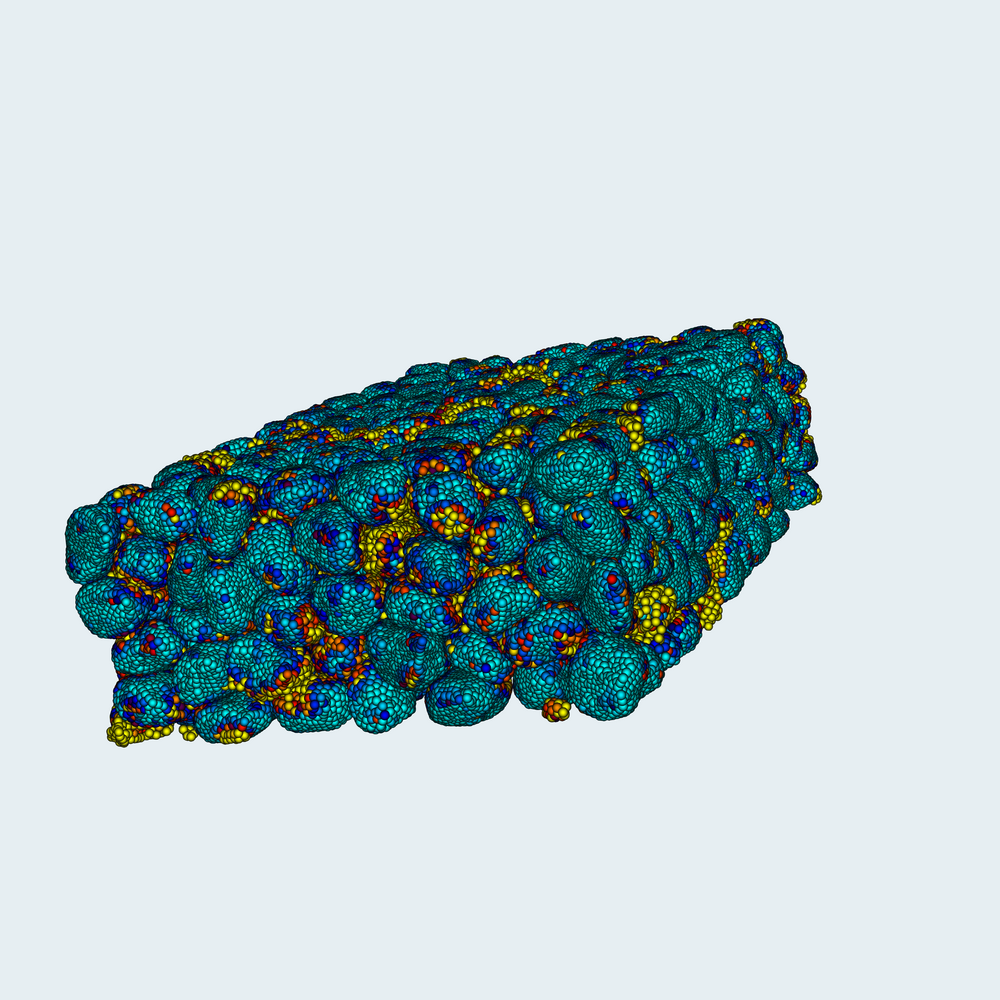}
    }
    \subfigure[800 particles]{
        \includegraphics[width=0.45\textwidth]{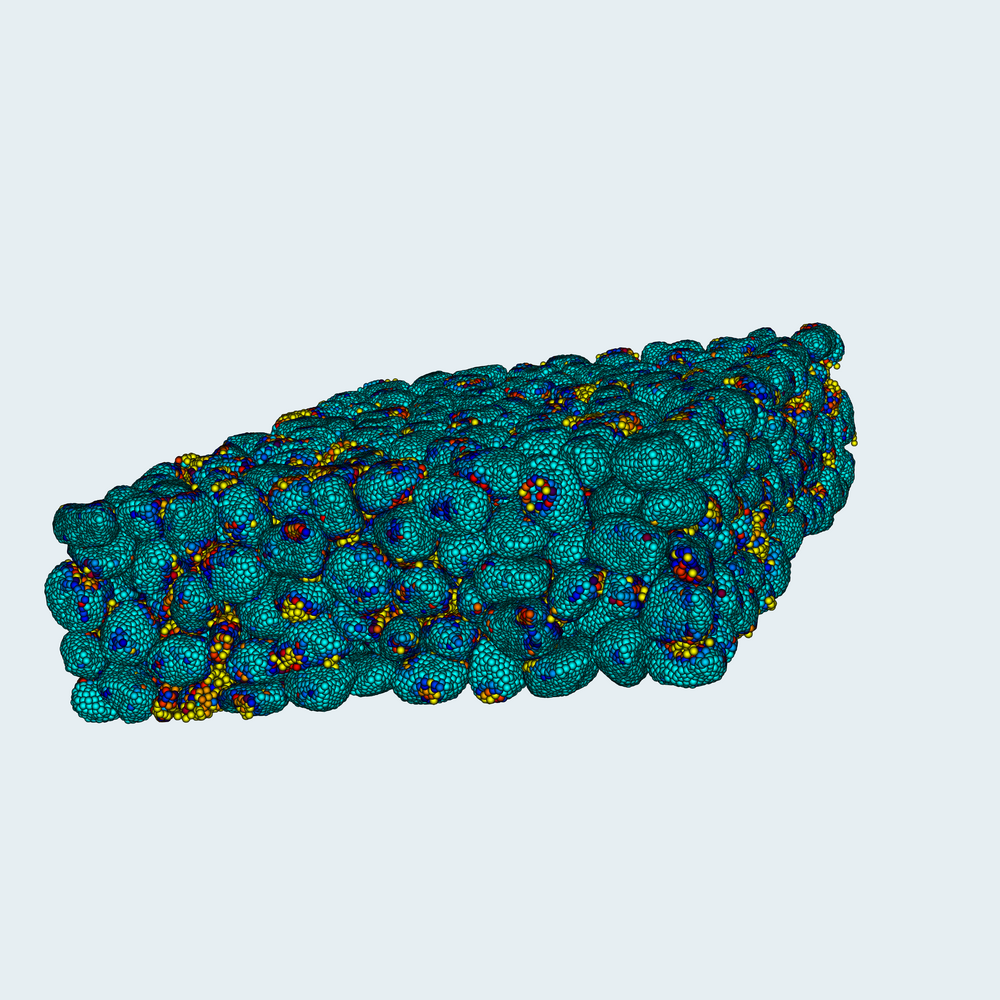}}
    \subfigure[1000 particles]{
        \includegraphics[width=0.45\textwidth]{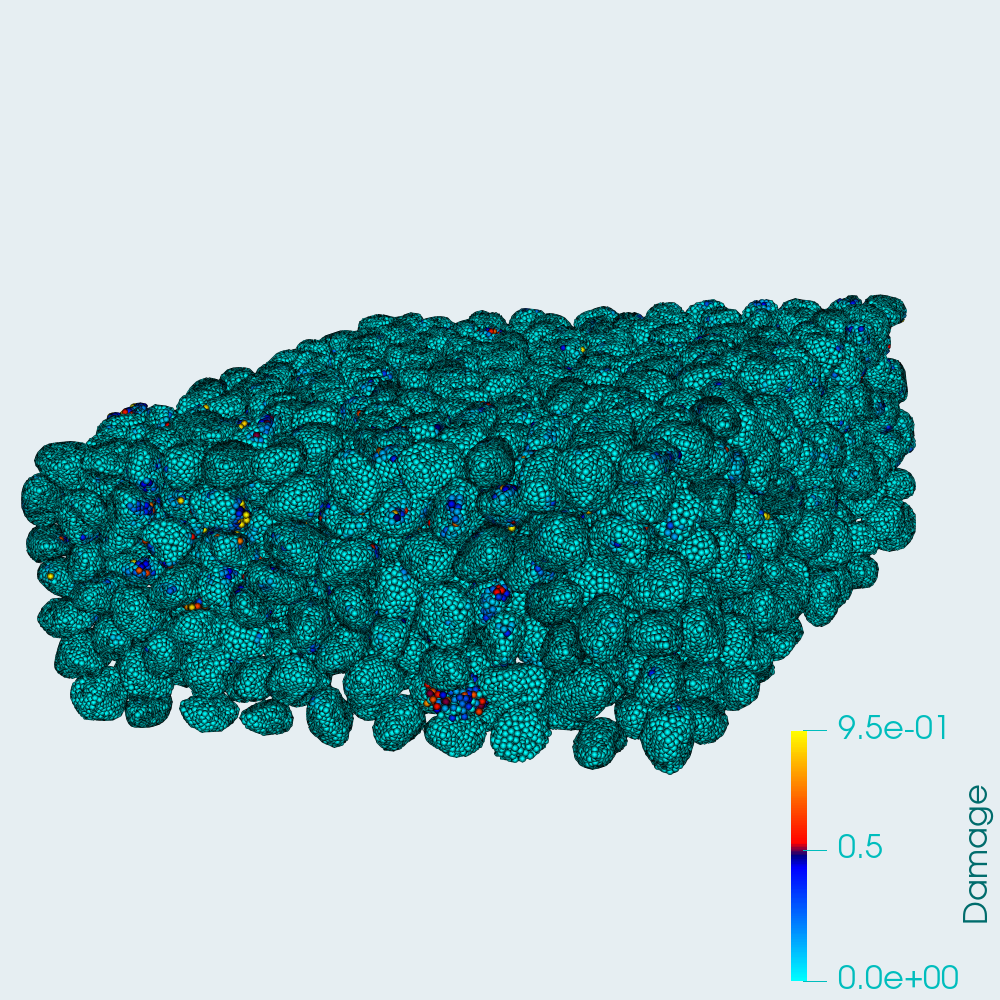}
    }
    \caption{ {Snapshots of particle crushing using PeriDEM at time step t=880, shown for aggregates containing (a) 400, (b) 600, (c) 800, and (d) 1000 particles. The container size is adjusted for each case based on the number of particles.}}\label{fig:aggregate-crushing}
  
\end{figure}

 {Fig.~\ref{fig:aggregate-crushing} shows snapshots from PeriDEM simulations at time step t=880, illustrating the crushing behavior of granular assemblies composed of (a) 400, (b) 600, (c) 800, and (d) 1000 particles. For each case, the size of the cubic container is adjusted according to the number of particles to maintain a comparable initial packing density. Despite the similar initial conditions, the results reveal that aggregates with fewer particles experience more pronounced damage and fragmentation under compression. In contrast, assemblies with higher particle counts tend to exhibit reduced visible damage, indicating a more efficient distribution of forces and enhanced collective resistance to breakage as the system becomes denser.}

\begin{figure}
    \centering
        \subfigure{\includegraphics[width=0.8\linewidth]{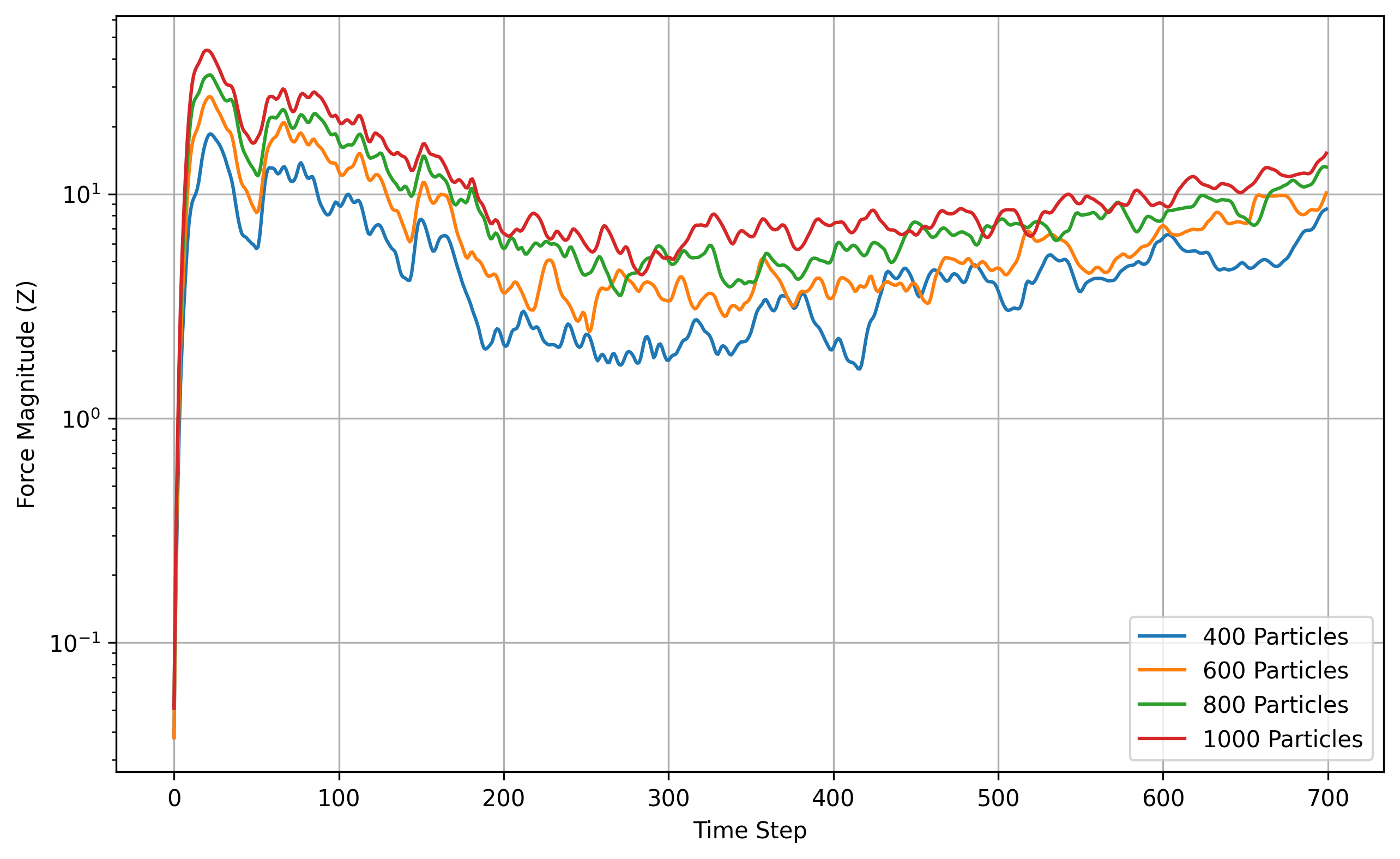}}
        \caption{ {Time evolution of the normalized compressive force on the top wall during PeriDEM simulations of particle crushing. The force is normalized by the top surface area of the container to enable comparison across systems of varying particle counts. An exponential moving average is applied to smooth the data. Assemblies with higher particle counts exhibit larger sustained normalized forces, indicating enhanced load-bearing capacity and reduced fragmentation.}}
    \label{fig:top-force-grain}
\end{figure}

 {Fig.~\ref{fig:top-force-grain} presents the time evolution of the compressive force measured on the top wall of the container during PeriDEM simulations. To account for variations in container size across different particle counts, the recorded force values were normalized by the top surface area of each container, resulting in a dimensionless force measure. This normalization enables a meaningful comparison of stress responses across systems of 400, 600, 800, and 1000 particles. To reduce noise and highlight overall trends, an exponential moving average (EMA) was applied to the raw force data. The resulting smoothed curves reveal how the force magnitude evolves over time for different particle configurations.}

 {The normalized force plot in Fig.~\ref{fig:top-force-grain} reveals distinct trends in the compressive response of particle assemblies with varying numbers of grains. All configurations show a sharp initial rise in force, corresponding to the moment of first contact between the top wall and the particles. This is followed by a rapid drop, reflecting particle rearrangement and initial fracture events as the system settles into a compressed state. After this transient phase, the force stabilizes and gradually increases, indicating a transition to a more compact and load-bearing structure. Notably, assemblies with more particles (800 and 1000) exhibit consistently higher normalized force values over time, suggesting a stiffer and more resilient collective response. In contrast, the 400-particle configuration maintains a lower force level, consistent with greater fragmentation and less efficient load distribution observed in the corresponding snapshots. These results support the conclusion that aggregates including more particles  enhance the mechanical strength of the granular assembly under compression.}
All curves in Fig.~\ref{fig:top-force-grain} exhibit the same overall shape and appear to converge as the number of particles increase.
\subsection{Effect of Particle Shapes on Bulk Strength}
\label{grain shape topology}
\begin{figure}
    \centering
     {\includegraphics[width=0.3\linewidth]{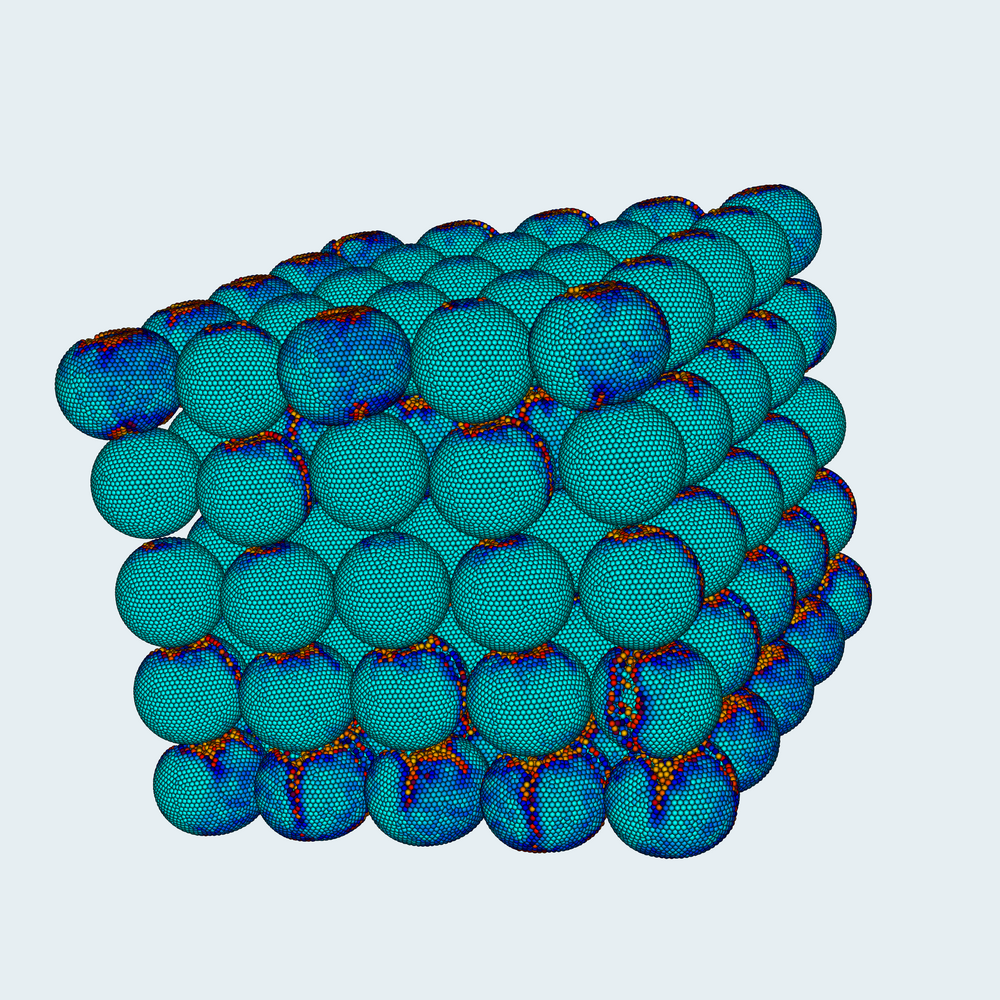}} 
     {\includegraphics[width=0.3\linewidth]{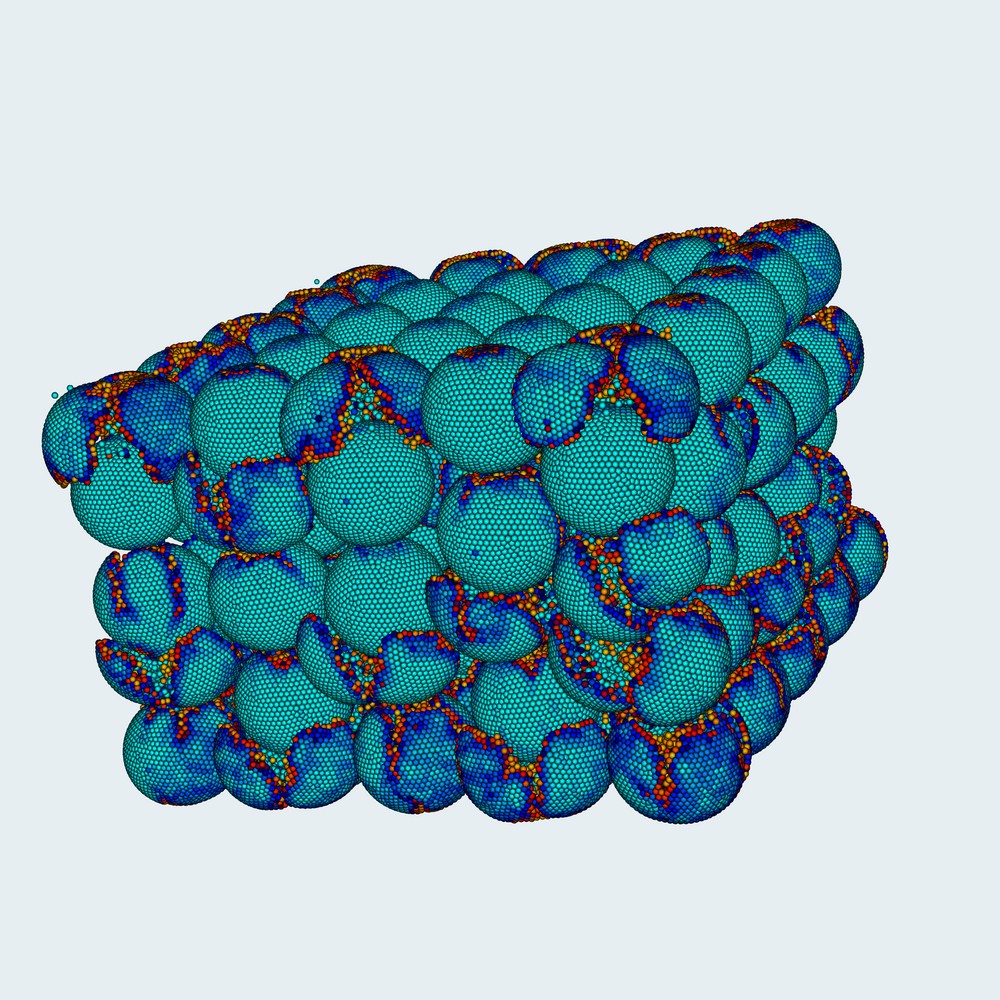}}
      {\includegraphics[width=0.3\linewidth]{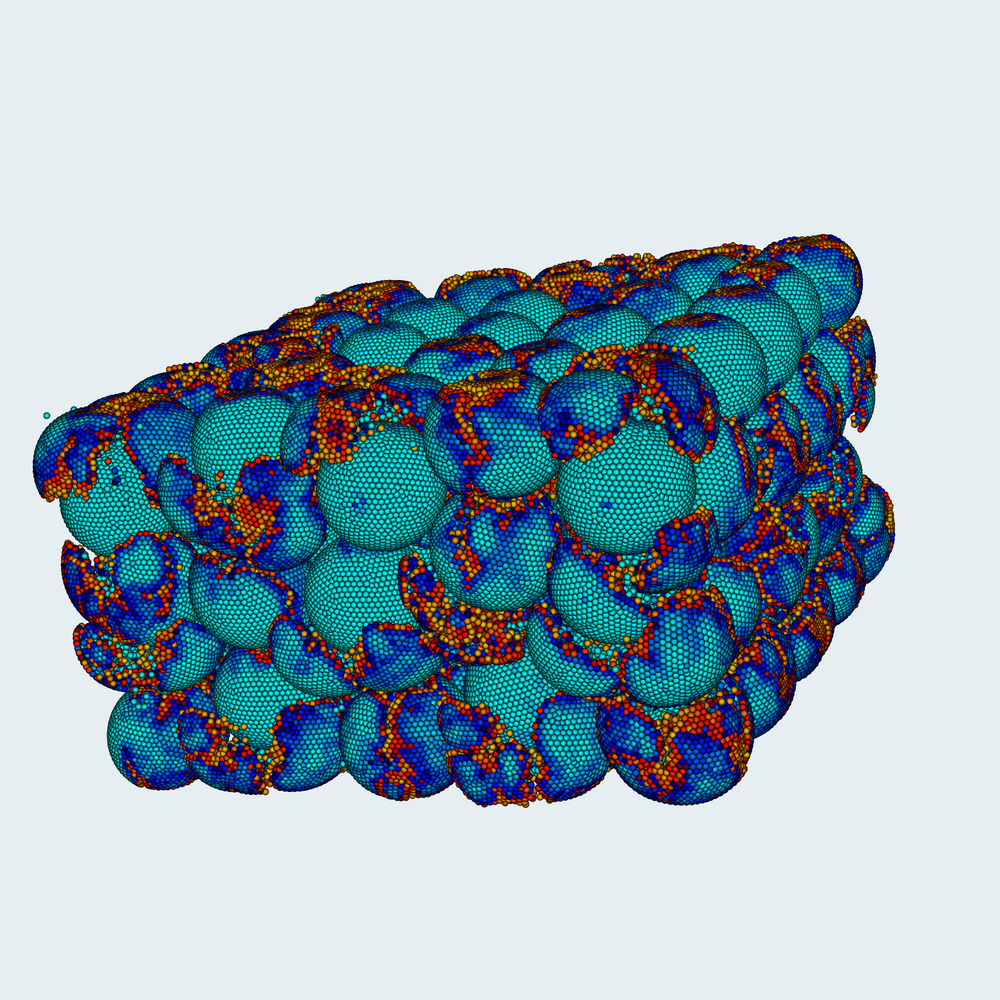}}
    \\
     {\includegraphics[width=0.3\linewidth]{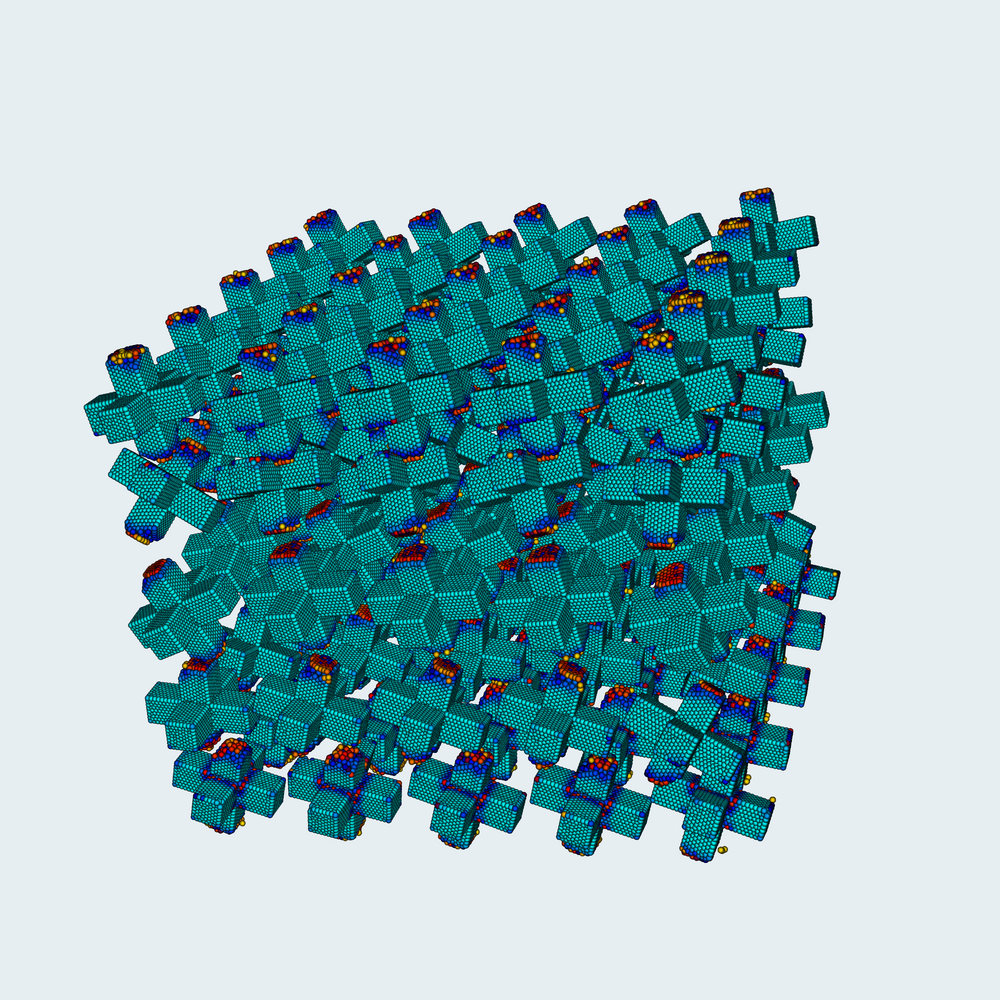}}
      {\includegraphics[width=0.3\linewidth]{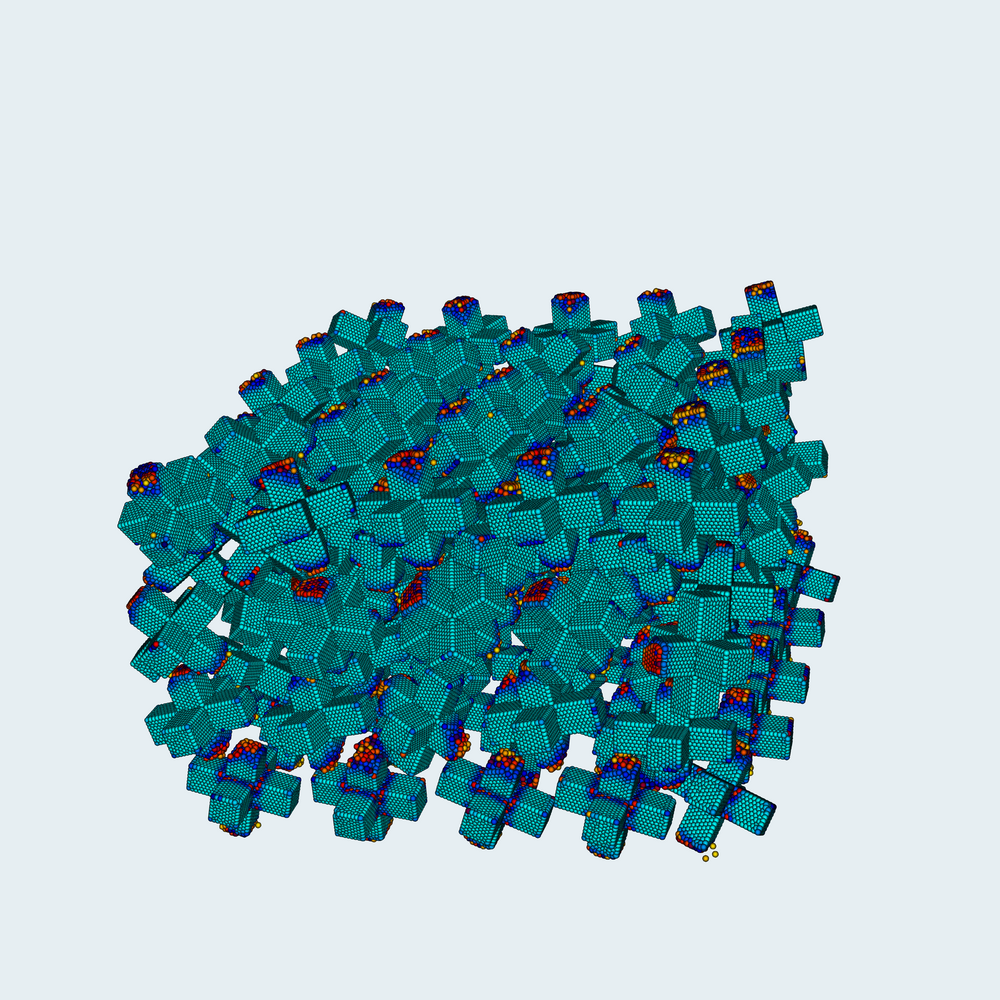}}
      {\includegraphics[width=0.3\linewidth]{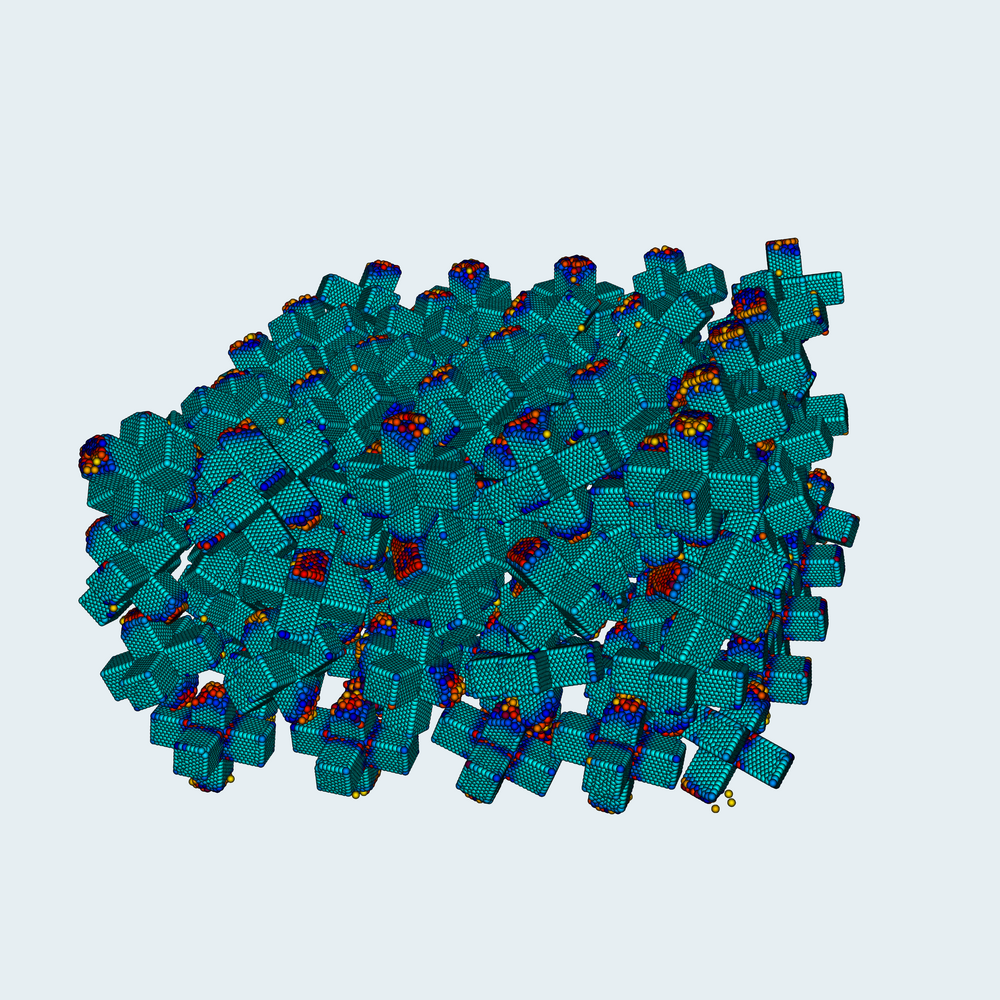}}
    \\
     \subfigure[$t=40$]{\includegraphics[width=0.3\linewidth]{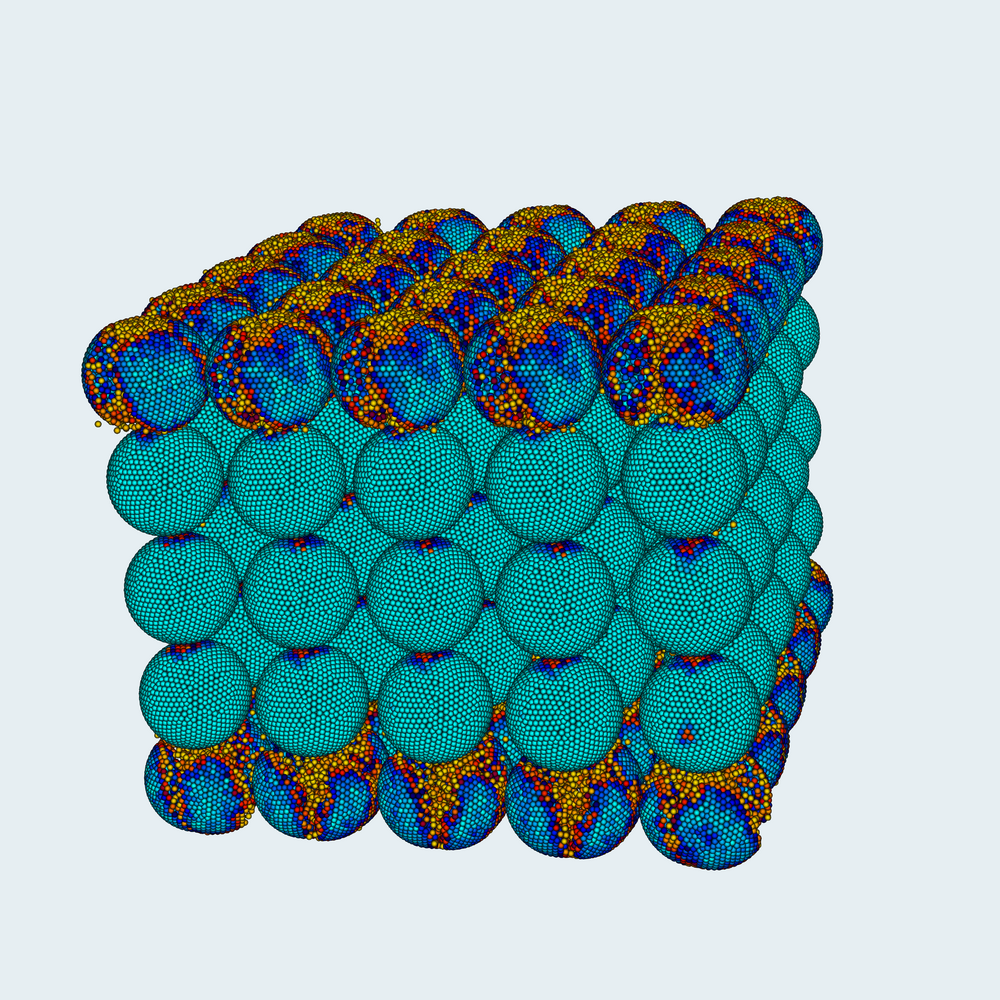}}
     \subfigure[$t=80$]{\includegraphics[width=0.3\linewidth]{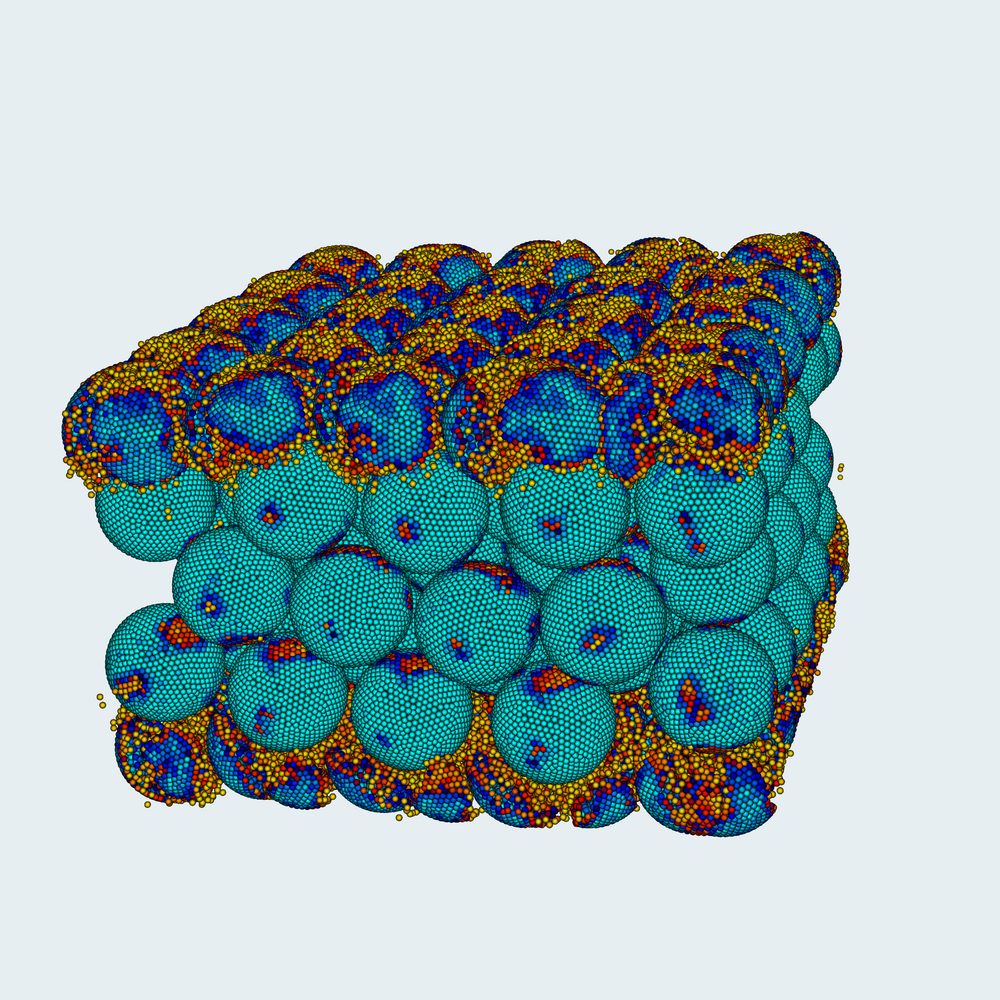}}
     \subfigure[$t=100$]{\includegraphics[width=0.3\linewidth]{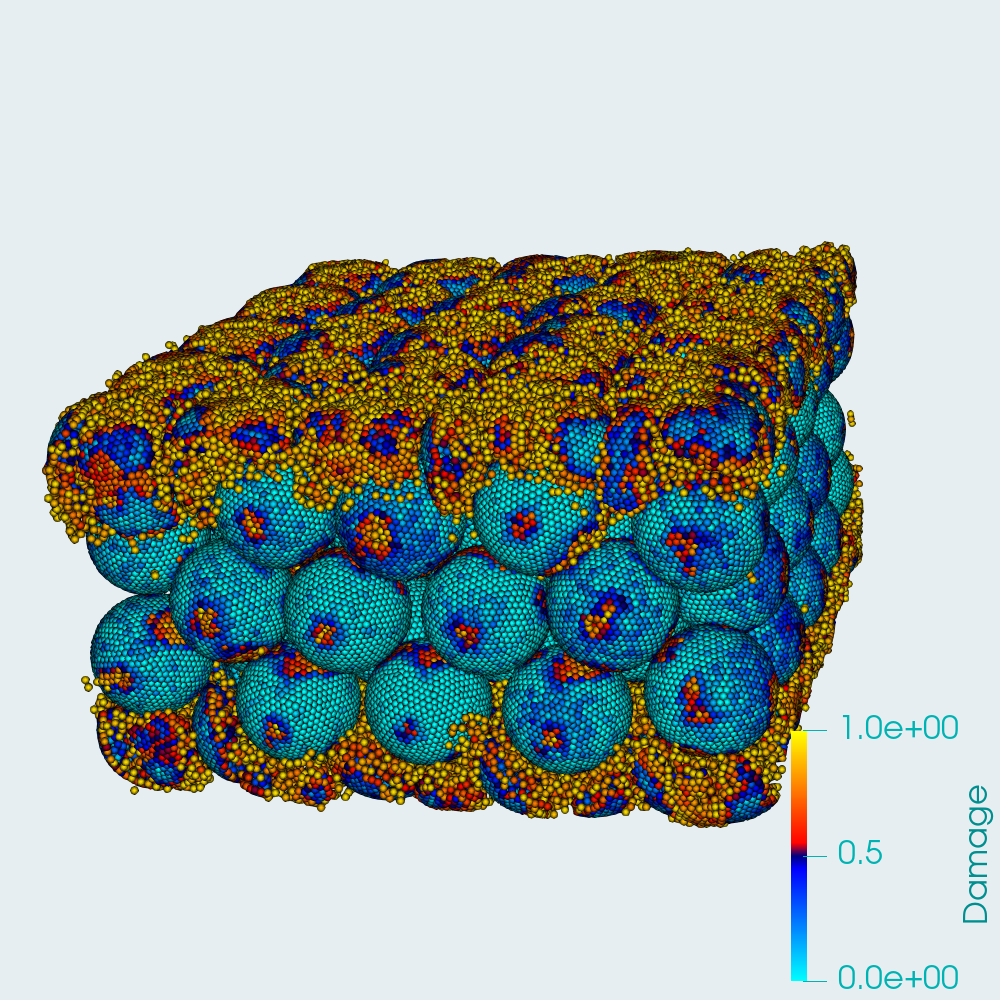}}
    \caption{ { Compression of particle assemblies with varying grain geometries. Each column corresponds to a different time step during compression: t=40 (left), t=80 (middle), and t=100 (right), given in the multiples of $5 \mu$s. Each row shows a different particle shape:  hollow spheres (top), jacks (middle), and solid spheres (bottom). The top wall compresses the aggregate at a constant speed of 10 m/s. Damage evolution and particle rearrangement vary significantly with shape, influencing the bulk strength response.}}
      \label{fig:initial}
\end{figure}

\begin{figure}
    \centering
    {\includegraphics[width=0.3\linewidth]{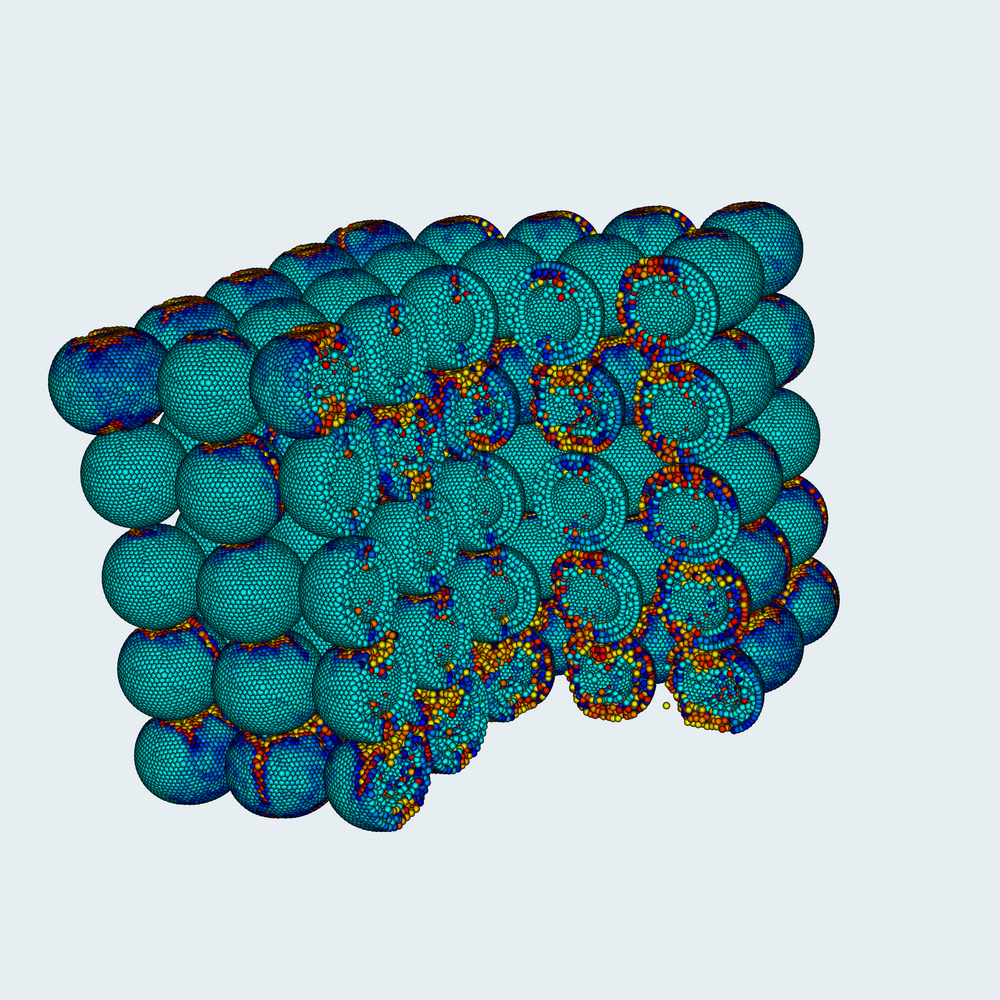}}
    {\includegraphics[width=0.3\linewidth]{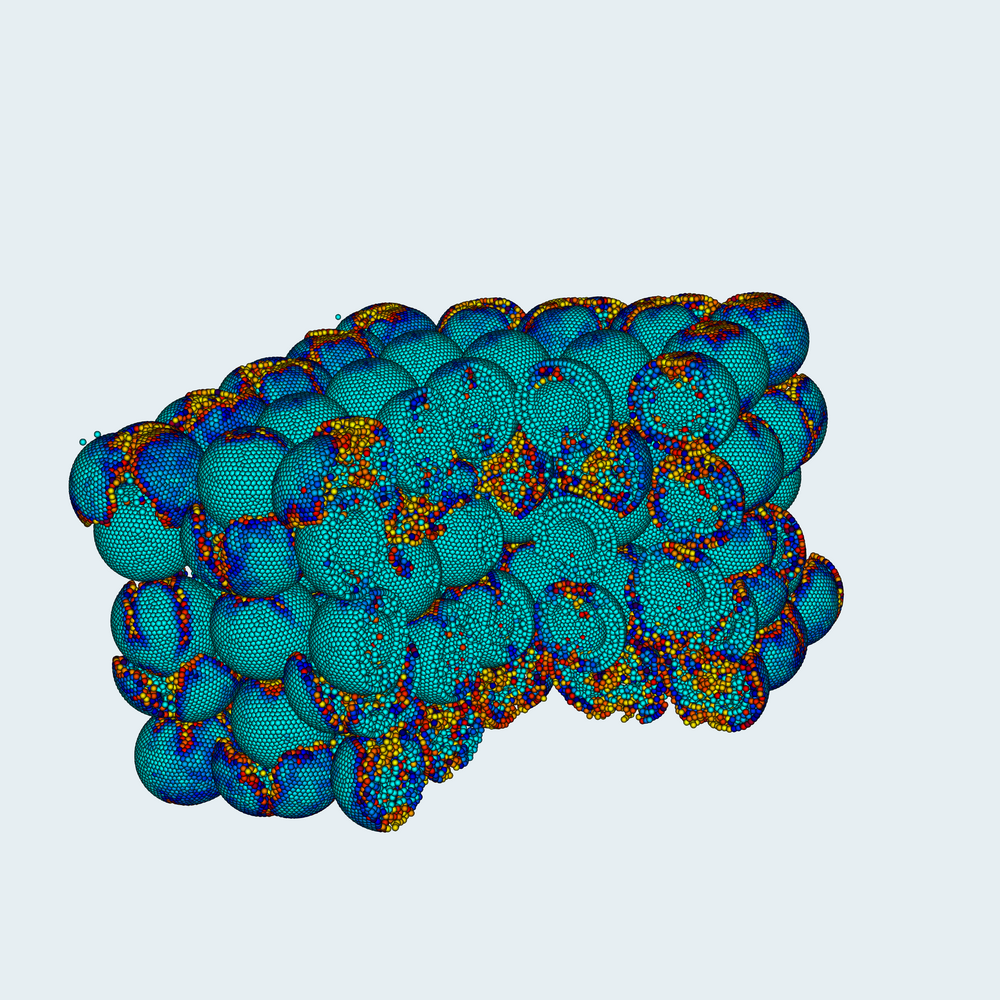}}
    {\includegraphics[width=0.3\linewidth]{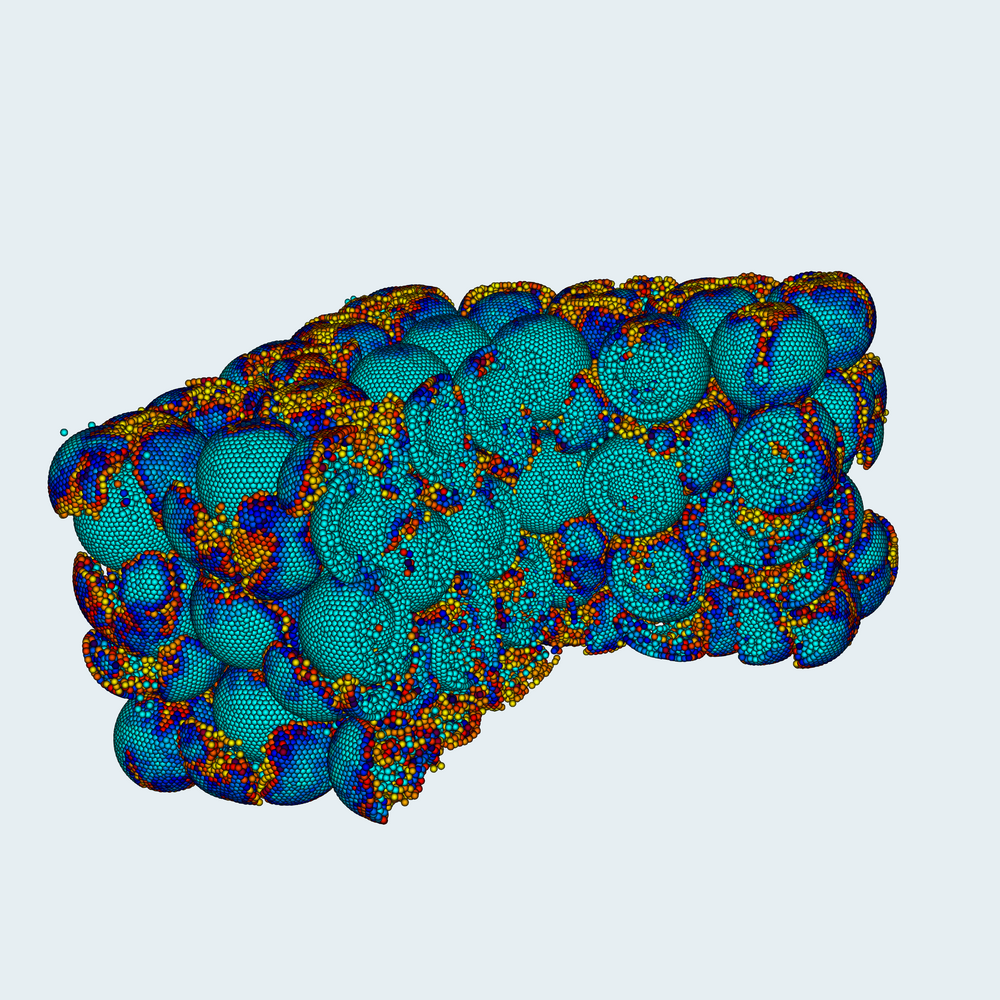}}
    \\
    {\includegraphics[width=0.3\linewidth]{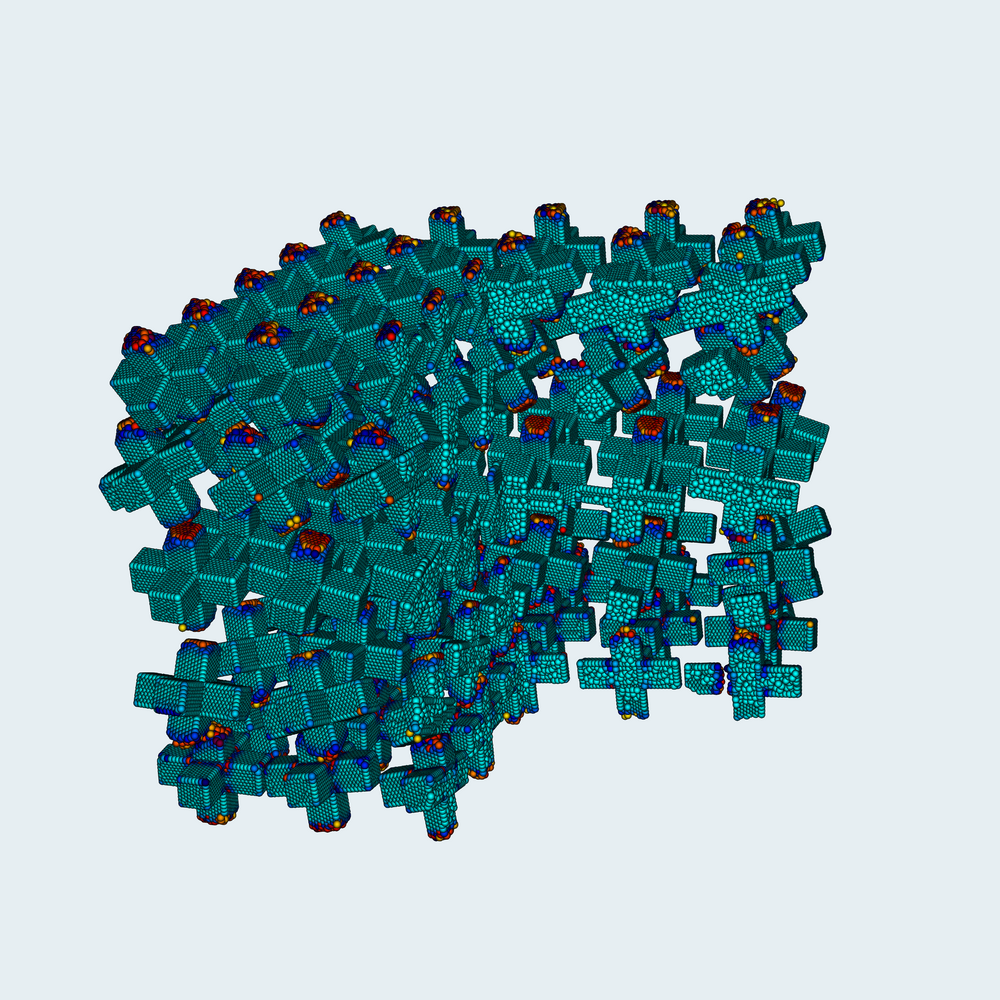}}
     {\includegraphics[width=0.3\linewidth]{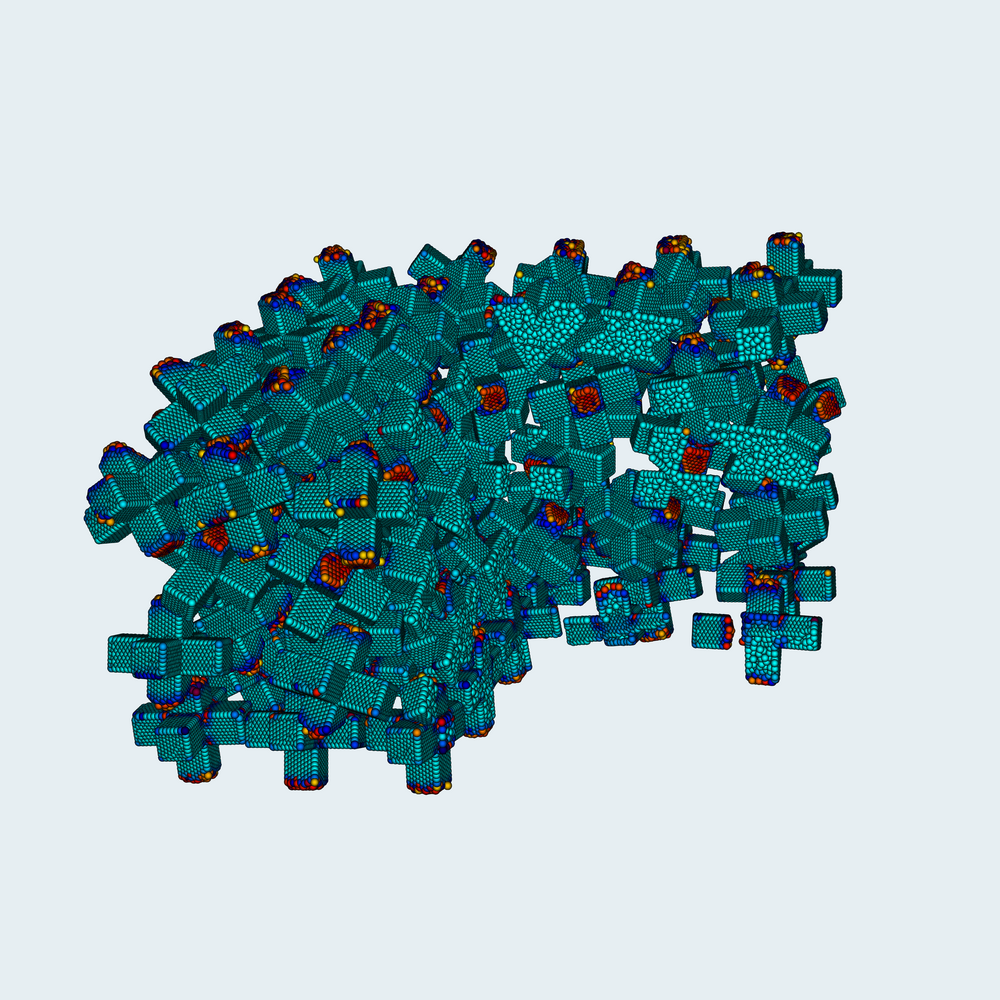}}
     {\includegraphics[width=0.3\linewidth]{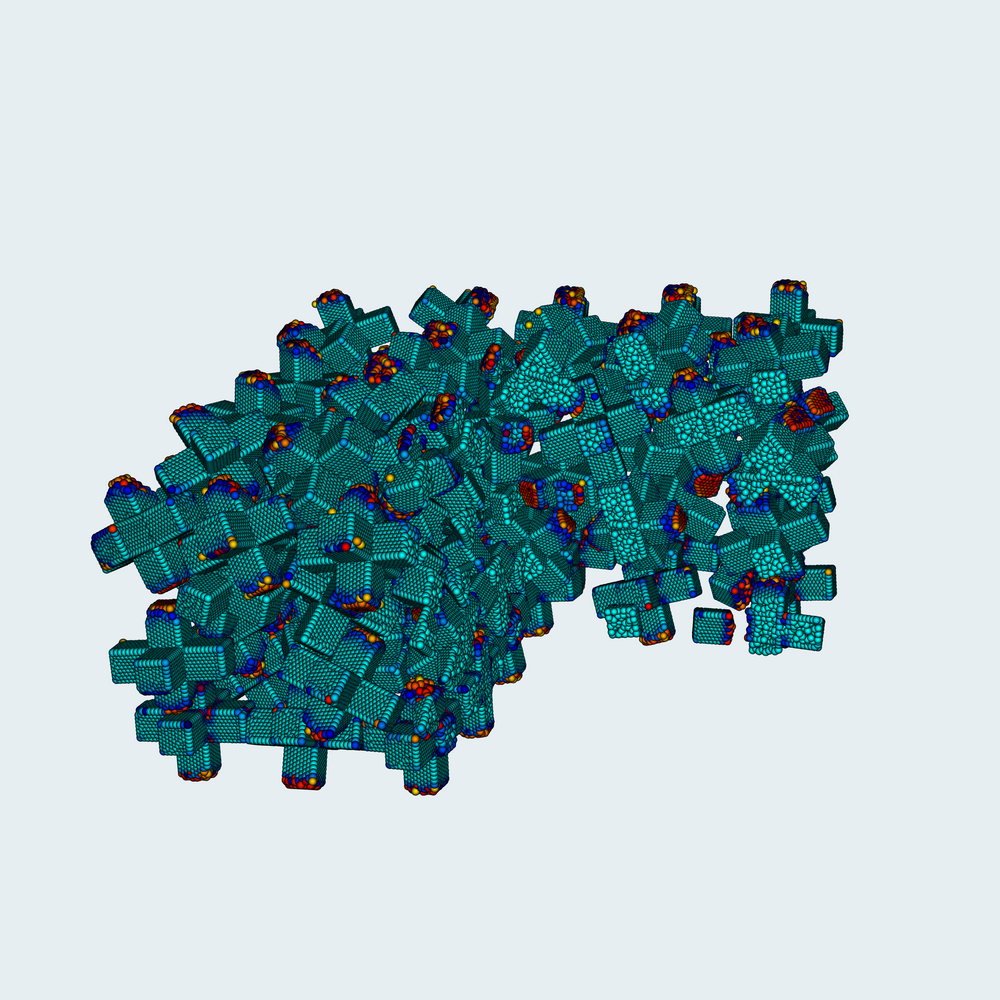}}
    \\
    \subfigure[$t=40$]{\includegraphics[width=0.3\linewidth]{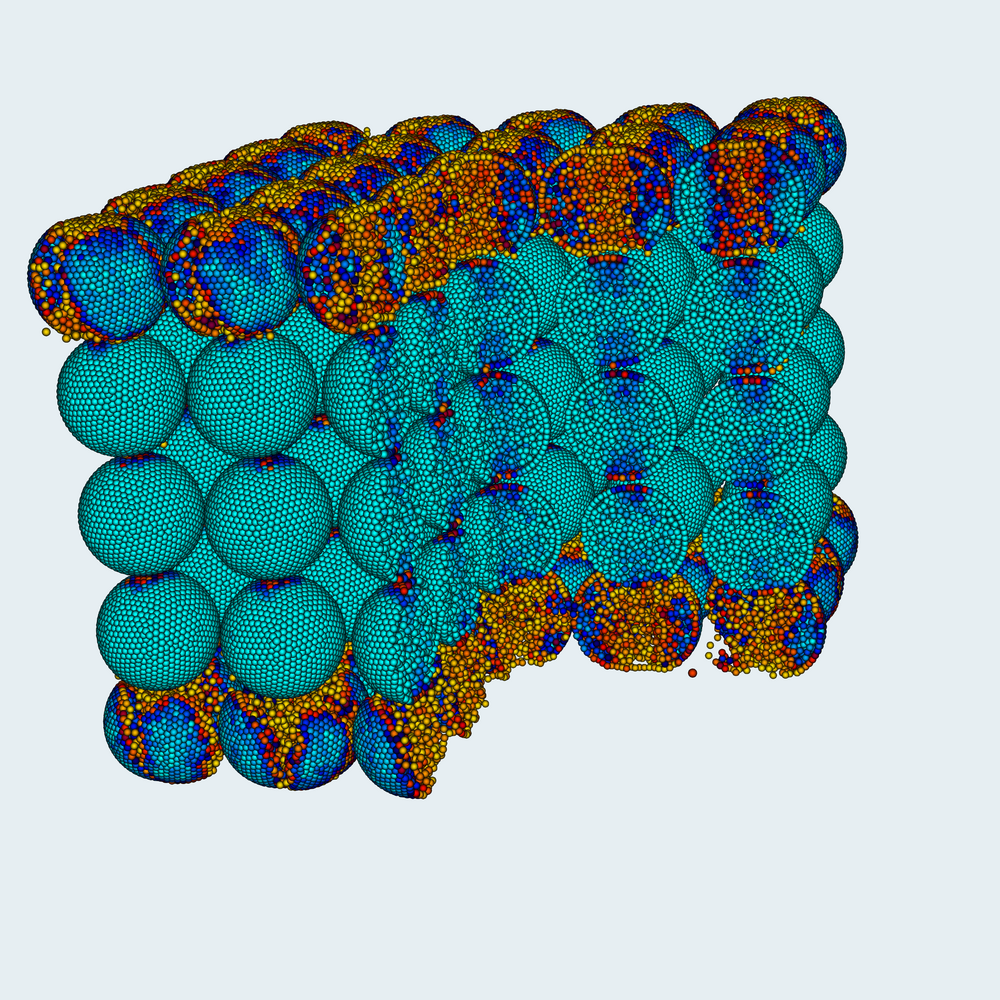}}
    \subfigure[$t=80$]{\includegraphics[width=0.3\linewidth]{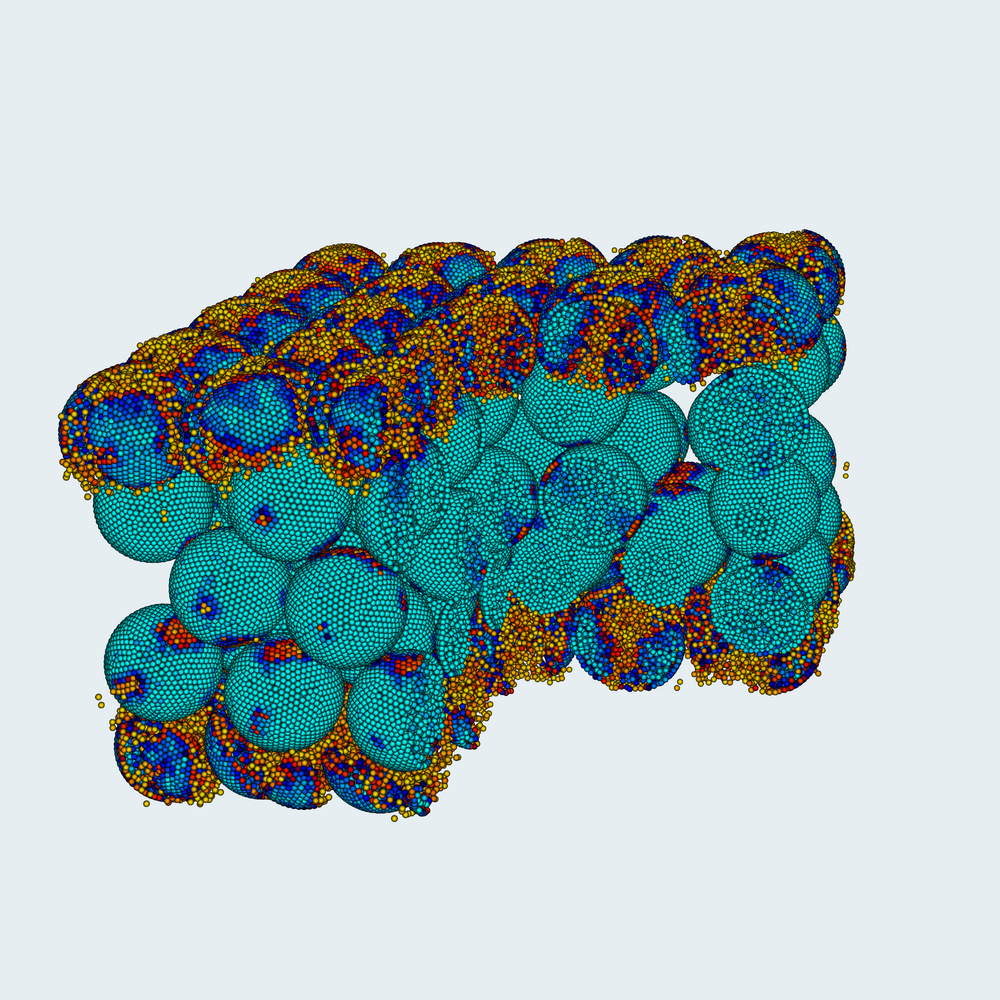}}
    \subfigure[$t=100$]{\includegraphics[width=0.3\linewidth]{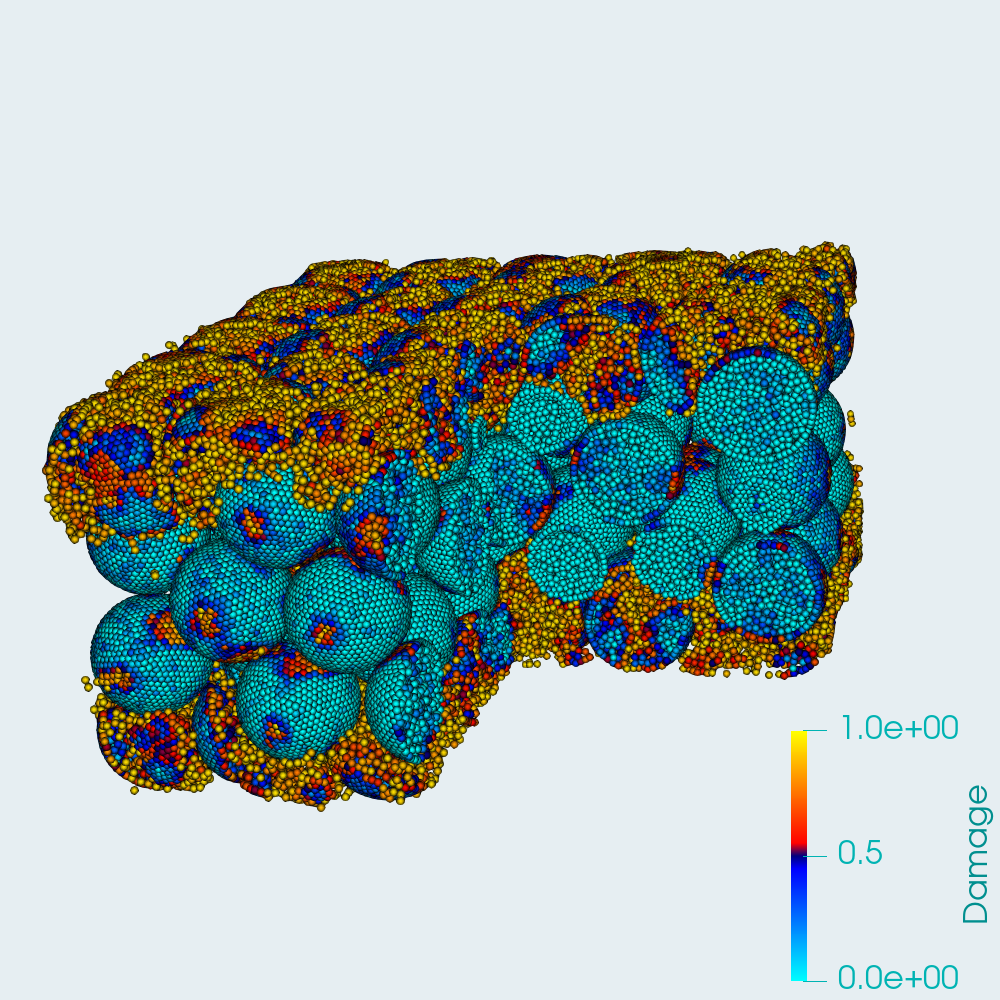}}
    \caption{ {Internal view of particle assemblies under dynamic compression, obtained by clipping the simulation domain to reveal interior particle behavior. Each column corresponds to a time step t=40, 80, and 100, expressed in multiples of $5 \mu$s. Each row represents a different particle geometry: solid hollow spheres (top), jacks (middle), and spheres (bottom). All assemblies begin with 125 particles arranged on a cubic grid, each with a 1 mm radius. A top wall compresses the system downward at 10 m/s. The cutaway views highlight the progression of internal fracture, contact evolution, and rearrangement patterns unique to each shape.}}
    \label{fig:initial-cross}
\end{figure}

\begin{figure}
    \centering
    \includegraphics[width=0.5\linewidth]{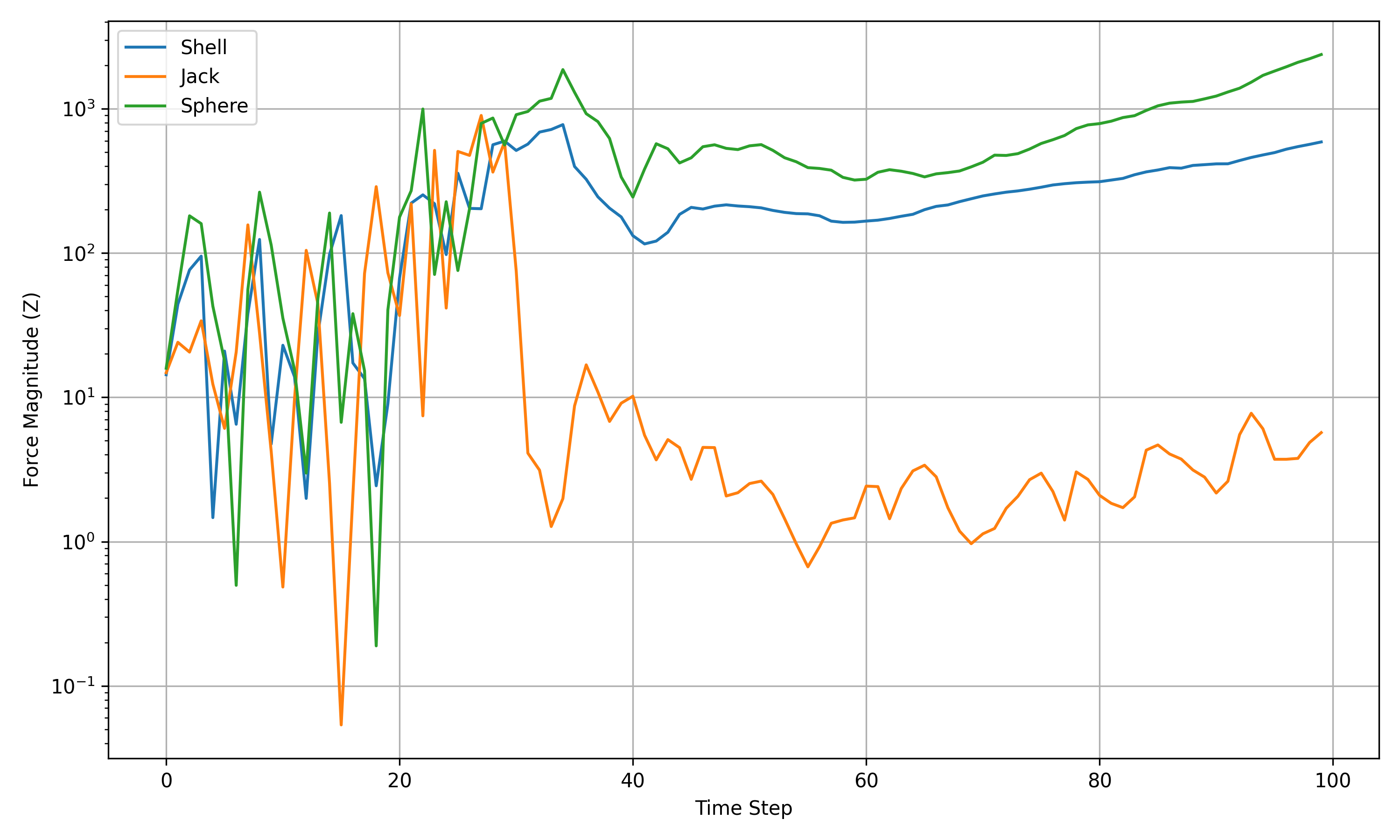}
    \caption{ {this plot got fixed, hollow sphrer in legend}Vertical bulk strength of aggregates of differently shaped particles under compaction. The x-axis represents time $t$ in the multiples of 5 $\mu$s. The y-axis represents the logarithm of the vertical force experienced by the particle bulk given in Newtons. The initial upward trend in the vertical wall force (until about t = 25) are observed when the confined particles experience increasing amount of compaction without displacement. After the collapse of particle columns due to breakage and rearrangement, a sudden drop in the vertical strength is observed. The jack-shaped particles go through the most significant amount of rearrangement and therefore exhibit the lowest amount of bulk strength. The shells exhibit lower amount of bulk strength compared to the spheres since the excluded volumes within shells are occupied by particle fragments as damage sets in.}
    \label{fig:vertical-bulk-strength}
\end{figure}

In this section, we focus on how individual grain shape and topology affect the strength of a granular aggregate. Nonspherical granular shapes, especially nonconvex shapes can exhibit complex inter-granular contact modes and in turn influence the bulk strength. Granular topology, especially the existence of holes within grains can reduce the volume fraction while maintaining the bulk strength. For brittle granular aggregates, granular damage influences and is influenced by granular shape and topology, thus affecting the bulk strength in yet another way. Effect of shapes on compressive strength and mobility of wheeled vehicles is studied in two dimensions \cite{bhattacharyalipton2023, bhattacharya2025macroscopic}. Our goal is to consider three   {prototypical }shapes and understand the interconnected mechanisms of shape, topology, and damage and their effect on granular strength.
A bulk compaction experiment is carried out numerically to determine the effect of individual particle shapes on the bulk strength. We consider the following grain shapes: solid spheres,  hollow spheres or shells, and jacks or 3D plus shapes.
An initial arrangement of 125 particles are considered and are placed on a cubic grid. The radius of each particle is taken to be 1 mm, irrespective of their shape. The top wall is moved downwards at a speed of 10 m/s, thereby compressing the particles quickly. \Cref{fig:initial,fig:initial-cross} show snapshots of the bulk at various levels of compression. We define the (vertical) bulk strength of the aggregate as the difference between the reaction forces experienced by the top and the bottom walls. The bulk strength of each of the aggregates as a function of time is shown in \Cref{fig:vertical-bulk-strength}.


\begin{figure}
    \centering
     {\includegraphics[width=0.3\linewidth]{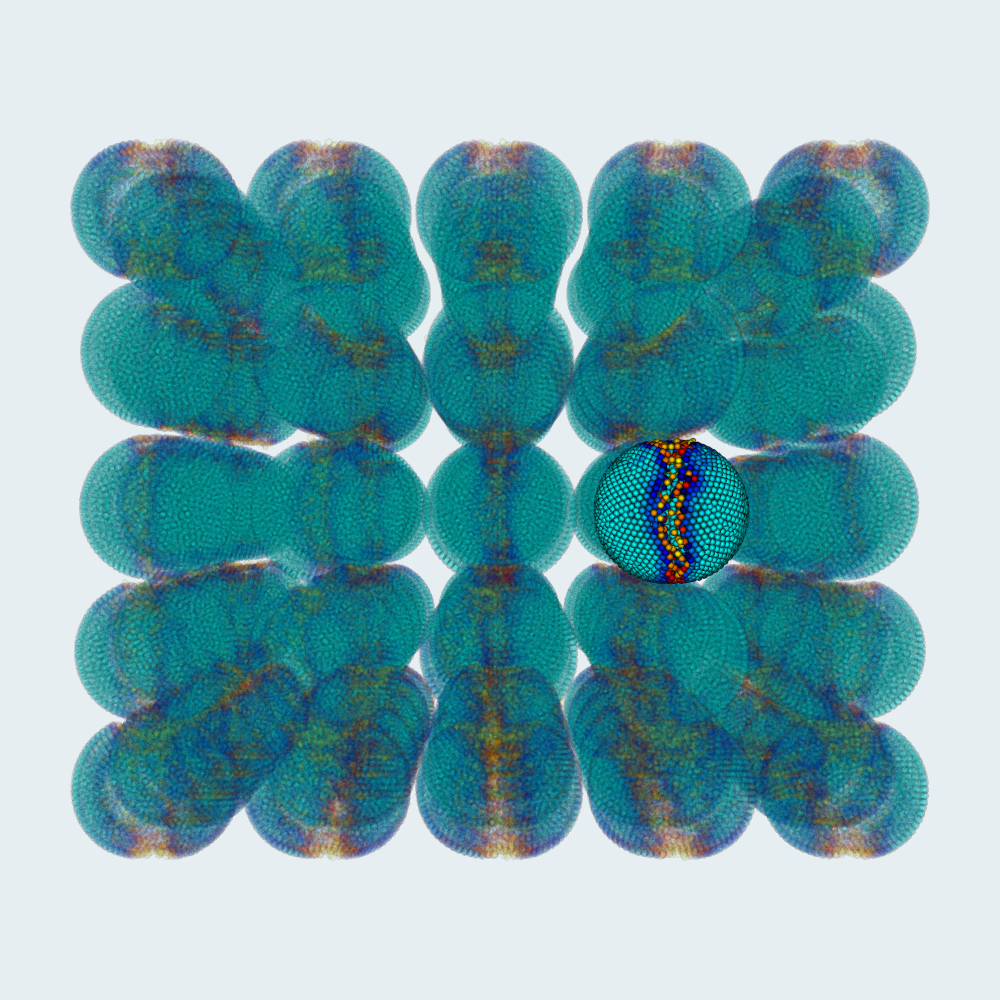}}
     {\includegraphics[width=0.3\linewidth]{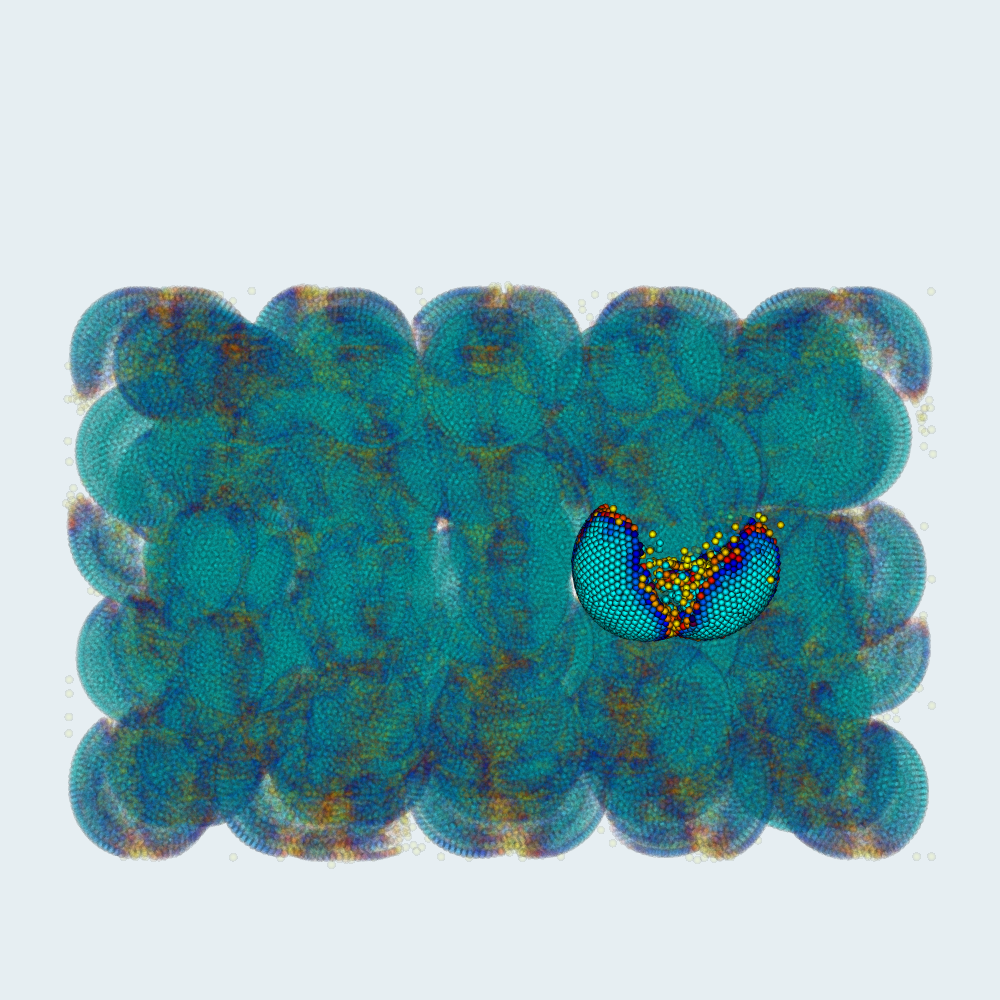}}
     {\includegraphics[width=0.3\linewidth]{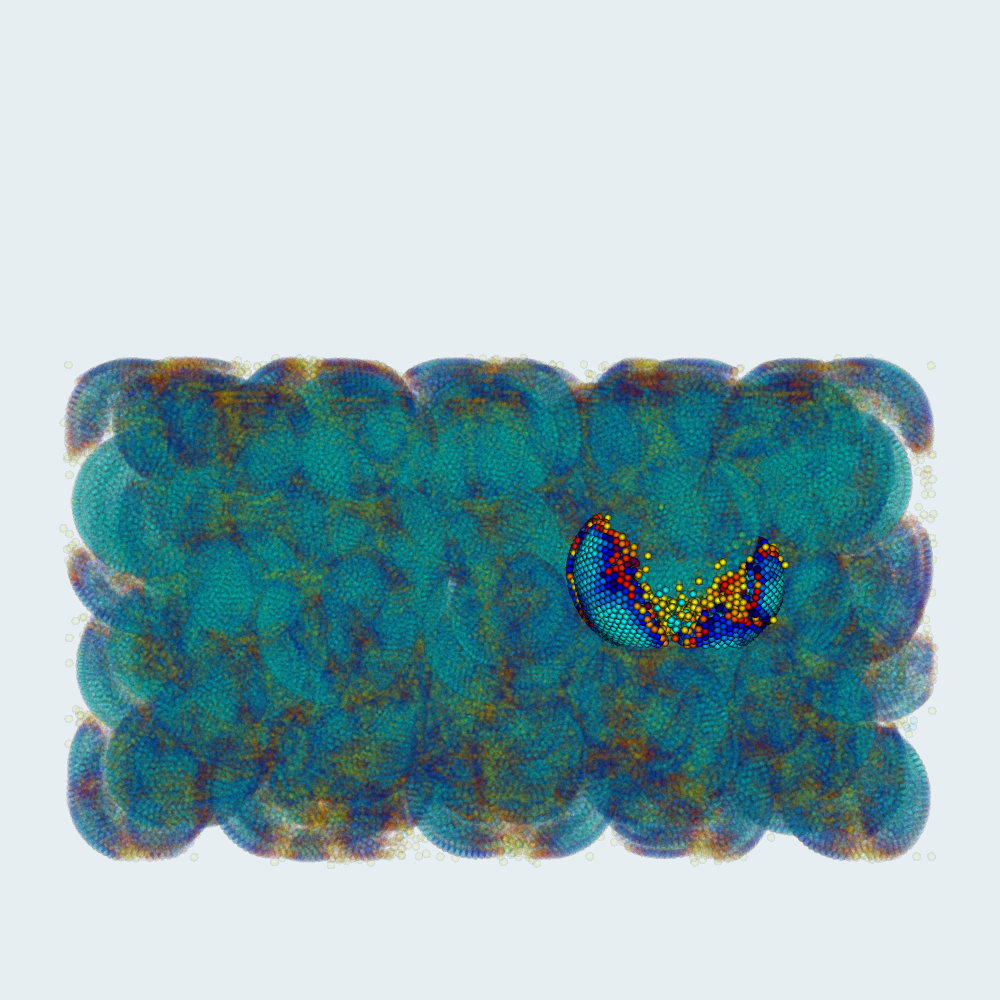}}
    \\
     \subfigure[$t=40$]{\includegraphics[width=0.3\linewidth]{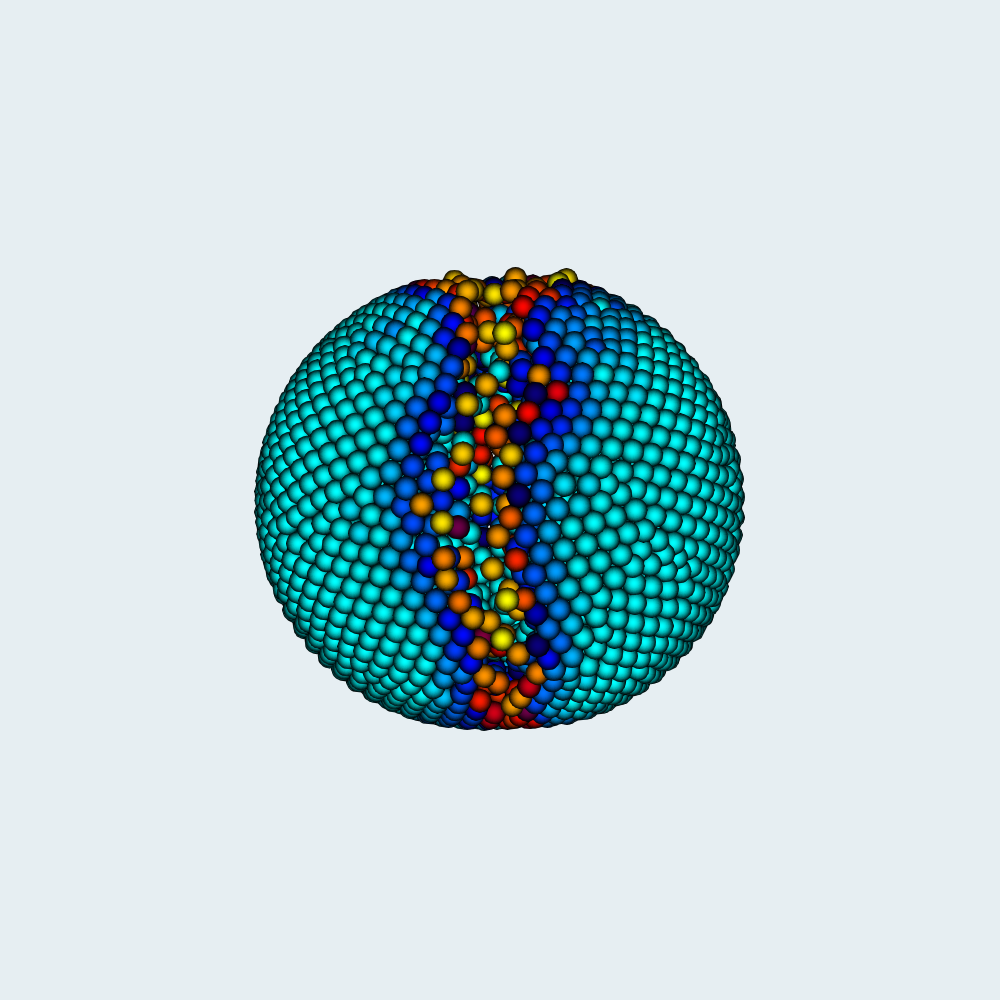}}
     \subfigure[$t=80$]{\includegraphics[width=0.3\linewidth]{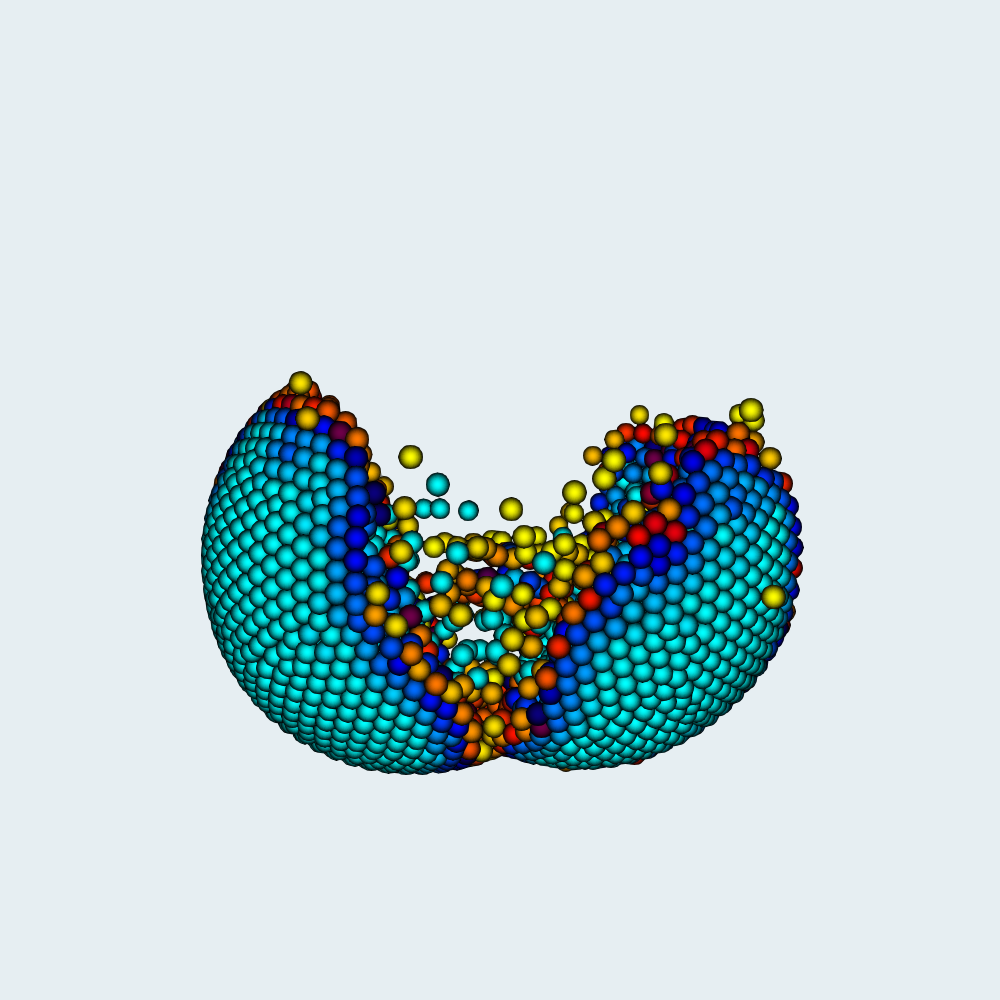}}
     \subfigure[$t=100$]{\includegraphics[width=0.3\linewidth]{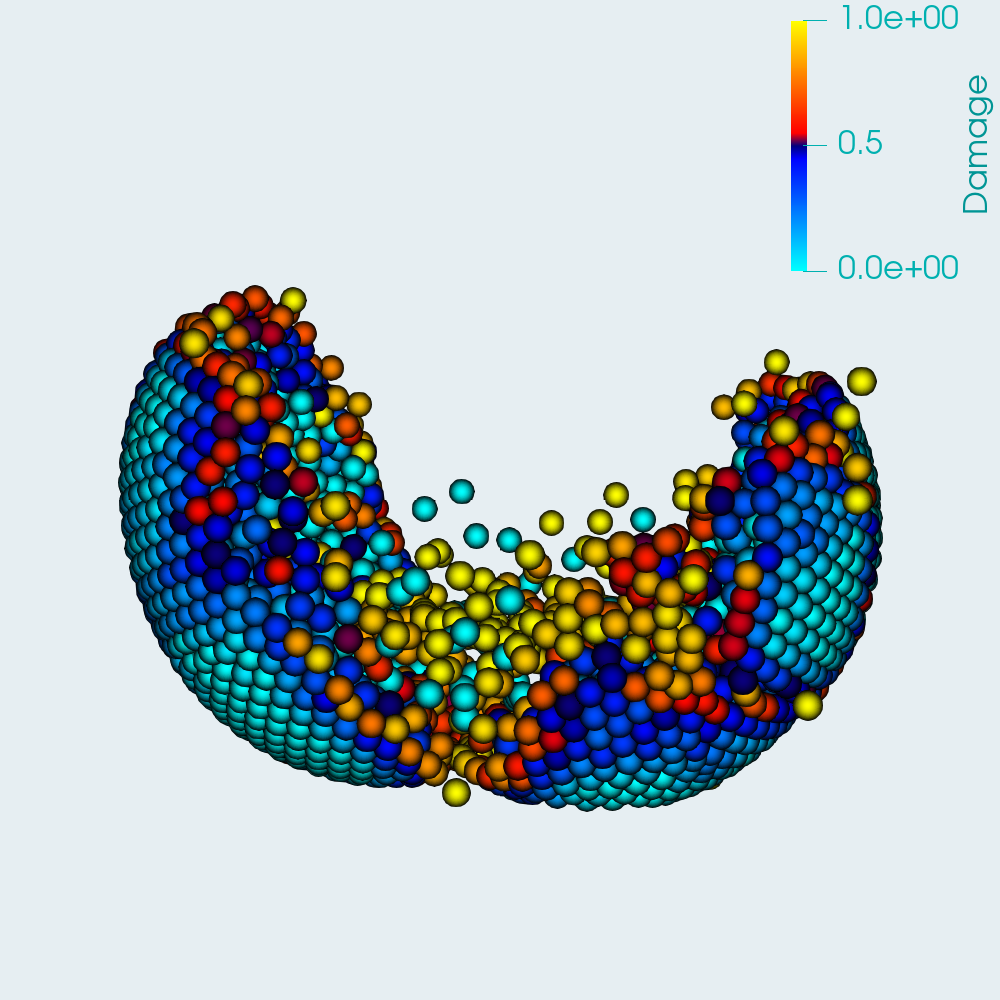}}
    \caption{ {Fracture evolution of a centrally located hollow spherical grain under dynamic compression. The top row shows the full particle assembly with transparency applied to visualize the tracked grain in the center. The bottom row isolates the tracked grain at each time step (t=40, 80, and 100, in units of $5 \mu$s), highlighting the progressive damage and crack propagation within the hollow shell.}}
    \label{fig:signle-sphere-bulk}
\end{figure}

\begin{figure}
    \centering
       {\includegraphics[width=0.3\linewidth]{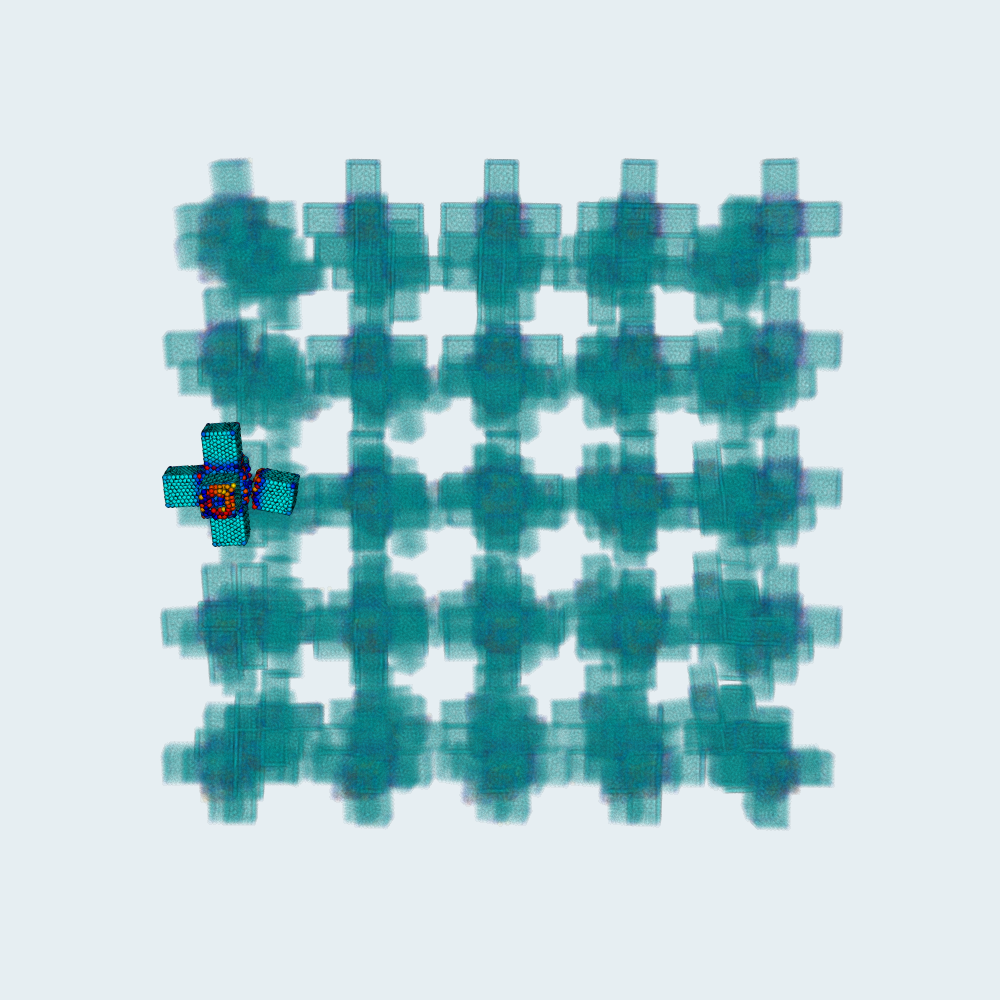}}
     {\includegraphics[width=0.3\linewidth]{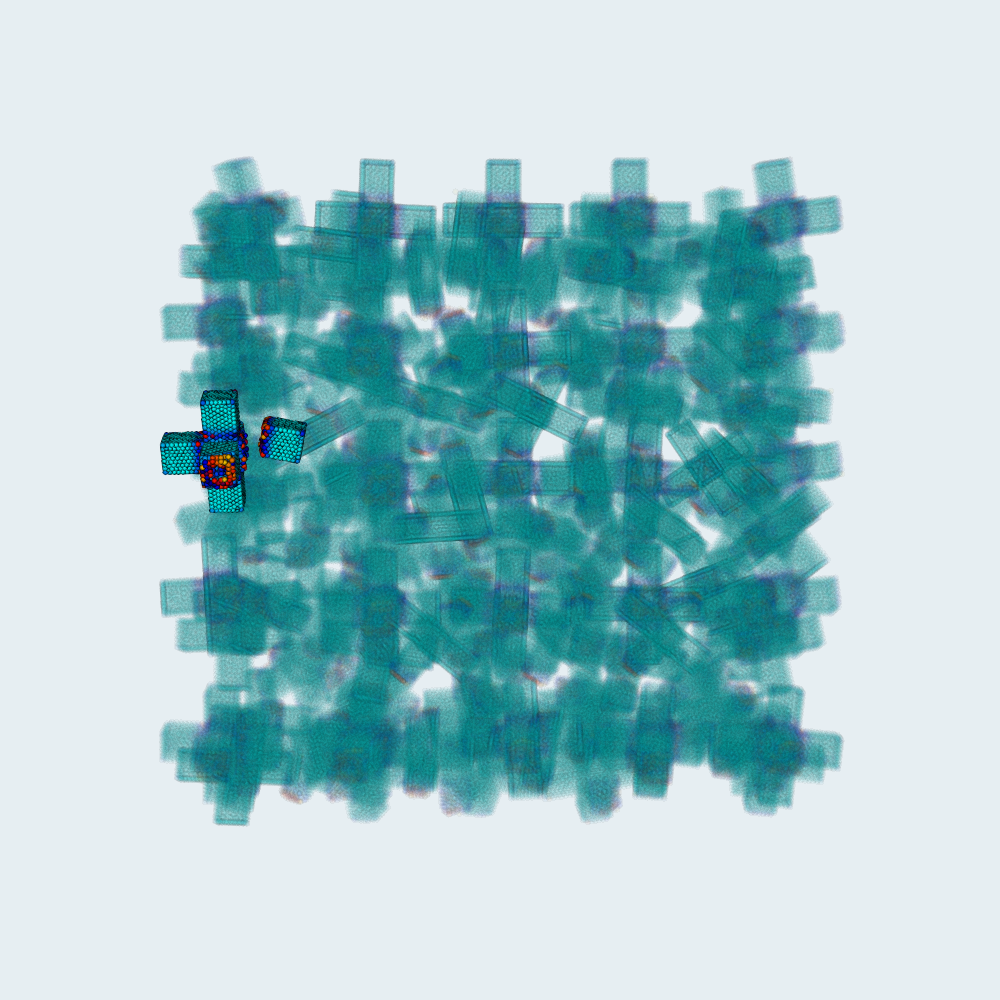}}
     {\includegraphics[width=0.3\linewidth]{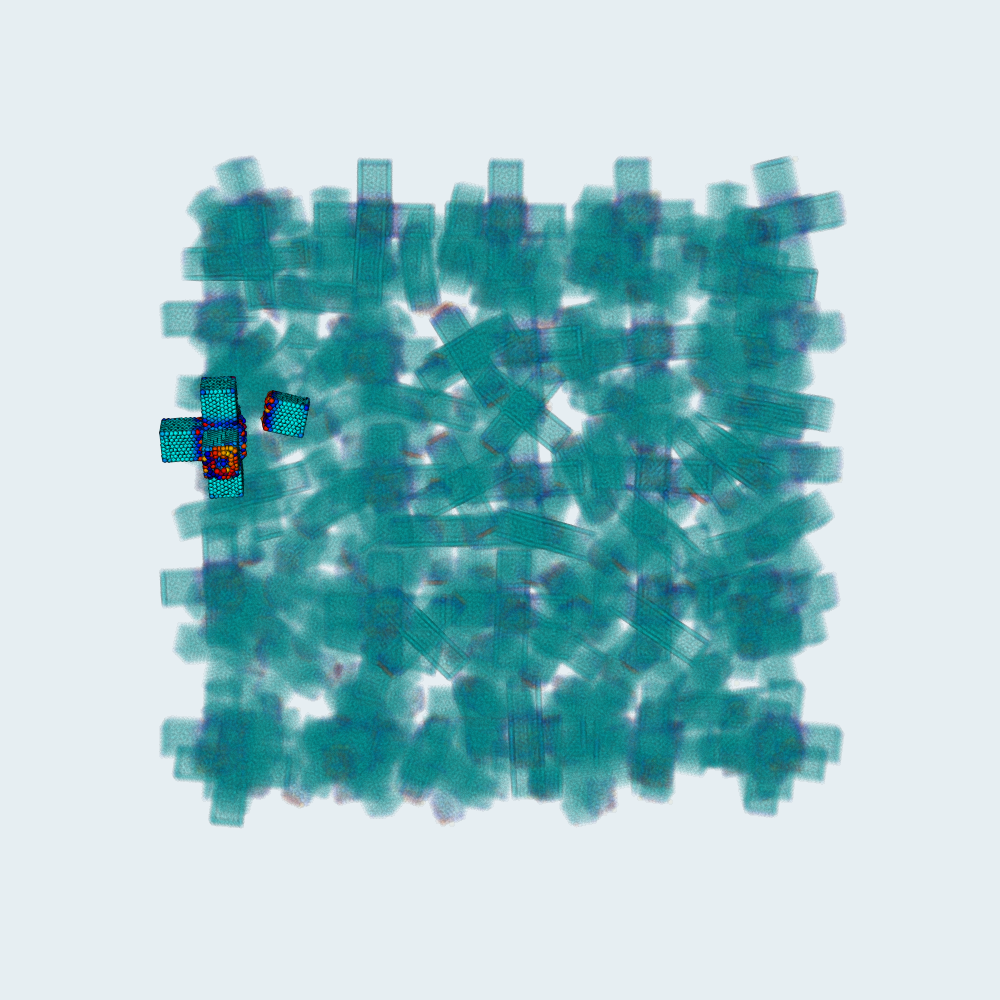}}
    \\
     \subfigure[$t=40$]{\includegraphics[width=0.3\linewidth]{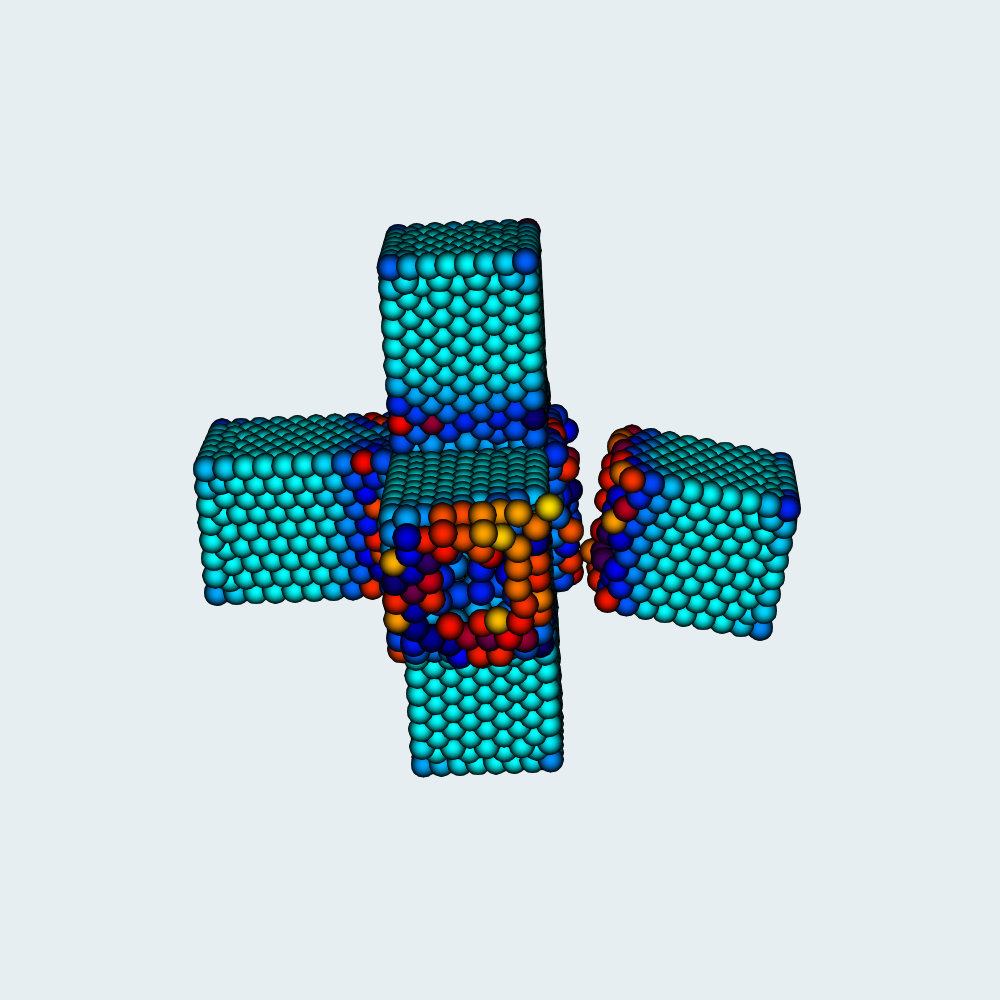}}
     \subfigure[$t=80$]{\includegraphics[width=0.3\linewidth]{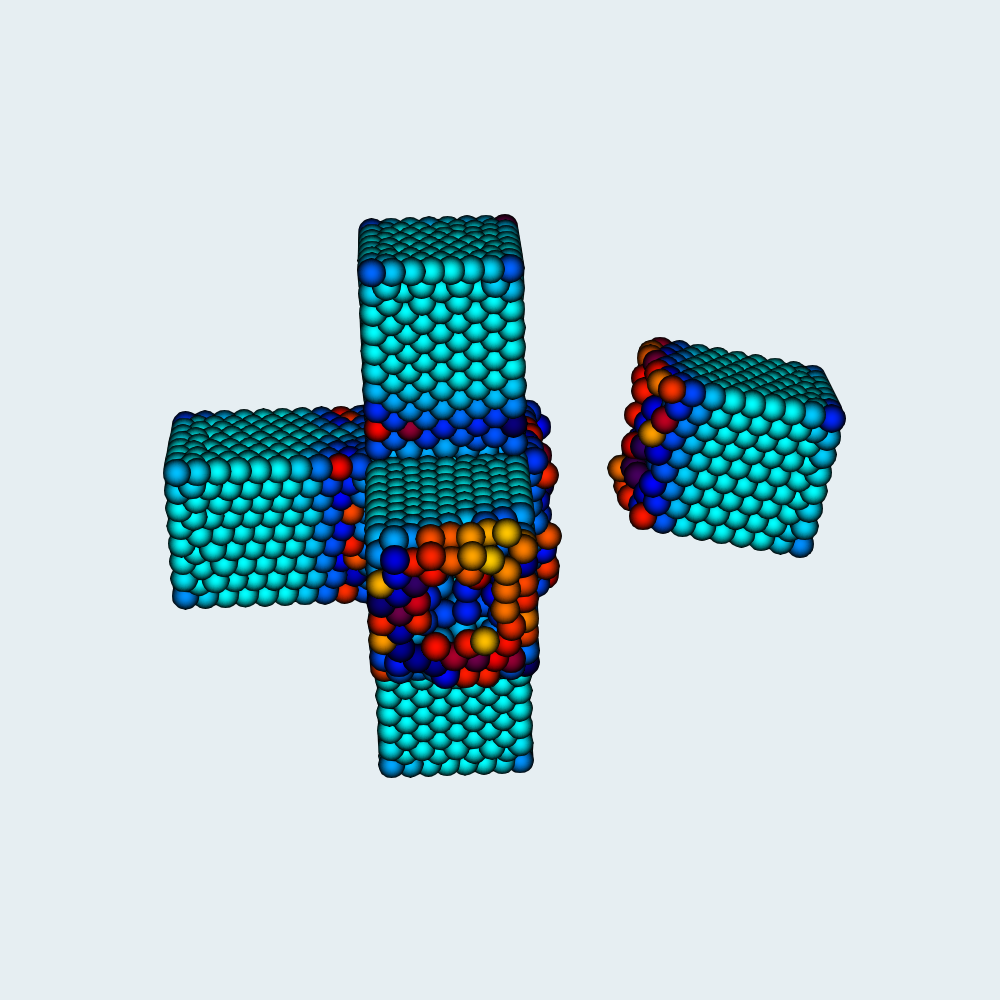}}
     \subfigure[$t=100$]{\includegraphics[width=0.3\linewidth]{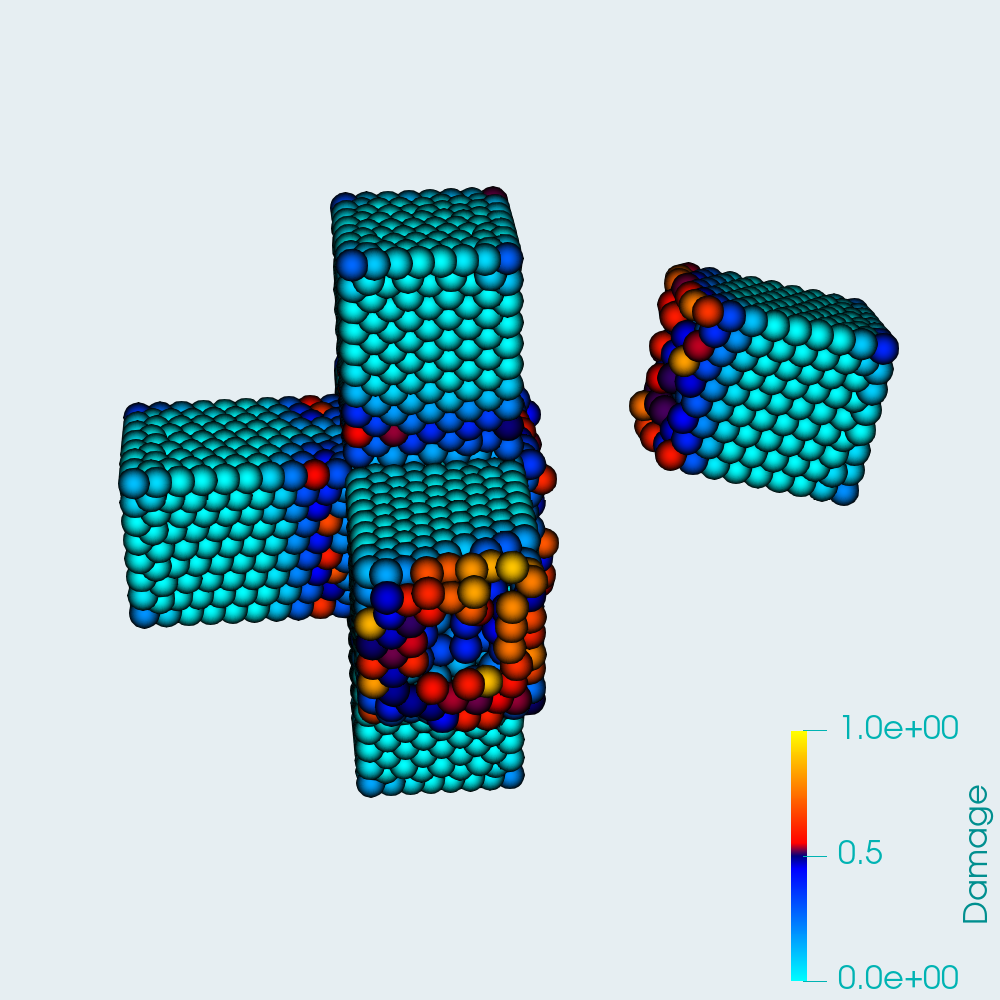}}
    \caption{ {Deformation and failure of a jack-shaped grain within a packed aggregate of identical jacks. The top row shows the grain's environment with the tracked grain visible through transparency, and the bottom row presents isolated views of the grain. The snapshots at t=40, 80, and 100 illustrate the onset of damage at the arms and eventual fragmentation along high-stress regions.}}
    \label{fig:signle-jack-bulk}
\end{figure}

\begin{figure}
    \centering
    {\includegraphics[width=0.3\linewidth]{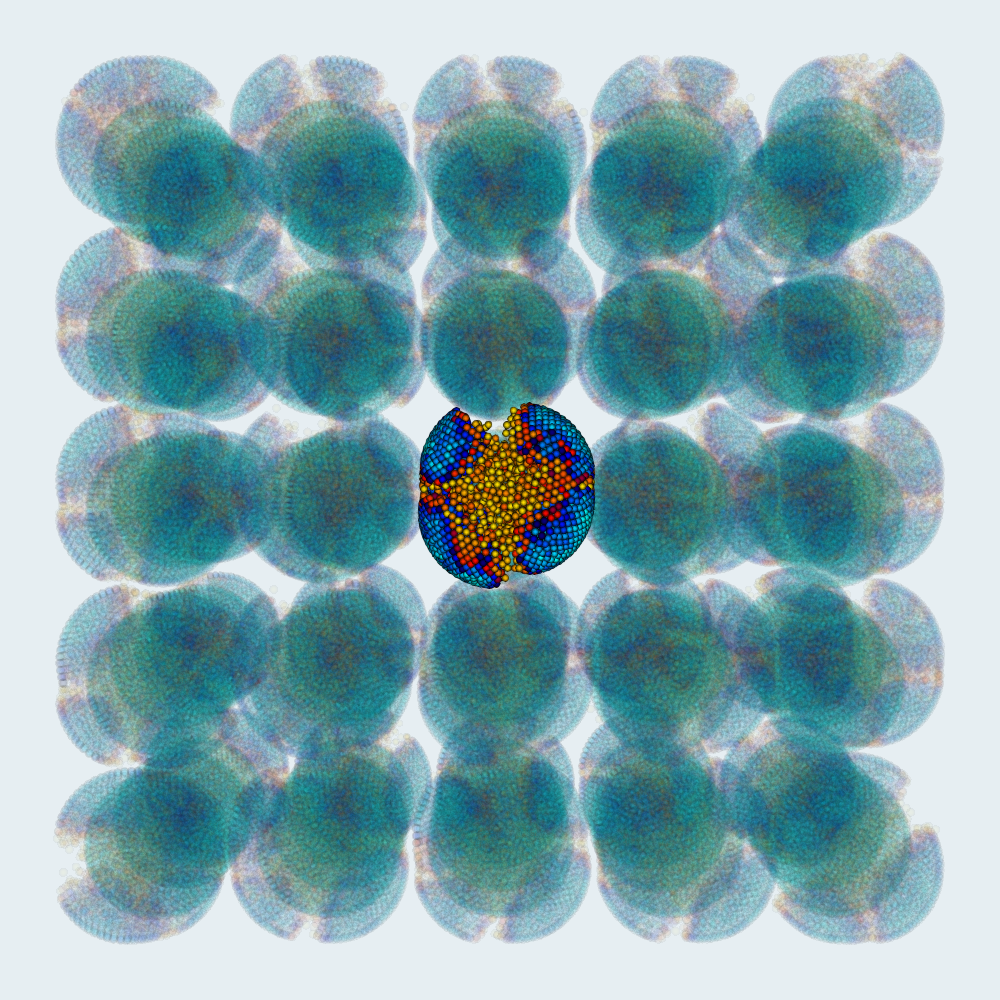}}
     {\includegraphics[width=0.3\linewidth]{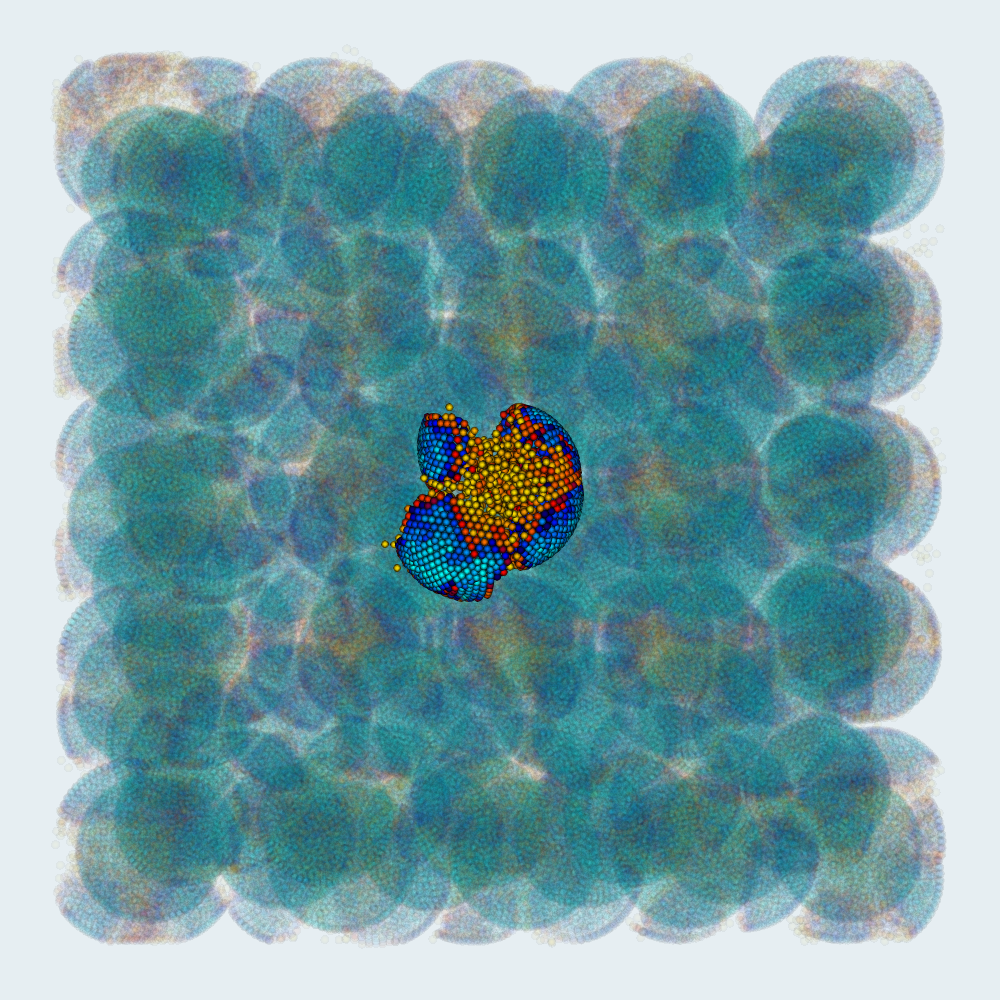}}
     {\includegraphics[width=0.3\linewidth]{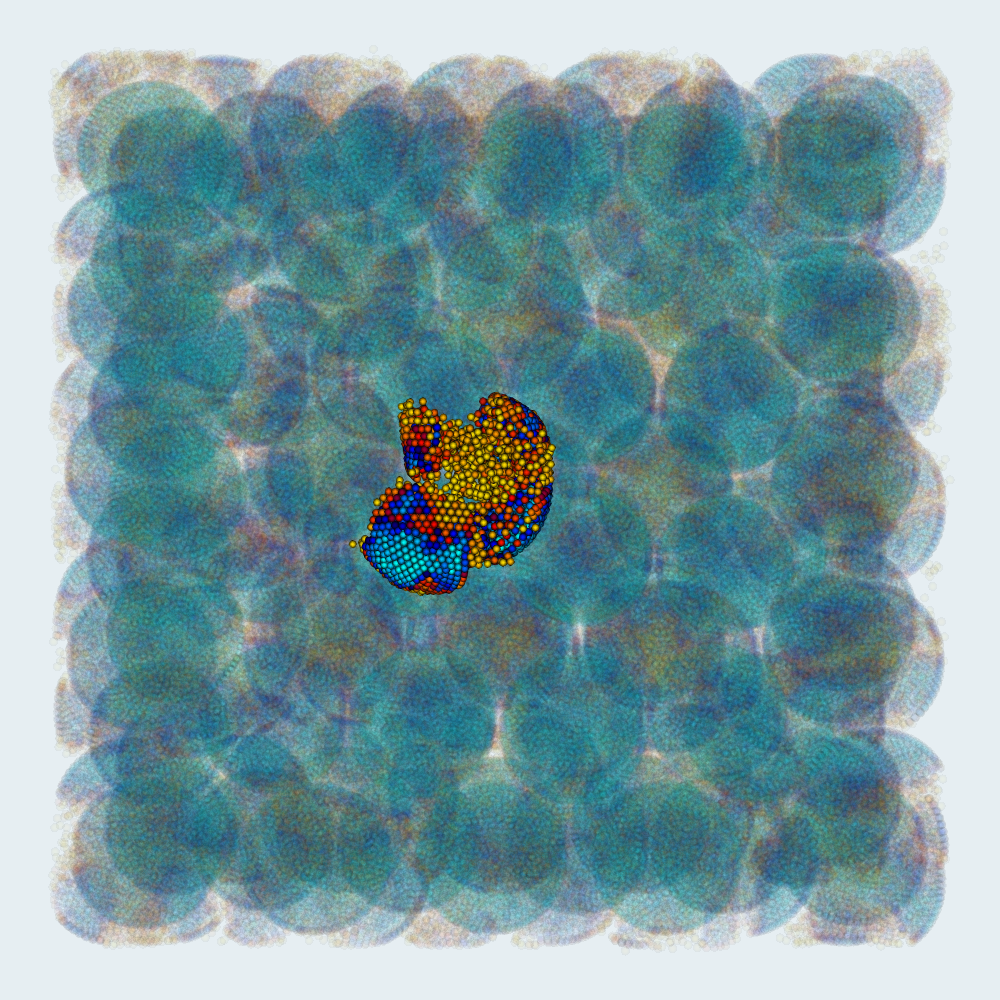}}
    \\
     \subfigure[$t=40$]{\includegraphics[width=0.3\linewidth]{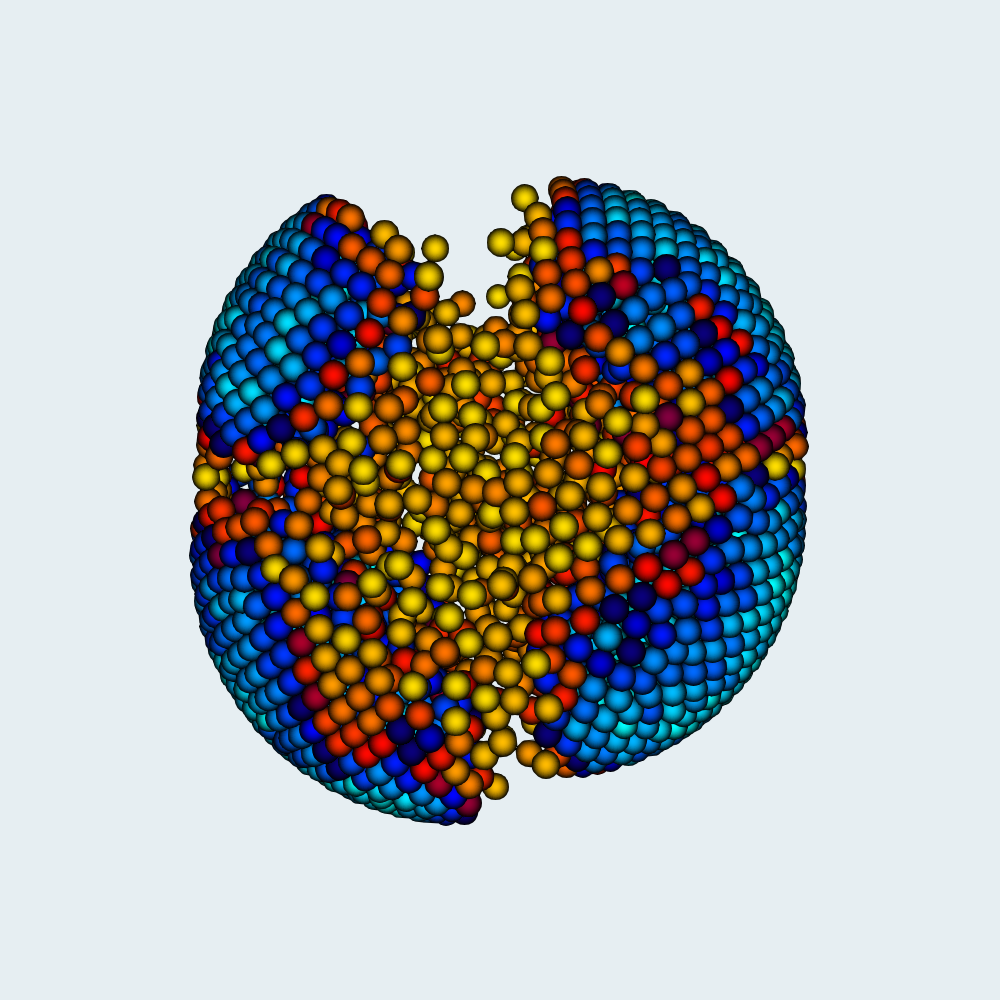}}
     \subfigure[$t=80$]{\includegraphics[width=0.3\linewidth]{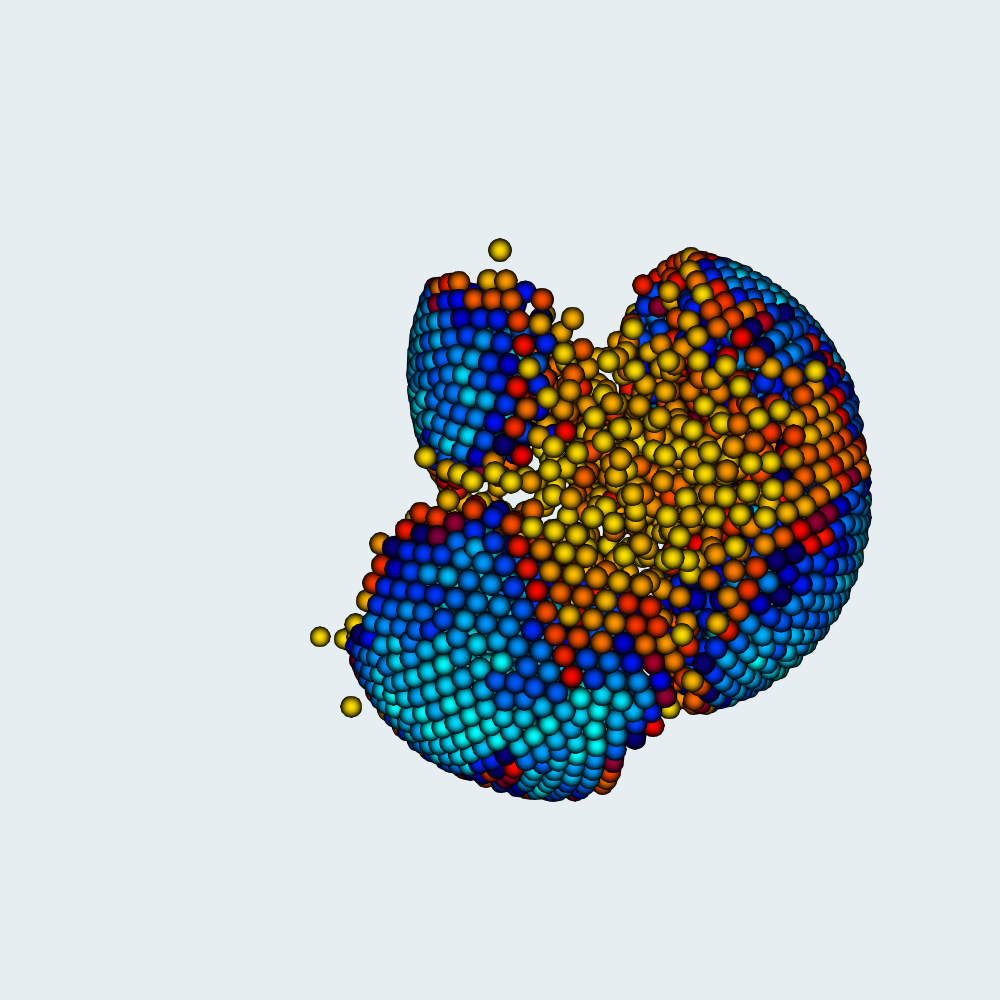}}
     \subfigure[$t=100$]{\includegraphics[width=0.3\linewidth]{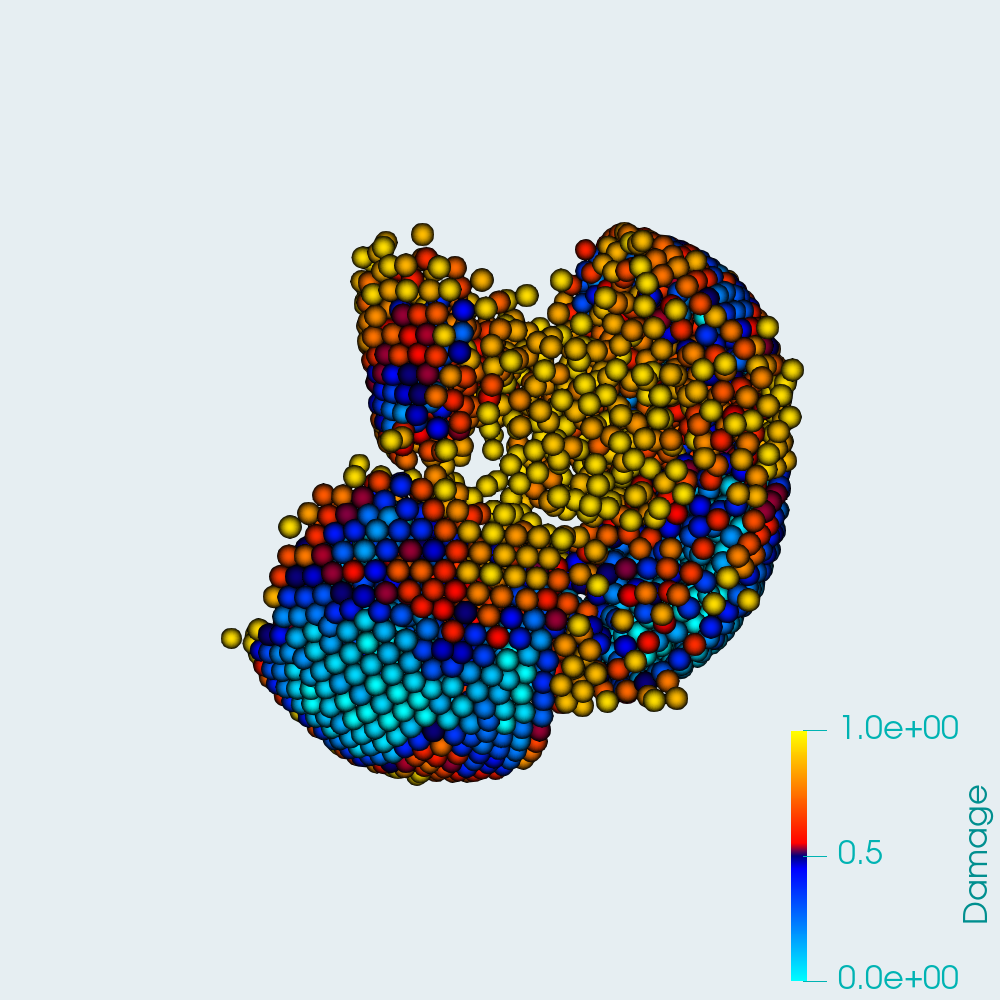}}
    \caption{ {Crack growth and fragmentation of a central solid spherical grain surrounded by identical particles under dynamic compression. Top views show the full aggregate, while the bottom row focuses on the tracked grain. The simulation captures the transition from initial stress localization to complete fracture and dispersal of fragments over time.}}
    \label{fig:signle-sphere-bulk-2}
\end{figure}

When the grains are intact, the strength of the bulk is mainly determined by the interconnected columns of grains \Cref{fig:initial}. In this regime, as compression increases, two processes become important: granular rearrangement and granular damage.
The jack-shaped particles go through a drastic rearrangement, mainly via rotation, leading to a large drop in the wall reaction force in the z-direction. However, neither the spheres nor the hollow spheres are able to exhibit this rearrangement behavior due to their radial symmetry. However, the hollow spheres (shells) exhibit less bulk strength compared to solid spheres since the excluded volume within shells become available for particle fragments when damage sets in. The spheres exhibit the most bulk strength when particles are fully confined.
Increased compression leads to broken particle fragments, which fill up the container, and breaks further. For sufficiently fragmented particle aggregate, the bulk strength is determined by the volume fraction of the particles within the container, which is observed near the tail of the plot.

A main advantage of our modeling approach is the visualization capability of individual particles within the bulk. This allows us to investigate the complex mechanism in which a particle deforms and breaks within the aggregate due to contact forces from its neighbors and particle fragments. To demonstrate this,  {in Fig.~\ref{fig:signle-sphere-bulk}, \ref{fig:signle-jack-bulk}, and \ref{fig:signle-sphere-bulk-2} }we highlight a few particles within the bulk.
For jack-shaped particles, most damage is initiated near the sharp corners. However, extreme deformation of the long arms connected to the center lead to the concentration of damage near the junction of the arms to the center. During the early onset of the damage, the distinction between the fracture modes of the spherical particles and the jacks is notable. Due to the lack of symmetry, the jacks pack in a `random' arrangement and exhibit a variety of damage modes whereas the spheres and shells pack in a square lattice, exhibiting a damage path that runs between the two contact points with neighbors (the north and south poles). The initial damage patterns in spheres and shells are similar to \ref{double-hollow-sphere}, where they break into two, three, four disjoint segments.

\section{ {Bulk-Scale Simulation of Hollow Granular Material}}
\label{sec:hollow}
 {We perform a set of PeriDEM simulations using hollow spherical particles to investigate the effect of internal voids on the collective crushing response. The setup mirrors the configuration used for the solid particle experiments: a cubic container is used for each case, and its size is adjusted according to the number of particles to maintain a comparable initial packing density. Rigid body motions are applied to each hollow grain to generate the initial packing configuration, and the top wall of the container moves downward at a constant velocity of 10m/s to induce compression. All particles are modeled as hollow spheres with an inner radius of $r=0.7R$, where R is the outer radius. Simulations are conducted for aggregates consisting of 400, 600, 800, and 1000 hollow particles.}

\begin{figure}[htbp]
    \centering
    \subfigure[$t=80$]{
        \includegraphics[width=0.45\textwidth]{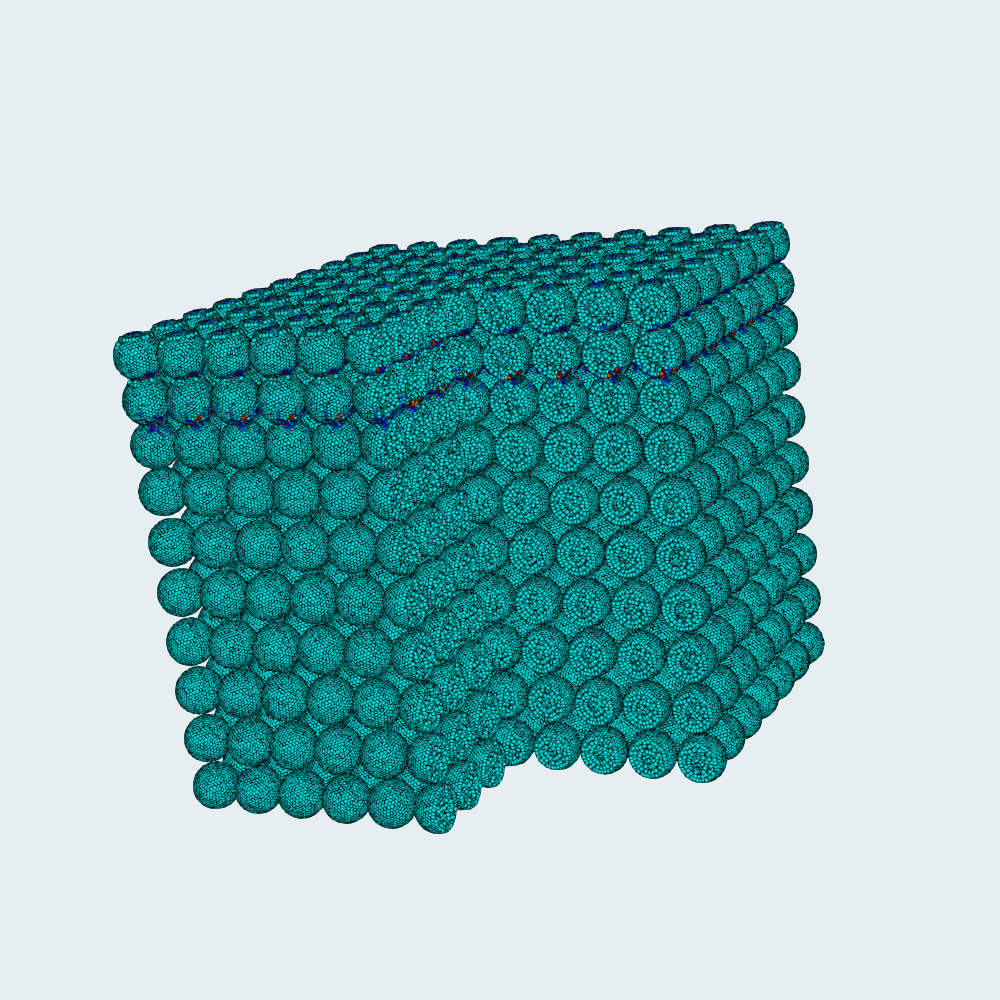}
    }
    \subfigure[$t=300$]{
        \includegraphics[width=0.45\textwidth]{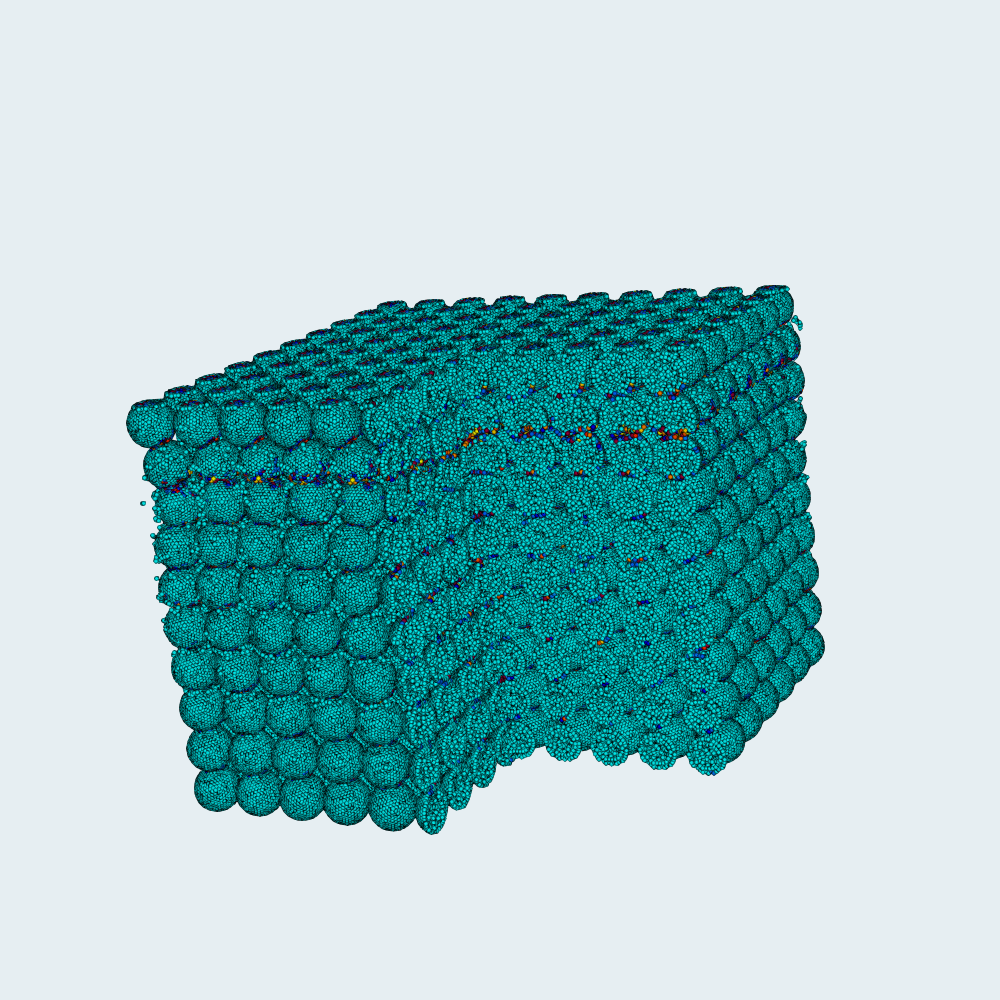}
    }
    \subfigure[$t=600$]{
        \includegraphics[width=0.45\textwidth]{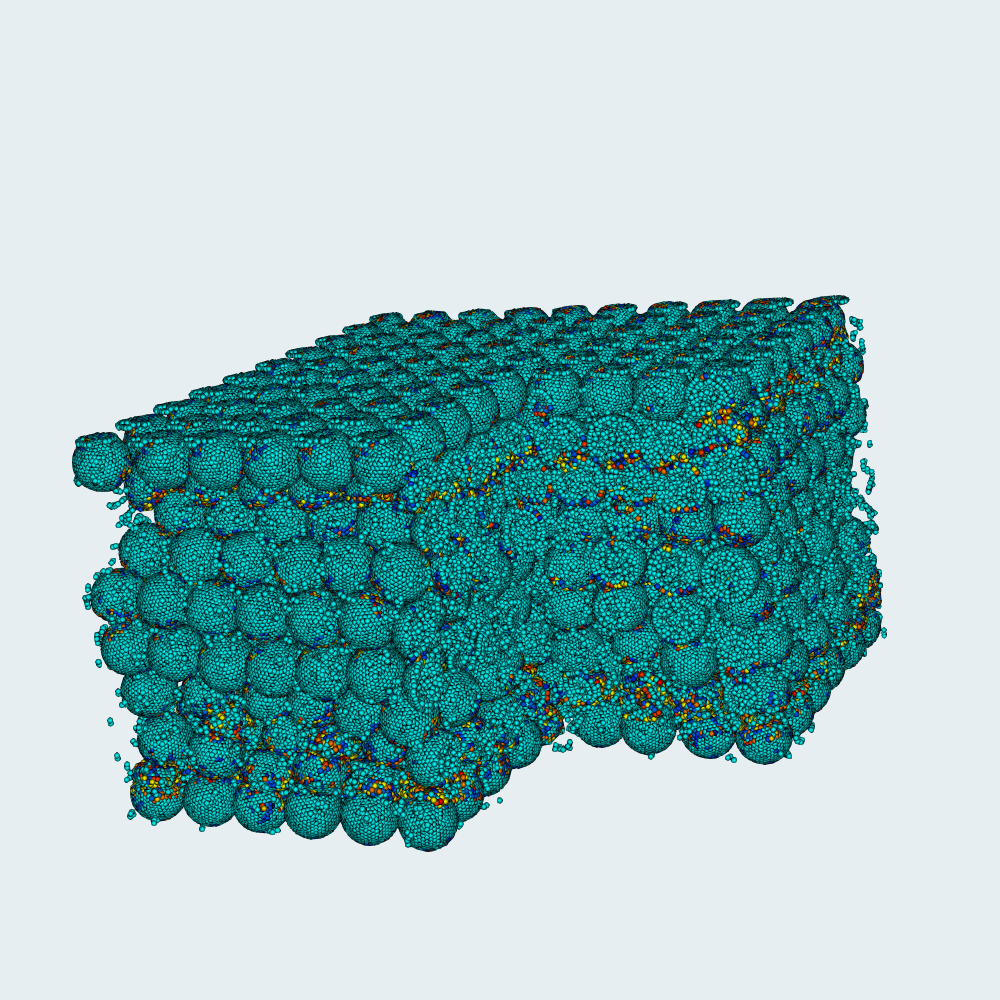}
        }
    \subfigure[$t=800$]{
        \includegraphics[width=0.45\textwidth]{{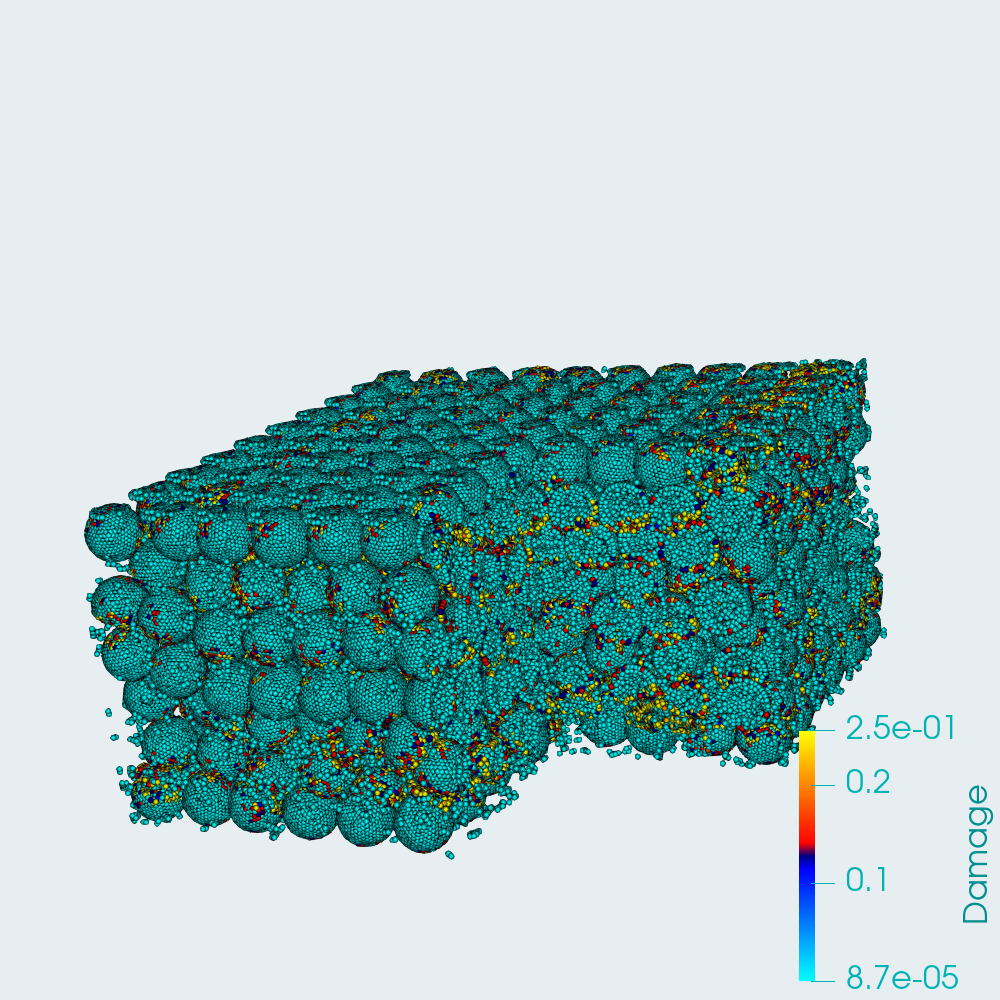}}
    }
    \caption{ {Snapshots of the PeriDEM simulation showing the crushing of 1000 hollow spherical particles with inner radius r=0.7R at time steps (a) t=80, (b) t=300, (c) t=600, and (d) t=800. The plots illustrate the progressive damage and fragmentation of the particle assembly under compression, with breakage initiating near contact regions and propagating inward as the loading continues.}
   }\label{fig:1000-hollow-timesteps}
  
\end{figure}
 {Fig.~\ref{fig:1000-hollow-timesteps} shows the time evolution of a PeriDEM simulation involving 1000 hollow spherical particles with inner radius r=0.7R, subjected to compression in a cubic container. The four subfigures display the state of the system at time steps (a) t=80, (b) t=300, (c) t=600, and (d) t=800. Initially, the particles are arranged in a densely packed configuration, as seen in (a), with no visible damage. As the simulation progresses, damage initiates and propagates primarily near the top and bottom contact regions due to the applied compressive loading. By t=300, fragmentation begins to appear, and by t=600, a significant portion of the internal structure shows signs of breakage. At t=800, extensive damage is observed throughout the aggregate, with widespread particle fragmentation and the collapse of internal voids, highlighting the progressive crushing behavior of the hollow particle assembly under sustained loading.}
\begin{figure}
    \centering
      \subfigure{\includegraphics[width=0.8\linewidth]{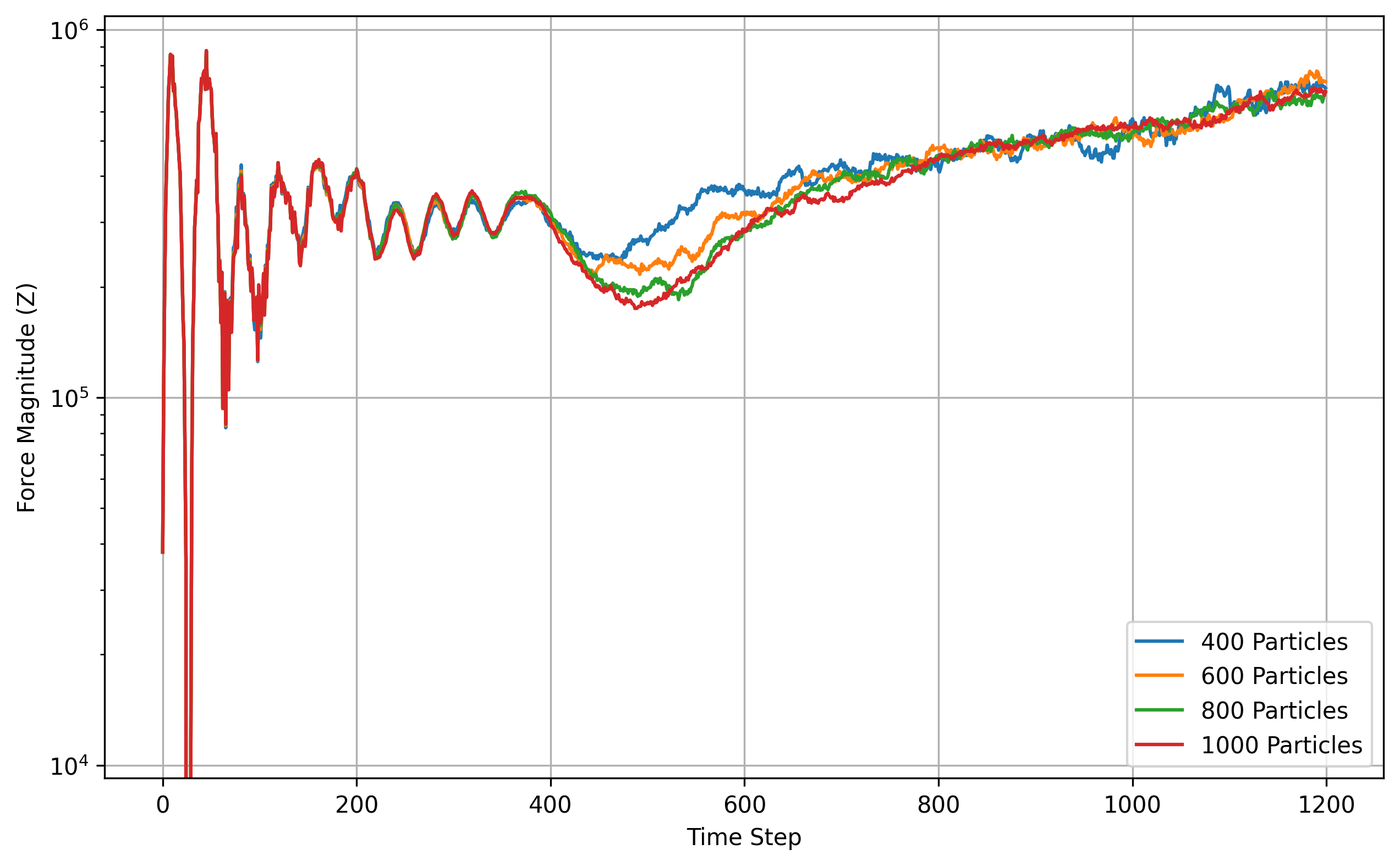}}
        \caption{ {: Time evolution of the normalized compressive force on the top wall during PeriDEM simulations of hollow spherical particle assemblies with inner radius r=0.7R. The force is normalized by the top surface area of the container to allow comparison across different particle counts. Despite differences in particle number (400, 600, 800, and 1000), the force curves gradually converge over time, indicating that the hollow geometry limits the load-bearing capacity, and increasing particle count yields diminishing gains in resistance.}}
    \label{fig:force-hollow}
\end{figure}

 {To evaluate the compressive response of hollow sphere aggregates, we measure the reaction force on the top wall during PeriDEM simulations of systems containing 400, 600, 800, and 1000 particles. All particles have a hollow structure with an inner radius of Rr=0.7R, and the simulation setup matches that of the solid particle experiments: particles are confined in a cubic container with adaptive size based on particle count, and the top wall is displaced downward at a constant velocity of 10m/s. The raw force data recorded on the top wall is normalized by the top surface area of the container to ensure consistent comparison across different system sizes. The plot in Fig.~\ref{fig:force-hollow} shows the resulting normalized force magnitude as a function of time step for each configuration.}

 {The normalized force plot in Fig.~\ref{fig:force-hollow} reveals the evolution of compressive resistance for hollow sphere aggregates with varying particle counts. Initially, all configurations exhibit rapid oscillations in force as the particles establish contact and begin to rearrange under the applied loading. Compared to the solid particle case, these oscillations are more pronounced, especially for the 1000-particle system, reflecting the structural complexity and reduced stiffness of the hollow grains. After the initial transients, the force responses stabilize and then gradually increase, indicating densification and compaction of the aggregate. Interestingly, unlike the solid particle simulations, the differences between samples of aggregate made up of different particle counts are less significant over time, and the normalized force curves tend to converge. This suggests that the hollow structure limits the overall load-bearing capacity, and remains roughly the same beyond a certain point, so increasing the number of grains yields diminishing returns in terms of collective resistance.}

\section{Scaling Analysis of PeriDEM Simulations}\label{sec:scale}

To evaluate the performance and scalability of the PeriDEM framework, we conduct a set of simulations using hollow spherical particles. The goal is to assess both the computational behavior on moderate-sized configurations and the viability of scaling to large core counts. This section is organized as follows: we first describe the computational resources, then present strong and weak scaling studies conducted on up to 80 cores, and finally analyze large-scale runs on up to 1600 cores.

All simulations were performed on the {SuperMIC} cluster at Louisiana State University. SuperMIC is a high-performance computing system composed of 360 compute nodes, each equipped with two 10-core Intel Xeon 64-bit processors running at 2.8 GHz, and 64 GB of memory per node. Every node is also accelerated with an Intel Xeon Phi 7120P coprocessor. The system runs on Red Hat Enterprise Linux v6 and delivers a peak performance of 1000 TFLOPS. Its scalable architecture and high-throughput capabilities make it well-suited for large-scale PeriDEM simulations.
\subsection{Efficiency Metrics}

To quantify the performance of parallel simulations, we report two key efficiency metrics, {CPU Efficiency}, and \textit{Memory Efficiency}. CPU efficiency measures how effectively the allocated CPU cores were utilized throughout the job duration. It is defined as the ratio of the total CPU time used to the maximum possible CPU time available:

\begin{equation}
\text{CPU Efficiency (\%)} = \left( \frac{\text{CPU Time Used}}{\text{Core Count} \times \text{Wall Time}} \right) \times 100
\end{equation}

\noindent
where CPU Time Used is the cumulative compute time across all cores, Core Count is the total number of allocated cores, and Wall Time is the actual runtime of the job. Memory efficiency quantifies the proportion of allocated memory that was actually used during the simulation:

\begin{equation}
\text{Memory Efficiency (\%)} = \left( \frac{\text{Memory Used}}{\text{Memory Allocated}} \right) \times 100
\end{equation}

\noindent
where Memory Used is the peak or average memory consumed, and Memory Allocated is the total available memory based on the number of nodes and memory per node (e.g., 64 GB per node on SuperMIC). These metrics help identify resource utilization patterns and guide the optimization of computational setups for future simulations.

\subsection{Scaling Experiments on SuperMIC (Up to 80 Cores)}

Before launching large-scale production runs, we carried out controlled strong and weak scaling studies using up to 4 nodes (80 cores). These experiments help assess baseline efficiency and inform performance expectations at larger scales.
The experiment set up consists of 225 cubic particles, which are allowed to fall on a plane surface under gravity and come to equilibrium. 
Each particle consists of 1,211 peridynamic nodes, leading to a total of 272,475 peridynamic nodes with three degrees of freedom and a total of 5.36 million peridynamic bonds.

\begin{figure}[htbp]
    \centering
    {
        \includegraphics[width=0.45\textwidth]{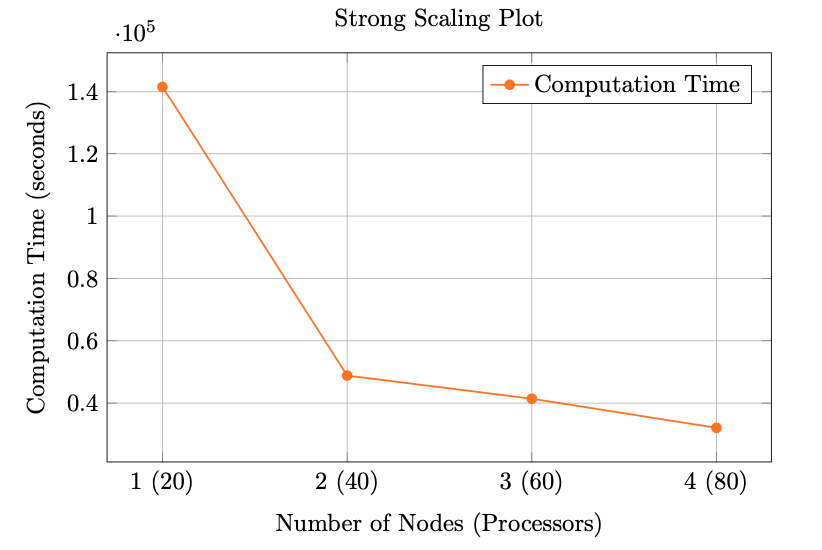}
    }
    {
        \includegraphics[width=0.45\textwidth]{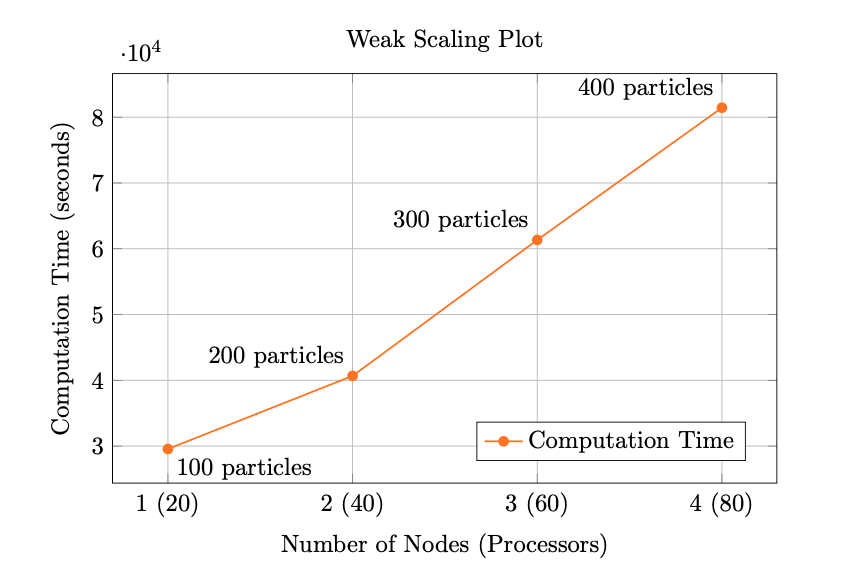}
    }
    \caption{Scaling plots comparing strong and weak scaling performance. Computation time is plotted against number of nodes (with total processors in parentheses).}
    \label{fig:scaling-plots}
\end{figure}

First, we fix the problem size and increase the number of compute nodes to obtain a strong scaling relation and evaluate how wall-clock time and CPU efficiency improve. As shown in Table~\ref{tab:strong-scaling} and Figure~\ref{fig:scaling-plots}, the wall-clock time decreases significantly, while CPU efficiency improves from 49.9\% at 20 cores to over 98\% at 40 cores and beyond.

Next, we focus on the weak scaling relation. We proportionally increase both the number of particles and the number of cores to maintain a constant workload per core. Table~\ref{tab:weak-scaling} and Figure~\ref{fig:scaling-plots} show that wall-clock time increases sub-linearly, and CPU efficiency remains high across all configurations.

\begin{table}[htbp]
\centering
\caption{Strong scaling results for a fixed-size PeriDEM simulation.}
\resizebox{\textwidth}{!}{\begin{tabular}{ccccccc}
\hline
\textbf{Cores} & \textbf{Nodes}  & \textbf{Wall Time (hrs)} & \textbf{CPU Time (hrs)} & \textbf{CPU Eff. (\%)} & \textbf{Memory (GB)} & \textbf{Mem. Eff. (\%)} \\
\hline
20 & 1 &  36.98 & 369.31 & 49.93 & 5.10 & 8.42 \\
40 & 2 &  13.57 & 536.32 & 98.83 & 5.07 & 4.19 \\
60 & 3 &  11.50 & 680.47 & 98.62 & 4.92 & 2.71 \\
80 & 4 &  8.91  & 699.07 & 98.11 & 5.06 & 2.09 \\
\hline
\end{tabular}}
\label{tab:strong-scaling}
\end{table}

\begin{table}[htbp]
\centering
\caption{Weak scaling results for PeriDEM simulations with increasing problem sizes.}
\resizebox{\textwidth}{!}{\begin{tabular}{cccccccc}
\hline
\textbf{Particles} & \textbf{Nodes} & \textbf{Cores} & \textbf{Wall Time (hrs)} & \textbf{CPU Time (hrs)} & \textbf{CPU Eff. (\%)} & \textbf{Memory (GB)} & \textbf{Mem. Eff. (\%)} \\
\hline
100 & 1 & 20 &  8.21 & 163.02 & 99.27 & 3.68 & 6.08 \\
200 & 2 & 40 &  11.30 & 450.61 & 98.83 & 4.59 & 3.79 \\
300 & 3 & 60 &  17.04 & 1009.30 & 98.73 & 6.09 & 3.35 \\
400 & 4 & 80 &  22.62 & 1774.39 & 98.04 & 6.01 & 2.48 \\
\hline
\end{tabular}}
\label{tab:weak-scaling}
\end{table}

These small-scale studies demonstrate the high parallel efficiency of PeriDEM even at modest node counts. This provides a robust foundation for justifying and analyzing large-scale production runs described below.

\subsection{Large-Scale Scaling Analysis (Up to 1600 Cores)}
To study the bulk mechanical behavior of granular aggregates under compression, we use the numerical experiment described in \Cref{sec:hollow}, where we consider 400, 600, 800, and 1000 hollow spheres. Each hollow sphere consists of 1,333 peridynamic nodes and 3,847 bonds, which implies a total of 1.33 million peridynamic nodes with three degrees of freedom and 3.84 million peridynamic bonds in the 1600-particle simulation. These simulations were run on 60 to 80 nodes (1200–1600 cores), and performance metrics are summarized in Table~\ref{tab:scaling-results}.

\begin{table}[htbp]
\centering
\caption{Summary of PeriDEM compression simulations with hollow spheres at large scale.}
\resizebox{\textwidth}{!}{\begin{tabular}{ccccccc}
\hline
\textbf{Particles} & \textbf{Cores} & \textbf{Wall Time} & \textbf{CPU Time} & \textbf{CPU Eff.} & \textbf{Mem.} & \textbf{Mem. Eff.} \\
\hline
400  & 1200 & 36:39:12 & 05:56:13 & 98.94\% & 6.03 GB & 0.17\% of 3.55 TB \\
600  & 1600 & 34:52:43 & 05:35:57 & 98.32\% & 7.70 GB & 0.16\% of 4.73 TB \\
800  & 1600 & 26:29:05 & 11:25:11 & 97.22\% & 5.68 GB & 0.12\% of 4.73 TB \\
1000 & 1600 & 33:22:22 & 05:12:25 & 97.14\% & 7.98 GB & 0.16\% of 4.73 TB \\
\hline
\end{tabular}}
\label{tab:scaling-results}
\end{table}

 We compute total core-hours by multiplying core count and wall time:

\begin{center}
\begin{tabular}{cccc}
\hline
\textbf{Particles} & \textbf{Wall Time (hrs)} & \textbf{Cores} & \textbf{Core-Hours} \\
\hline
400  & 36.65 & 1200 & 43,980 \\
600  & 34.88 & 1600 & 55,808 \\
800  & 26.48 & 1600 & 42,368 \\
1000 & 33.37 & 1600 & 53,392 \\
\hline
\end{tabular}
\end{center}

The 800-particle case is the most efficient in terms of core-hour usage. Deviations from ideal linear scaling may be attributed to memory overhead, load imbalance, or parallel I/O. All cases maintain CPU efficiencies above 97\%, underscoring strong parallel performance at scale.


\section{Conclusion}%
\label{sec:conclusion}
 {We have developed and demonstrated a robust three dimensional peridynamic framework for simulating  extreme deformations characterized by fracture, crushing, and comminution of particles resulting in  the failure of granular aggregates. By resolving both inter-particle contact and intra-particle fracture within a unified nonlocal formulation, the method overcomes limitations of classical DEM approaches. Through a series of validation and application cases—including hollow sphere crushing and large-scale aggregates reconstructed from tomography data—we show that the framework captures complex fracture patterns and converging bulk mechanical responses under compression. The results highlight the significance of particle geometry and internal structure on macroscopic strength and support the use of such simulations in constructing representative volume elements for multiscale modeling of granular materials. {Furthermore, the framework exhibits strong and weak scaling behavior on high-performance computing platforms, maintaining high parallel efficiency up to 1600 cores, thus enabling large-scale simulations necessary for statistically meaningful analysis of granular systems.}}
 
\subsection{Acknowledgements}
\label{sec:ack}
This material is based upon work supported by the U. S. Army Research Laboratory and the U. S. Army Research Office under Contract/Grant Number W911NF-19-1-0245 and W911NF-24- 2-0184.
Portions of this research were conducted with high performance computing resources provided by Louisiana State University (http://www.hpc.lsu.edu). 

\bigskip

\noindent {\bf Data availability} The code used to generate the results in this study is available at\\
 https://github.com/DavoodDamircheli/PeriDEM.
\bibliographystyle{asmems4}

\bibliography{main}

\end{document}